\documentclass[preprint,12pt]{elsarticle}
\usepackage[colorlinks,citecolor=red,urlcolor=blue,bookmarks=false,hypertexnames=true]{hyperref} 
\usepackage{amsmath,esvect}
\usepackage{bbold}
\usepackage{subcaption}
\usepackage[english]{babel}
\usepackage{csquotes}
\usepackage[pagewise]{lineno}
\journal{Computer Physics Communications}

\begin{document}

\begin{frontmatter}

\title{A Kalman Filter for track reconstruction in very large time projection chambers}

\author[inst1]{Federico Battisti}
\ead{federico.battisti@physics.ox.ac.uk}
\affiliation[inst1]{organization={University of Oxford},
            addressline={Wellington Square}, 
            city={Oxford},
            postcode={OX1 2JD}, 
            country={United Kingdom}}

\author[inst2]{Marian Ivanov}
\ead{marian.ivanov@cern.ch }
\affiliation[inst2]{organization={GSI},
            addressline={Planckstraße 1}, 
            city={Darmstadt},
            postcode={64291}, 
            country={Germany}}

\author[inst3]{Xianguo Lu}
\ead{Xianguo.Lu@warwick.ac.uk}
\affiliation[inst3]{organization={University of Warwick},
            city={Coventry},
            postcode={CV4 7AL}, 
            country={United Kingdom}}


\begin{abstract}
This study introduces a Kalman Filter tailored for homogeneous gas Time Projection Chambers (TPCs), adapted from the algorithm utilized by the ALICE experiment. In order to describe semi-circular paths in the plane perpendicular to the magnetic field, we introduce a novel mirror rotation technique into the Kalman Filter algorithm, enabling effective tracking of trajectories of varying lengths, including those with multiple circular paths within the detector, also known as ``loopers''. Demonstrated relative improvements of up to 80\% in electron momentum resolution and up to 50\% in muon and pion momentum resolution underscore the significance of this enhancement. Significant improvements in the reconstruction efficiency for relatively short low momentum ``looper'' tracks are also shown. Such advancements hold promise not only for the future of the ALICE TPC but also for neutrino high-pressure gas TPCs, where loopers become significant owing to the randomness of production points and their relatively low energies in neutrino interactions. In particular, an improvement in low energy electron reconstruction, for which the production of \enquote{looping} tracks is likely and the impact of the new algorithm is directly demonstrated, could significantly impact the quality of flux determination, which in accelerator neutrino experiments relies on the measurement of $\nu_e$ electron scatterings.
\end{abstract}

\end{frontmatter}

\section{Introduction}
\label{sec:Introduction}

The time projection chamber (TPC) has enjoyed ample success in high-energy particle physics. Since its original proposal by Nygren in 1975~\cite{NygrenTPC}, it has been utilized in various experiments and setups~\cite{ATTIE200989,Hilke:2010zz}. In a TPC, signal and track formation are achieved through the production of ionization electrons induced by the energy deposition of passing charged particles. The electrons then drift towards a sensor region in an electric field produced through an electrode plane. Subsequently, the electrons undergo multiplication through electromagnetic avalanches and are read out using technologies such as multi-wire proportional chambers (MWPCs)~\cite{Charpak:1968kd} or gas electron multipliers (GEMs)~\cite{SAULI1997531}. Additionally, the TPC is usually equipped with a magnetic field, enabling  momentum measurement by curvature and charge identification. The avalanche-induced signals provide hit coordinates in two dimensions, while the drift time provides the third.

The ALICE TPC at the LHC stands out as the most notable among those currently operational~\cite{ALICE:2008ngc}. ALICE is a nucleus-nucleus collision experiment, designed to study the physics of strongly interacting matter at extreme values of energy density and temperature. The gas TPC technology was chosen by the ALICE collaboration due to its robustness in providing charged-particle momentum measurements with good two-track separation, particle identification, and vertex determination, even at the extreme levels of occupancy reached in Pb-Pb collisions. A similar TPC, but relatively smaller, has been used by the STAR experiment at RHIC~\cite{STAR:2002eio}. Recently, ALICE has undergone a significant upgrade~\cite{ALICE:2023udb}, sparking renewed interest in TPC R\&D. 

The TPC technology is also heavily discussed in the realm of accelerator neutrino experiments, where it is typically used in the form of liquid argon TPCs~\cite{Rubbia:1977zz} as an interaction target and a tracking device, such as those employed in the Short-Baseline Neutrino program (SBN)~\cite{Machado:2019oxb} and as the Deep Underground Neutrino Experiment (DUNE) Far Detector~\cite{abi2020deep}. Alternatively, gas TPCs like those of the T2K Near Detector~\cite{T2KND280TPC:2010nnd} serve as trackers for particles produced in neutrino interactions in the upstream denser components of the detector. DUNE will include the Gaseous Argon Near Detector (ND-GAr) in its near detector complex. ND-GAr will feature a high-pressure gas TPC (HPgTPC), heavily inspired by ALICE's design~\cite{DUNE:2021tad}.

ND-GAr's TPC will have a cylindrical shape with the same dimensions of the ALICE TPC: a radius of 250 cm and a length of 500 cm. It will also incorporate the recently decommissioned MWPCs used by the ALICE experiment up to Run-3~\cite{Adolfsson_2021}, which achieved hit resolutions of approximately 1 mm~\cite{LIPPMANN2012}. However, ND-GAr will not feature an internal tracking system; instead, its central region will be filled with additional MWPCs, making it the largest gas TPC ever built. Furthermore, its gas mixture will be argon-based and maintained at a pressure of 10 atm, whereas ALICE operates at atmospheric pressure. ND-GAr's design is unique in that its TPC will have sufficient mass to provide its own sample of neutrino interactions while maintaining relatively low tracking thresholds. These characteristics will make it an ideal laboratory for studying neutrino interactions on gas, while also providing charge separation and full 4$\pi$ acceptance. ND-GAr's physics program will be centered on the reduction of systematic uncertainties in the neutrino oscillation measurement. A major source of systematics derives from the nuclear medium effects in neutrino interactions, which are less understood for heavier nuclei than carbon~\cite{Mosel:2016cwa}. ND-GAr has the potential to be extremely useful in the study of nuclear effects, using a variety of techniques, including transverse kinematic imbalance~\cite{Lu:2015hea, PhysRevC.94.015503, PhysRevC.99.055504, Cai:2019jzk, PhysRevD.102.033005}. The efficacy of these studies depends heavily on the detector's reconstruction resolution.

The Kalman Filter, an iterative Bayesian technique, facilitates estimating the state of a dynamic system by reconciling discrete measurements with predictions derived from prior knowledge of the system. Introduced by Kalman in 1960~\cite{Kalman_OG} and independently discovered by Stratonovich a year prior~\cite{stratonovich}, the technique has been the standard in TPC track fitting since its introduction by the DELPHI experiment~\cite{Kalman_app}, and remains the method with the best overall performance for most applications~\cite{RevModPhys.82.1419}. The Kalman Filter developed by the ALICE experiment for track formation and reconstruction can be considered the state of the art in the field~\cite{Ivanov:2003yr, Arslandok:2022dyb}, but it has some limitations which make its direct application to a neutrino experiment such as ND-GAr problematic.

The paper will be divided into four sections. In Sec.~\ref{sec:Principles}, we provide a general introduction to the Kalman Filter technique and present a Kalman Filter application developed for a homogeneous cylindrical gaseous TPC, which is based on and expands the track fitting algorithm developed by the ALICE experiment. Part of the code is directly taken from \texttt{AliExternalTrackParam}, the ALICE TPC Kalman Filter framework~\cite{aliroot, Belikov:1997ska, carminati2003simulation}. A limitation of the parametrization used by the ALICE experiment's Kalman Filter is that it can only follow tracks that describe at most a semicircle in the plane perpendicular to the magnetic field, introducing non-physical breaking points in the reconstruction. Using a simple mirror rotation operation, the new algorithm is capable of following the track indefinitely, especially in the case of low-energy, low-mass (i.e. low energy loss) particles which form several circular trajectories inside the detectors, also known as ``loopers''. The application of this novel technique could be particularly relevant for a neutrino experiment detector such as ND-GAr, for which particles are relatively low energy and are produced in neutrino interactions on gas at random points in the TPC volume. In Sec.~\ref{sec:Simulation}, we introduce a toy Monte Carlo simulation tool capable of generating and propagating arbitrary particle tracks in a simplified detector geometry. This tool, which will be referred to as \texttt{fastMCKalman}, has been used to develop and test the algorithm~\cite{fastMCKalman}. Two samples have been produced for this study. The first includes a spectrum of different detector characteristics and particle properties and is used to validate the algorithm across a wide range of parameter space. To analyze this sample, a recently developed interactive data visualization tool called \texttt{ROOTInteractive} has been used~\cite{RootInt}. The second sample is designed to produce performance estimates for a HPgTPC similar to ND-GAr. Finally, in Sec.~\ref{sec:Conclusions}, we discuss the results of the study and the possible application of the algorithm.

\section{The Kalman Filter}
\label{sec:Principles}

In this Section we offer a brief review of the Kalman Filter technique, specifically in the context of track fitting~\cite{Bishop1995,Kalman_app}.
Details related to ALICE's \texttt{AliExternalTrackParam} can be found in Ref.~\cite{Belikov:1997ska}. 
Track fitting consists in estimating track parameters, while filtering involves analyzing linear dynamic systems. By viewing a track in space as a dynamic system, we can utilize filtering techniques for track fitting, including Kalman Filters. This can be achieved by uniquely describing the conditions of the particle with a number of parameters grouped into a true state vector, $s^{\textrm{true}}$---a function of a suitable coordinate, $x_k$, known as the free parameter---at each trajectory point $k$, $s^{\textrm{true}}(x_k) \equiv s_k^{\textrm{true}}$. 

Assuming that the system is linear, the propagation of $s_k^{\textrm{true}}$ can be described by a linear transformation, $F_{k}$. The propagation of the system can be corrupted by inherent processes, such as multiple scattering for a charged particle moving across a medium. This random disturbance can be encapsulated in a process noise vector, $w_k$, and can affect all or only some of the state vector variables. The propagation of the system can then be written as:
\begin{equation} \label{eq:ev}
     s_k^{\textrm{true}} = F_{k-1}s_{k-1}^{\textrm{true}}+w_{k-1}.
\end{equation}

By using a detector we are able to measure some properties of the particle at specific intervals of $x_k$, where the trajectory and the detector intersect. We can encapsulate these properties in a measurement vector,  $m_k$, which is a linear combination of the properties in $s_k^{\textrm{true}}$. If the detection process is affected by noise, $m_k$ will also be corrupted by a measurement noise vector, $\epsilon_k$. The whole measurement operation can be written as :
\begin{equation}
    m_k = H_k s_k^{\textrm{true}} + \epsilon_k,
\end{equation}
where $H_k$ is a linear transformation. 

We assume that all components of $w_k$ and $\epsilon_k$ are Gaussian distributed, unbiased, and uncorrelated. The expectation values and covariances for the $k^\textrm{th}$ are defined as:
\begin{align}
    \text{E}\left[w_k\right]&=\vv{0}, \;\text{Cov}\left[w_k\right]=Q_k, \\  
    \text{E}\left[\epsilon_k\right]&=\vv{0}, \;  \text{Cov}\left[\epsilon_k\right]=R_k.
\end{align}

\begin{figure}[t]
     \centering
     \includegraphics[width=0.99\textwidth]{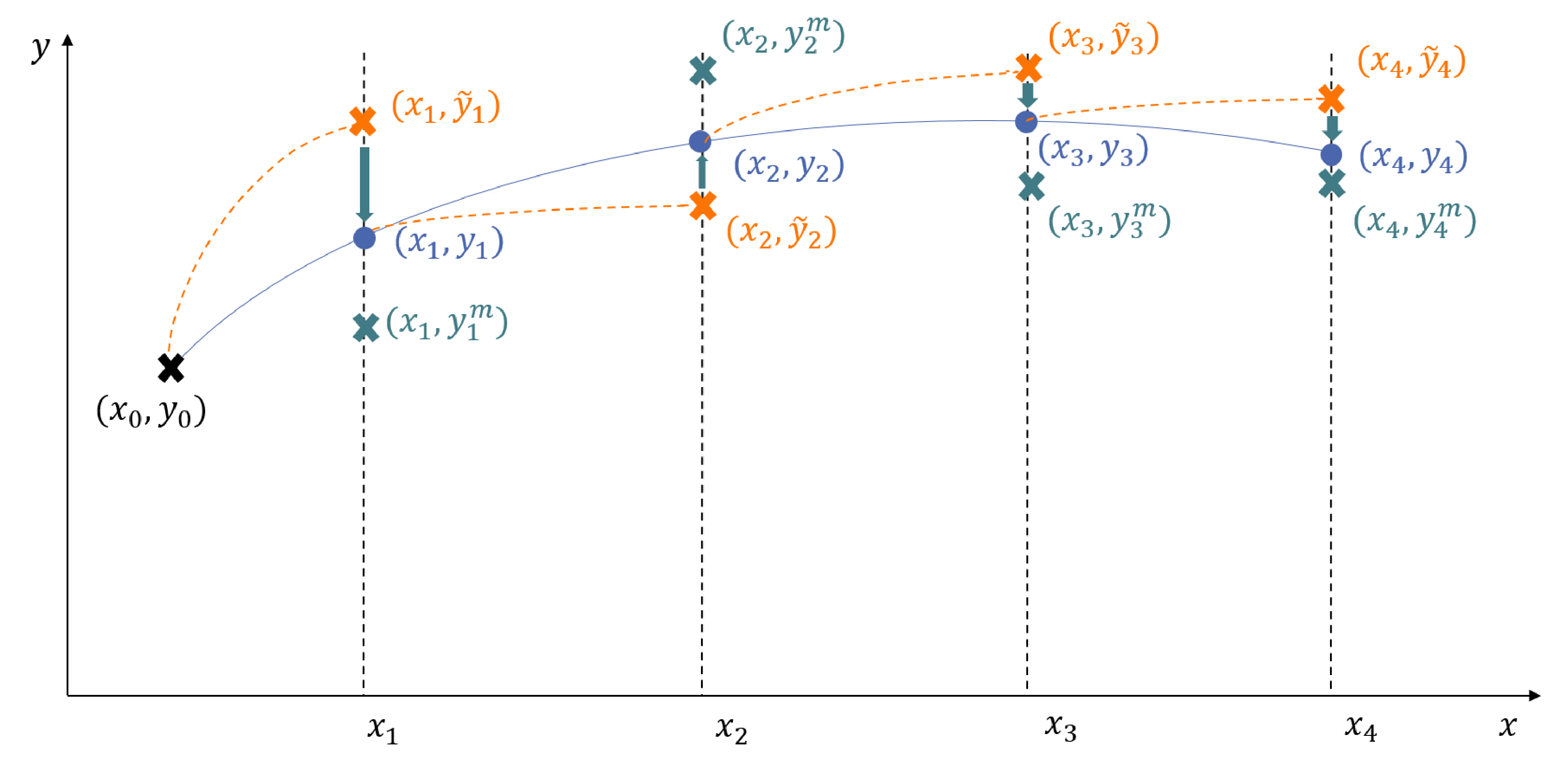}
     \caption{Schematic representation of a Kalman Filter: $y$ represents one of the variables of the state vector $s$ which is a function of the free parameter $x$. The coordinates of the free parameter $x_k$ are taken at the points of intersection between the detector and the particle trajectory. Starting from the first estimate $(x_0,y_0)$, which is obtained from a seeding algorithm, the Kalman Filter produces an \textit{a priori} estimate at the following point $(x_1,\widetilde{y}_1)$, shown in orange. The result is compared with the measurement $(x_1,y_1^\textrm{m})$ shown in green and the filtering step is applied, producing an updated estimate $(x_1,y_1)$. The procedure is repeated until no more track points are available.}
        \label{fig:KFdiagram}
\end{figure}

The Kalman Filter is a Bayesian iterative algorithm which produces an estimate,  $s_k$, of the true state vector, $s_k^{\textrm{true}}$, at each trajectory point. It combines \textit{a priori} knowledge of the system, condensed in the track propagator, $F_k$, and the measurement information from $m_k$. The covariance matrix associated with the estimated state vector, $s_k$, is defined as:
\begin{equation} \label{eq:covdef}
    \text{Cov}\left[s_k\right] = C_k.
\end{equation}
The Kalman Filter procedure can be divided into discrete operational steps, that are applied iteratively:
\begin{enumerate}
    \item Seeding: Produce an initial estimate for the state vector and covariance matrix, $s_0$ and $C_0$, respectively, using a certain technique.
    \item Propagation: Produce an \textit{a priori} estimate for the state vector and the covariance matrix, $\widetilde{s}_k$ and $\widetilde{C}_k$, respectively, at the next $x_k$ ($k\ge1$) step, using only the track propagator and no measurement knowledge:
    \begin{align}
        \widetilde{s}_k&=F_{k-1}s_{k-1},\label{eq:prop} \\
        \widetilde{C}_k&=F_{k-1}C_{k-1}F_{k-1}^T+Q_{k-1},\label{eq:pcov}
    \end{align}
    where the process noise matrix $Q_k$ is added as a correction to the covariance matrix. Note that $T$ stand for transpose.
    \item Update: Produce an updated estimate for the state vector and the covariance matrix, $s_k$  and $C_k$, respectively, using information from the current measurement. The update is performed requiring that the covariance for the new estimate is minimized. This is done through the use of the so called Kalman gain, which is defined as:
    \begin{equation}
        K_k\equiv\widetilde{C}_k H_k^T\left(R_k+H_k \widetilde{C}_k H_k^T\right)^{-1}.\label{eq:kgain}
    \end{equation}
    The update operation, also known as filtering, then follows as:
    \begin{align}
        s_k&=\widetilde{s}_k+K_k\left(m_k-H_k \widetilde{s}_k\right),\label{eq:updates}\\
        C_k&=\left(\mathbb{1}-K_kH_k\right)\widetilde{C}_k.\label{eq:updatec}
    \end{align}
 To proceed to the next point, the algorithm is repeated from the propagation step, using the current updated estimate, $s_k$, as the input. 
\end{enumerate}
An illustration of the basic functioning of the Kalman Filter algorithm is provided in Fig.~\ref{fig:KFdiagram}.

The machinery described so far, assumes that the evolution of the dynamic system is determined by linear transformations. However, the propagation of a charged particle in a magnetic field is non-linear. Equation~\ref{eq:prop} in this case takes the more general form:
\begin{equation}\label{eq:EKF}
    \widetilde{s}_k=f_{k-1}\left(s_{k-1}\right),
\end{equation}
where $f_{k-1}$ is a non-linear function. In order to apply the Kalman Filter technique, the track propagator in Eq.~\ref{eq:pcov}, $F_{k-1}$, has to be approximated by the Taylor expansion coefficient defined as follows:
\begin{align}
    f_{k-1}\left(s^*\right)& \simeq f_{k-1}\left({s}_{k-1}\right) + F_{k-1}\cdot\left(s^*-{s}_{k-1}\right), \label{eq:jacobian} \\
    F_{k-1}&=\frac{\partial f_{k-1} }{ \partial s^*},\label{eq:jacobian2}
\end{align}
where $s^*$ is a generic state vector coordinate near the point of expansion, $s_{k-1}$. All the other Kalman Filter steps (Eqs.~\ref{eq:kgain}-\ref{eq:updatec}) remain identical to the linear procedure. This technique is known as extended Kalman Filter. Every Kalman Filter discussed in this paper is an extended Kalman Filter unless stated otherwise. 

\subsection{The custom Kalman Filter}
\label{sec:Loop}

\begin{figure}[!ht]
     \centering
     \begin{subfigure}[b]{0.48\textwidth}
         \centering
         \includegraphics[width=\textwidth]{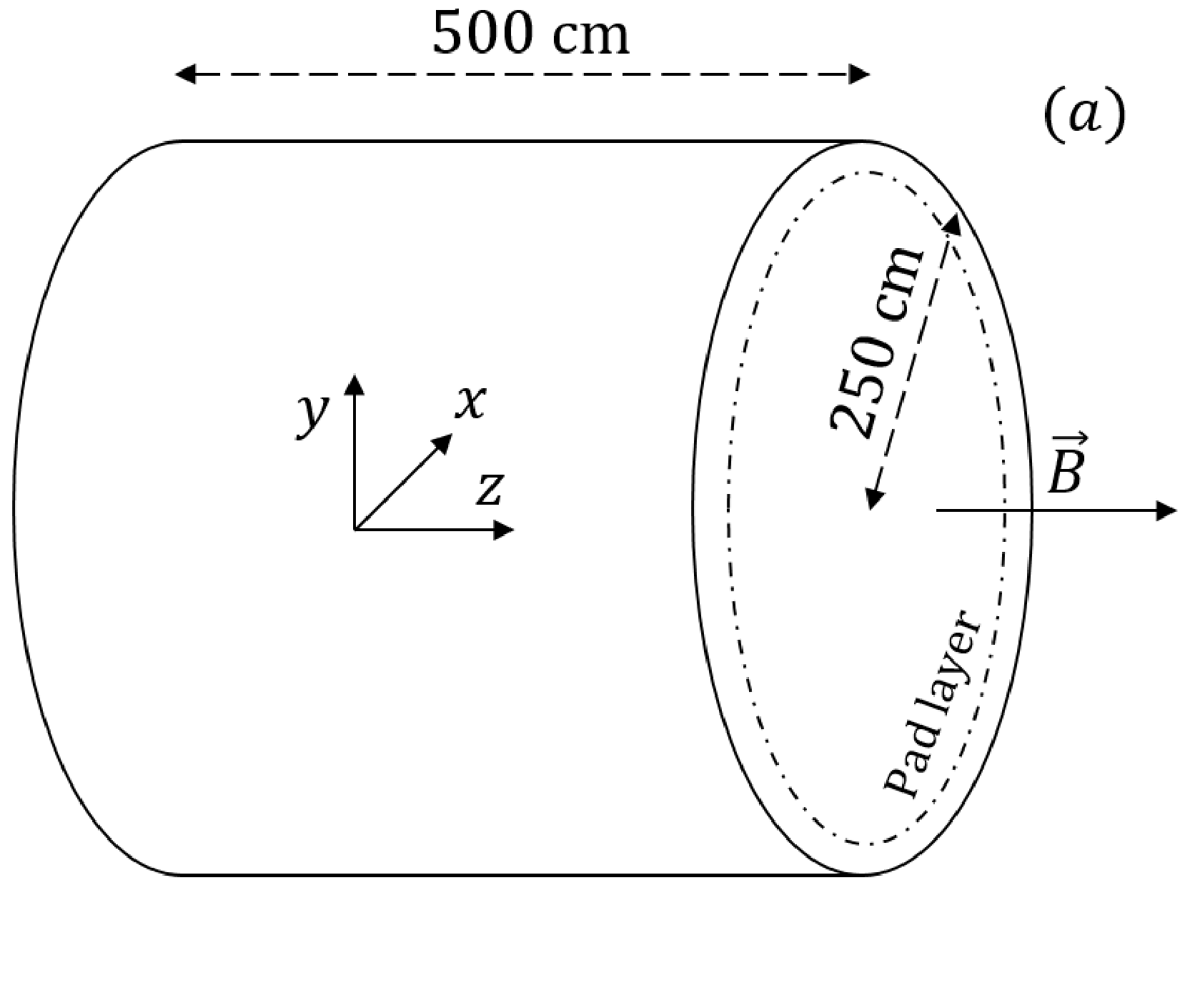}
         \caption{}
         \label{fig:Detector}
     \end{subfigure}
     \begin{subfigure}[b]{0.48\textwidth}
         \centering
         \includegraphics[width=\textwidth]{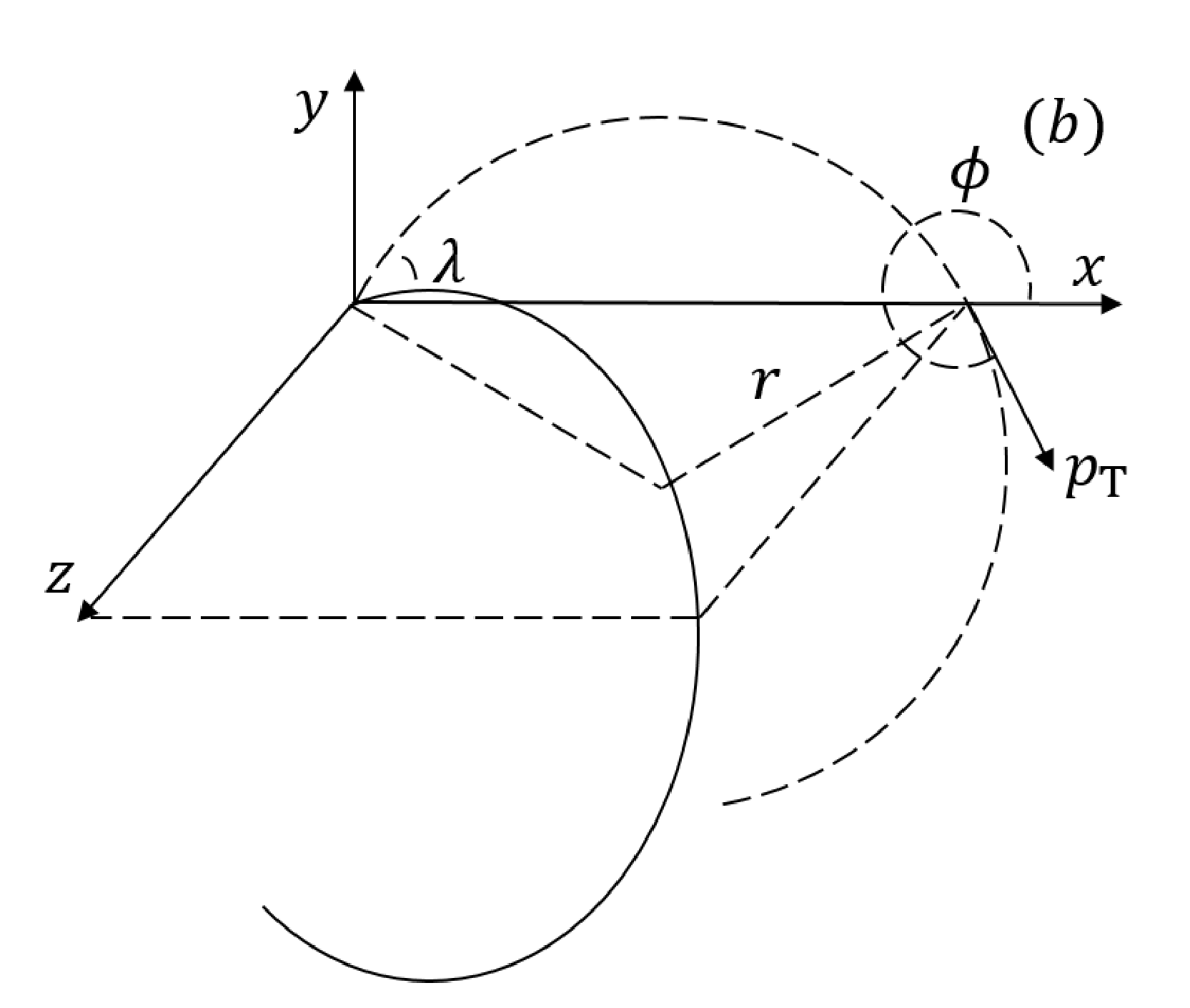}
         \caption{}
         \label{fig:Var}
     \end{subfigure}
        \caption{(a) Diagram of the simplified detector geometry, showing the direction of the magnetic field and the position of one of the radial pad layers. (b) Diagram illustrating the definition of the coordinates defining the evolution of the custom Kalman Filter. } \label{fig:Detector_var}
\end{figure}

The Kalman Filter described in this work has been developed to be used in an homogeneous cylindrical gas TPC. We assume that an ideal magnetic field is applied along the drift direction identified by the coordinate $z$. Deviations from the ideal mono-directional magnetic field lines can be simulated and be accounted for using the infrastructure available in \texttt{AliExternalTrackParam}, but were not implemented in this study. The spatial information in the perpendicular $xy$ plane is given by detector elements disposed in radial layers on the two sides of the cylinder. The detector elements will be referred to as pads and no assumption on the underlying technology is made. The $x$ coordinate identifies the horizontal direction, while the $y$ coordinate identifies the vertical. A diagram of the detector cylinder is shown in Fig.~\ref{fig:Detector}. 

The algorithm is evolved along the free parameter, $x$, and its state vector is defined as:
\begin{equation}\label{eq:state}
    s(x) = \left(y,z,\sin{\phi},\tan{\lambda}, q/p_{\text{T}}\right),
\end{equation}
where $y$ is the vertical direction; $z$ is the drift direction; $\phi$ is the azimuthal angle of the transverse momentum i.e. the component of the momentum vector transverse to the drift direction; $\lambda$ is the ``dip angle''  between the transverse momentum and the total momentum vector; q is the charge sign of the particle and $p_{\text{T}}$ is the module of the transverse momentum. Note that the inverse transverse momentum can also be written in terms of track curvature $1/r$. The conversion is easily obtained using the standard formula for charged particles moving in a magnetic field:
\begin{equation}\label{eq:curvatureconv}
    p_{\text{T}} \ \left(\text{GeV}/c\right) =0.3 \ B \left(\text{T}\right) \ r \left(\text{m}\right).
\end{equation}
A visual representation of the coordinates is given in Fig.~\ref{fig:Var}.

The evolution of the state vector is divided into two steps: a rotation of the global coordinates to a local frame and a propagation along the helix trajectory. The rotation is applied in the $xy$ plane around the center of the TPC cylinder. The rotation angle $\alpha = \arctan (y/x)$ is defined so that the $x$ coordinate becomes the radial distance from the center of the TPC and the $y$ coordinate is $\sim 0$. After the rotation the state vector is moved along the trajectory using a propagator function, as described in Eq.~\ref{eq:EKF}:
\begin{equation} \label{eq:func}
    \widetilde{s}_k = f_{k-1}(s_{k-1}) =
        \left\{
        	\begin{aligned}
        		& \widetilde{y}_k  =  y_{k-1}+ \frac{\sin{\phi}_{k-1}+\sin\widetilde{\phi}_k} 
                                     {\cos{\phi}_{k-1}+\cos\widetilde{\phi}_k}  \Delta x_k,  \\
        		& \widetilde{z}_k  =  z_{k-1}+\left(\widetilde{\phi}_k-\phi_{k-1}\right)\frac{r}{q}_{k-1}\tan{\lambda}_{k-1} ,\\ 
                    & \sin \widetilde{\phi}_k =  \sin \phi_{k-1} + \frac{q}{r}_{k-1}\Delta x_k,  \\               
                    & \tan \widetilde{\lambda}_k   =  \tan \lambda_{k-1}, \\   
                    & \widetilde{\frac{q}{p_{\text{T}}}}_k = \frac{q}{p_{\text{T}}}_{k-1} \times\frac{p_{k-1}}{\Delta p_k+p_{k-1}},
        	\end{aligned}
        \right.
\end{equation}
where $\Delta x_k$ is the distance in the $x$ direction between the previous and current points, $p_k$ is the total momentum and $\Delta p_k$ is the total momentum loss. In order to obtain the propagation matrix, $F_k$, one only needs to calculate the Taylor expansion coefficient $\partial f_k / \partial s_k$, as described in Eqs.~\ref{eq:jacobian} and~\ref{eq:jacobian2}, with the exception of the $q/p_{\text{T}}$ term, which is treated separately as discussed below.

In order to compute the momentum loss, $\Delta p_k$, at each trajectory point, the ionization energy loss, $-\textrm{d}E/\left(\rho\textrm{d}x\right)$ (where $\rho$ is the density of the material in $\text{g/cm}^3$), of the particle is evaluated using the standard Bethe-Bloch formula~\cite{PDG}:
\begin{equation} \label{eq:Bethe}
    -\frac{\textrm{d}E}{\rho\textrm{d}x} = 4\pi N_{A}r_e^2m_ec^2z^2\frac{Z}{A}\frac{1}{\beta^2}\left(\frac{1}{2}\ln{\frac{2m_ec^2\beta^2\gamma^2T_{\textrm{max}}}{I^2}}-\beta^2-\frac{\delta}{2}\right),
\end{equation}
where $N_{A}$ is Avogadro's number, $r_e$ is the classical electron radius, $m_ec^2$ is the electron mass energy, $z$ is the charge of the particle, $Z$ and $A$ are the atomic number and mass of the absorbing material, $\beta$ and $\gamma$ are the usual relativistic factors for the passing particle, $I$ is the material mean excitation energy, $T_{\textrm{max}}$ is the maximum kinetic energy which can be imparted to a free electron in a single collision and $\delta/2$ is a density effect correction factor. 

The differential energy loss, $-\textrm{d}E/\left(\rho\textrm{d}x\right)$, is calculated using the properties of the most abundant gas present in the gas mixture in standard conditions and then multiplied by the material's density to obtain a reasonable approximation of the $\textrm{d}E/\textrm{d}x$~\cite{STERNHEIMER1984261}. The total momentum loss between two steps is then calculated by numerical integration~\cite{Griffiths2010}. 

In the evaluation of $F_k$, the $q/p_{\text{T}}$ parameter is treated as if it were static. A correction term, $c_k$ is added to the $q/p_{\text{T}}$ diagonal element of the covariance matrix, $\widetilde{C}_k$, after the propagation step:
\begin{equation} \label{eq:eloss-factor}
    c_k=\left(a\cdot\frac{\Delta p_k}{p_{k-1}} \cdot\frac{q}{p_{\text{T}}}_{k-1}\right)^2,
\end{equation}
where $a=3.162\times 10^{-3}$ is a constant multiplicative factor which is directly taken from the ALICE TPC framework~\cite{carminati2003simulation}.

Multiple scattering is treated through the noise correction matrix, $Q_k$. At each step the scattering angle can be treated as emerging from a Gaussian distribution with a root mean square equal to the Molière angle $\theta_{\textrm{M}}$, which is calculated using the formula given by Lynch and Dahl~\cite{LYNCH19916}:
\begin{equation}
    \theta_{\textrm{M}} = \frac{13.6 \ \text{MeV}}{\beta pc}z\sqrt{\frac{\Delta d}{X_0}} \left[ 1+0.038\ln{\left(\frac{\Delta d}{X_0 }\frac{z^2}{\beta^2}\right)}\right], 
\end{equation}
where $\Delta d$ is the total distance traveled between two steps and $X_0$ the radiation length in cm. The $Q_k$ terms relative to $\sin \phi$, $\tan \lambda$ and $q/p_{\text{T}}$ are evaluated through error propagation and added to the covariance matrix as described in Eq.~\ref{eq:pcov}:
\begin{equation}\label{eq:Q}
    Q =\begin{bmatrix}
    0 & 0 & 0 & 0& 0 \\
    0 & 0 & 0 & 0& 0 \\
    0 & 0 & \theta_{\textrm{M}}^2\cdot\frac{\cos^{2}\phi}{\cos^{2}\lambda} & 0& 0 \\
    0 & 0 & 0 & \frac{\theta^2_{\textrm{M}}}{\cos^4\lambda}& \frac{q}{p_{\text{T}}}\cdot\theta^2_{\textrm{M}}\cdot\frac{\tan\lambda}{\cos\lambda} \\
    0 & 0 & 0 & \frac{q}{p_{\text{T}}}\cdot\theta^2_{\textrm{M}}\cdot\frac{\tan\lambda}{\cos\lambda}& \left(\frac{q}{p_{\textrm{T}}}\right)^2\theta^2_{\textrm{M}} \tan^2\lambda
    \end{bmatrix} .
\end{equation}

Each step in the evolution of the Kalman Filter can potentially fail, in which case the algorithm is stopped. This can happen mainly in two scenarios: $\sin \phi$ can be calculated to be out of range, i.e. $|\sin \phi|>(1-10^{-7})$ or the particle can lose all its remaining energy. Once the Kalman Filter is stopped, the information for each of the reconstructed points is saved. Flags are used to preserve information on which of the reconstruction steps have been successful and which have failed. 

\begin{figure}[!ht]
     \centering
     \includegraphics[width=\textwidth]{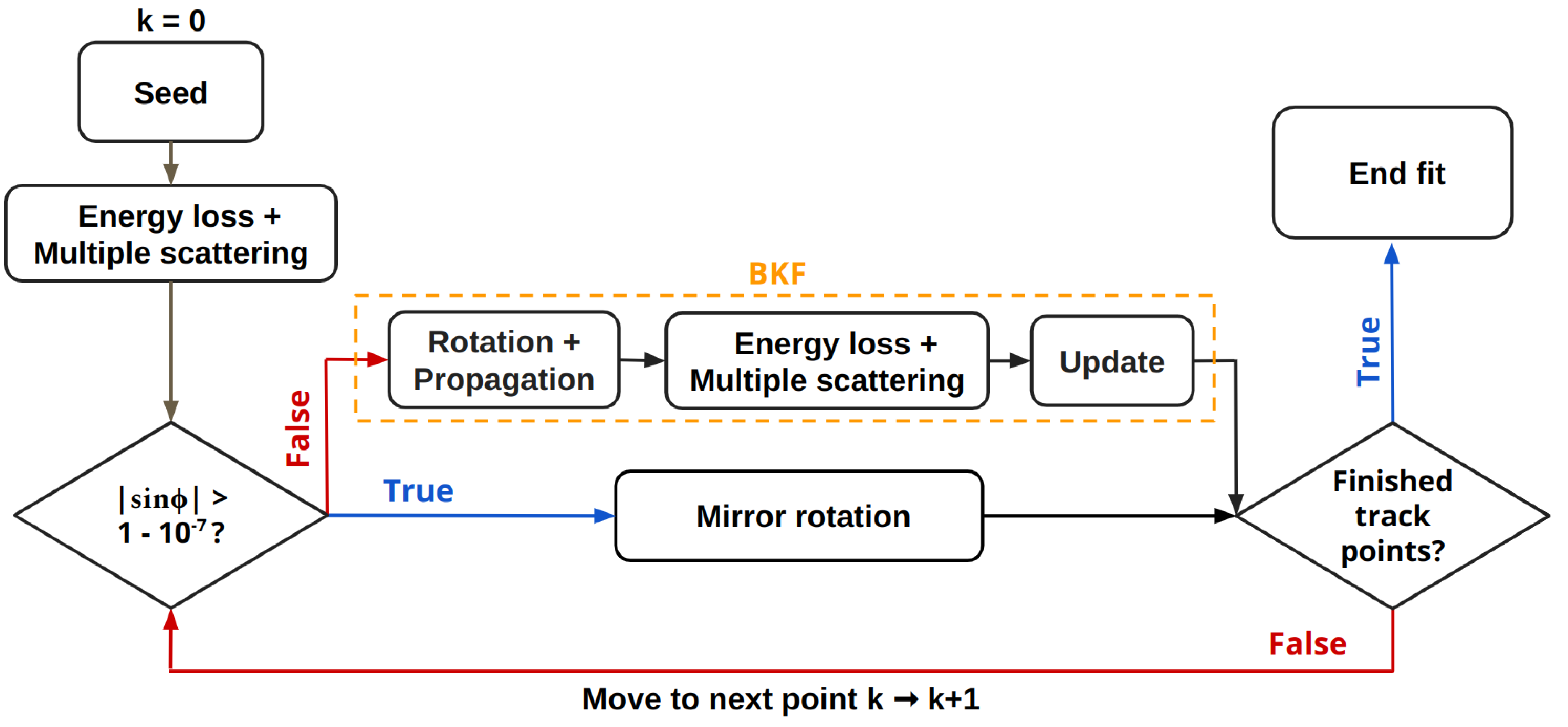}
     \caption{Flow chart describing the \texttt{CKF} algorithm. A seeding algorithm is used to obtain an estimate for the status of the system at the start of the trajectory $k=0$. Energy loss and multiple scattering corrections are applied to the estimate. The fit is then moved to the next point $k\rightarrow k+1$ either by applying the \texttt{BKF} procedure or by using the mirror rotation, in the case that the limits of the $\sin\phi$ range have been surpassed. The algorithm is iterated point by point until the end of the trajectory is reached. Other minor modifications have been made to the \texttt{CKF} algorithm compared to the \texttt{BKF} in order to make the mirroring operation more stable. }
        \label{fig:Logic}
\end{figure}

One inherent limitation exists in the propagator function in Eq.~\ref{eq:func}, specifically in the equation describing the evolution of $\sin \phi$. The formula can only be applied within the range of $\sin \phi \in [-1,1]$, which describes one semi-plane. For $|\sin \phi|\rightarrow1$ the uncertainty on the parameter tends to infinity and the operation is no longer well defined. In radial coordinates this coincides with the moment when the particle is moving parallel to a pad layer (see Fig.~\ref{fig:Detector}) and the radial direction of the propagation is inverted. In order to overcome this limitation and further evolve the Kalman Filter, one can apply a \enquote{mirror rotation} or reflection on the state vector~\cite{lay2003linear}. The mirror plane is the one perpendicular to the $xy$-plane,  which connects the coordinate frame's center (i.e. the center of the TPC) with the center of the circular motion of the particle. The application of this technique to a Kalman Filter for charged particle tracking in a magnetic field has no precedent in the literature. In the local coordinate frame the mirror rotation is linear and can be written as:
\begin{equation}\label{eq:mirror}
    \left\{
    \begin{aligned}
        &s_k^\textrm{M} = M s_k,\\
        &C_k^\textrm{M} = M C_k M^T,
    \end{aligned}
    \right.
    \;\; \textrm{where} \;\;
    M=\begin{bmatrix}
    1 & 0 & 0 & 0& 0 \\
    0 & 1 & 0 & 0& 0 \\
    0 & 0 & -1 & 0& 0 \\
    0 & 0 & 0 & -1& 0 \\
    0 & 0 & 0 & 0& -1 
    \end{bmatrix}.
\end{equation}
The angle $\alpha$, which defines the local coordinate frame, needs to be updated accordingly. This is done by finding the angle $\alpha_\textrm{C}$ corresponding to the mirror plane, and updating $\alpha$ as:
\begin{equation}
    \begin{aligned}
       \alpha _k^\textrm{M} &=\alpha_\textrm{C}-\Delta \alpha \\
                   &= \alpha_\textrm{C}-(\alpha_k-\alpha_\textrm{C}).
    \end{aligned}   
\end{equation} 
Finally, to update the $z$ position the angular displacement around the center of rotation is calculated as:
\begin{equation}
    \Delta \phi_C = 2\arcsin \left( \frac{\Delta_{xy}}{2r_k} \right),
\end{equation}
where $\Delta_{xy}$ is the distance between the two points in the $xy$ plane. From $\Delta \phi_C$, the correspondent circumference arch in the $xy$ plane, $a_{xy}$, can be found, and from it, the displacement in the drift direction, $\Delta z_k$, reads:
\begin{equation}
    \begin{aligned}
        \Delta z_k &= a_{xy} \cdot \tan{\lambda_k} \\
             &= \Delta \phi_C \cdot r_k \cdot \tan{\lambda_k} .
    \end{aligned}  
\end{equation}
Once all the mirror operations are complete the closest trajectory point is found and the Kalman Filter is further evolved from there. From this point on-wards we will refer to the Kalman Filter algorithm, not including the mirroring operation as the Basic Kalman Filter or \texttt{BKF}. We will refer to the full algorithm, which includes both the \texttt{BKF} and the mirroring operation as the Corrected Kalman Filter or \texttt{CKF}. A flow chart describing the algorithm is shown in Fig.~\ref{fig:Logic}.

The seeding strategy used for the \texttt{CKF} (as well as for the \texttt{BKF}) consists in a simple three-point circle finding algorithm and will be referred to from now on simply as \texttt{Seed}. In the plane perpendicular to the magnetic field, the trajectory of a charged particle is a circle. Since only one circumference will pass through any three points, one can find a point which is roughly at the start of the particle trajectory, one at the end and one in the middle and obtain the properties of the circle that passes though them. This equates to solving a system of three linear equations for three unknown variables: the coordinates of the center of the circumference $(x_C,y_C)$ and its radius $r$. From the circle properties and the coordinates of the three points, one can find an estimate for the state vector $s_0$ at the starting point $(x_0,y_0,z_0)$: $y_0$ and $z_0$ are taken as the measured values; $q/p_{T_0}$ is converted from the track curvature $1/r$; $\sin \phi_0=x_0/r$; $\tan \lambda_0$ is estimated as the ratio between the displacement in the drift direction between the first and middle point and the correspondent circumference arch in the transverse plane. The three-point method for the estimation of the initial-state vector $s_0$ can be written as: 
\begin{equation}
    s_0 = h\left[\left(x_0,x_1,x_2);(y_0,y_1,y_2,z_0,z_1, z_2\right)\right]\equiv h(\zeta;\eta) ,
\end{equation}
where $x_i,y_i$ and $z_i$ are the measured coordinates of the three points, all taken to be independent and uncorrelated. In order to compute an estimate for $C_0$ one can use the matrix expression for error propagation~\cite{Cov}:
\begin{equation}\label{eq:error_Prop}
    C_0 = gVg^T 
\end{equation}
\begin{equation}
   g_{ij}  = \frac{\partial h_i}{\partial \eta_j},
\end{equation}
where $V$ is the covariance matrix of the vector $\eta$, which is determined by the resolution of the detector in $y$ and $z$. The coordinate $x$ is taken to be the free parameter and thus is not considered in the error propagation. The partial derivatives are estimated numerically as:
\begin{equation}
    \frac{\partial h_i}{\partial \eta_j} \approx \frac{h(\zeta;\eta_{i\neq j},\eta_j+\sigma_{\eta_j})-h(\zeta;\eta_{i\neq j},\eta_j)} {\sigma_{\eta_j}},
\end{equation}
where $\sigma_{\eta_j}$ is the resolution of the vector element $\eta_j$. The \texttt{Seed} estimation for both the covariance matrix $C_0$ and the state vector $s_0$ is adjusted for energy loss and multiple scattering using the same method as the Kalman Filter. The $q/p_\texttt{T}$ ratio is corrected with the factor described in Eq.~\ref{eq:func}, and the relative covariance matrix element is updated by adding the $c_k$ factor from Eq.~\ref{eq:eloss-factor}. To handle multiple scattering, the $Q$ matrix calculated in Eq.~\ref{eq:Q} is added to the covariance. The total distance traveled, needed to calculate total energy loss and the scattering angle $\sigma_\texttt{M}$, is determined by summing the distances between the starting and midpoint, and the endpoint used for circle finding.

\section{Toy Monte Carlo Simulation}
\label{sec:Simulation}

To generate particle samples and validate the \texttt{CKF} algorithm, we employed a toy Monte Carlo (MC) tool called \texttt{fastMCKalman}~\cite{fastMCKalman}. This tool, stemming from the \texttt{AliExternalTrackParam} framework in the \texttt{AliRoot} code-base~\cite{aliroot}, was designed to be complemented by \texttt{RootInteractive}~\cite{RootInt}, an advanced statistical analysis tool. \texttt{fastMCKalman} has been developed with several objectives in mind: conducting rapid Monte Carlo (MC) simulations to evaluate tracking performance metrics, particle identification, and time-of-flight measurements across various detector setups. It was also designed to facilitate detailed studies on signal distortion in the ALICE detector and the derivation of performance metrics for its Run-3 upgrade and future iterations. In this discussion, we highlight the effectiveness of \texttt{fastMCKalman} in rapid simulation and tracking performance assessment within a TPC setting. It is important to note that this article represents the first use of this fast simulation tool and serves as part of its validation process. The use of fast Monte Carlo tools to complement more extensive generators such as Geant4 \cite{GEANT4:2002zbu,Allison:2006ve,Allison:2016lfl}, is commonplace in high energy physics. The ALICE experiment, for example, implements independent  Monte Carlo components in its simulation pipeline, to model detector effects that would be difficult to simulate using a more standard MC generator \cite{ALICE:2000jwd}. This is the envisioned future use of \texttt{fastMCKalman}. However, as of now, only basic effects such as multiple Coulomb scattering and energy loss through ionization have been implemented. In order to ensure the agreement of \texttt{fastMCKalman} with more traditional MC generators, the formulas used to simulate these effects are the same implemented by Geant4. Additionally the agreement of the \texttt{fastMCKalman} simulation with theoretical expectations is tested as part of the validation of the \texttt{CKF} algorithm in Sec. \ref{sec:Validation}. 

The first step in the toy Monte Carlo simulation consists in defining a simplified detector geometry. The radius and length of the TPC cylinder are specified, together with the number of pad rows, the spatial resolution of the detector in the radial and drift directions (defined as $\sigma_{r\phi}$ and $\sigma_Z$ respectively) and the gas properties (i.e. the radiation length in cm, $X_0$, the density in g/cm$^3$, $\rho$, and the gas pressure in atm, $P_\textrm{gas}$). Each simulated particle is defined by specifying its type, charge, transverse momentum, $p_{\text{T}}$, azimuth angle, $\phi$, dip angle tangent, $\tan\lambda$, and the starting position. From this information, the initial MC-true (henceforth ``true'') state vector $s_0^{\text{true}}$ is built. The true state vector is moved through the detector by applying the same operations of the propagation steps of \texttt{CKF} in reverse, moving from layer to layer of pads.

The propagation of the state vector is obtained by applying Eq.~\ref{eq:func}. At each step the energy loss is calculated using the Bethe-Bloch formula as described in Eq.~\ref{eq:Bethe} and multiplying by the distance traveled and the material density. The energy loss is then converted into a $q/{p_{\text{T}}}$ multiplicative factor described in Eq.~\ref{eq:func} and smeared with a Landau distribution having a width equal to the $c_k$ factor described in Eq.~\ref{eq:eloss-factor}. The multiple scattering effects are simulated by calculating the scattering angle distribution root mean square $\theta_{\textrm{M}}$ and from that the process noise matrix $Q$. The diagonal elements of the matrix are then used as the widths of smearing Gaussian distributions that are applied to parameters $s_2=\sin\phi,s_3=\tan\lambda$ and $s_4=q/p_{\text{T}}$. To reproduce the measurement noise encapsulated in matrix $R$, a Gaussian smearing is applied to the position parameters $s_0=y$ and $s_1=z$. The widths of the distributions are equal to the position resolutions $\sigma_{r\phi}$ and $\sigma_{z}$ respectively.

The propagation continues until any of the following happens: the particle reaches the edges of the detector cylinder; the particle looses all its remaining energy; the track has traversed a predefined maximum number of points; one of the propagation steps fails.  Once the track is fully generated, the track fit is done as described in Sec.~\ref{sec:Loop}. No element of track formation or particle identification is included. After saving all information, a new simulation and track fit start.  

\subsection{Sample definition}
\label{sec:Geometry}

The aforementioned \texttt{fastMCKalman} was used to produce two separate samples, simulated in the same simplified gas TPC geometry. The TPC has a cylindrical form with a radius $r=250 \ \text{cm}$ and the length of the cylinder is taken as $L=500 \ \text{cm}$. There are 250 circular layers of pads placed radially at each of the two end caps ($z=\pm250~\textrm{cm}$). A magnetic field of intensity $B=0.5 \ T$ is placed in the drift direction along the cylinder axis. 

\begin{figure}[!ht]
     \centering
     \begin{subfigure}[b]{0.48\textwidth}
         \centering
         \includegraphics[width=\textwidth]{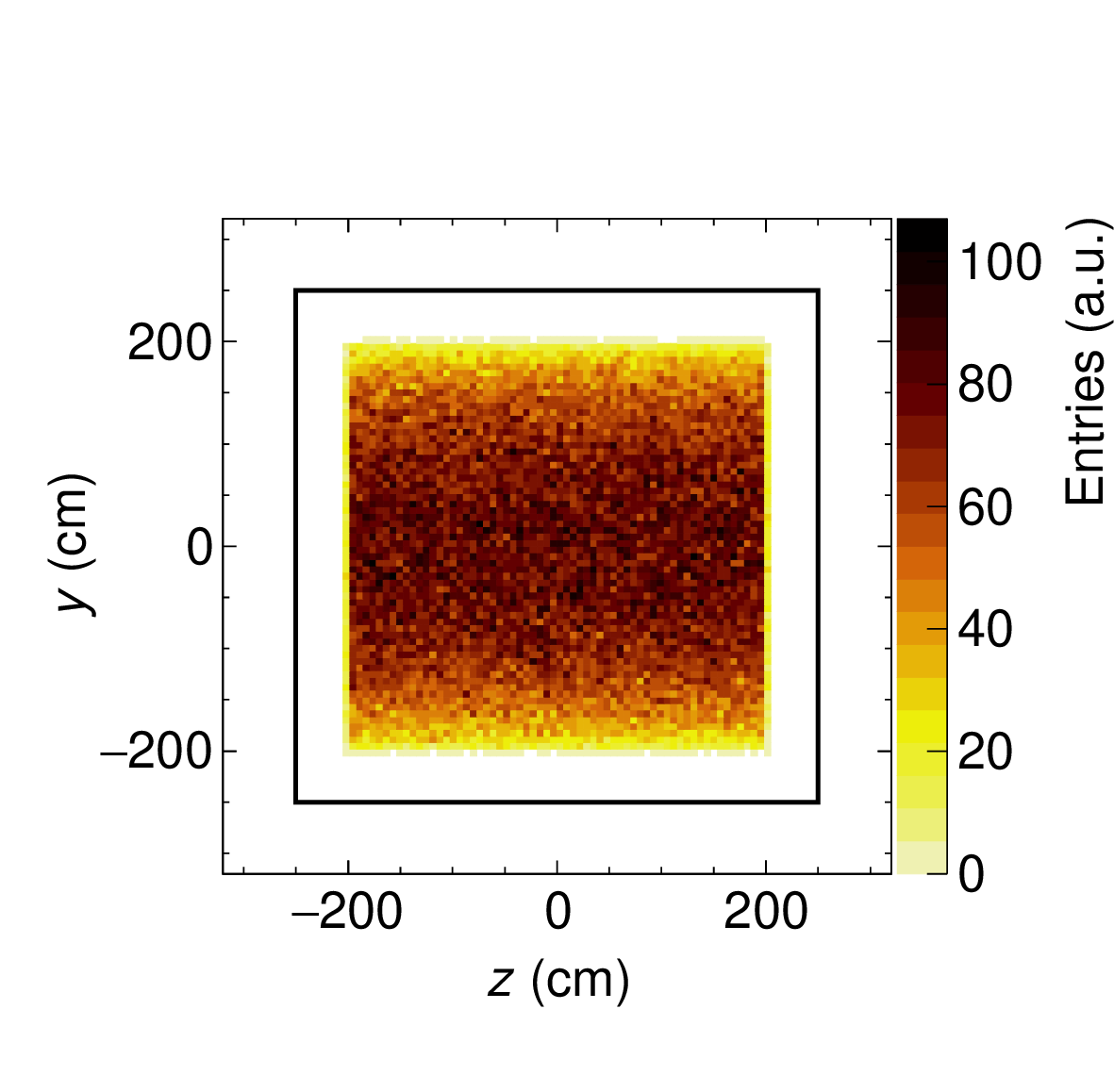}
         \caption{}
         \label{fig:YZViewGAr}
     \end{subfigure}
     \begin{subfigure}[b]{0.48\textwidth}
         \centering
         \includegraphics[width=\textwidth]{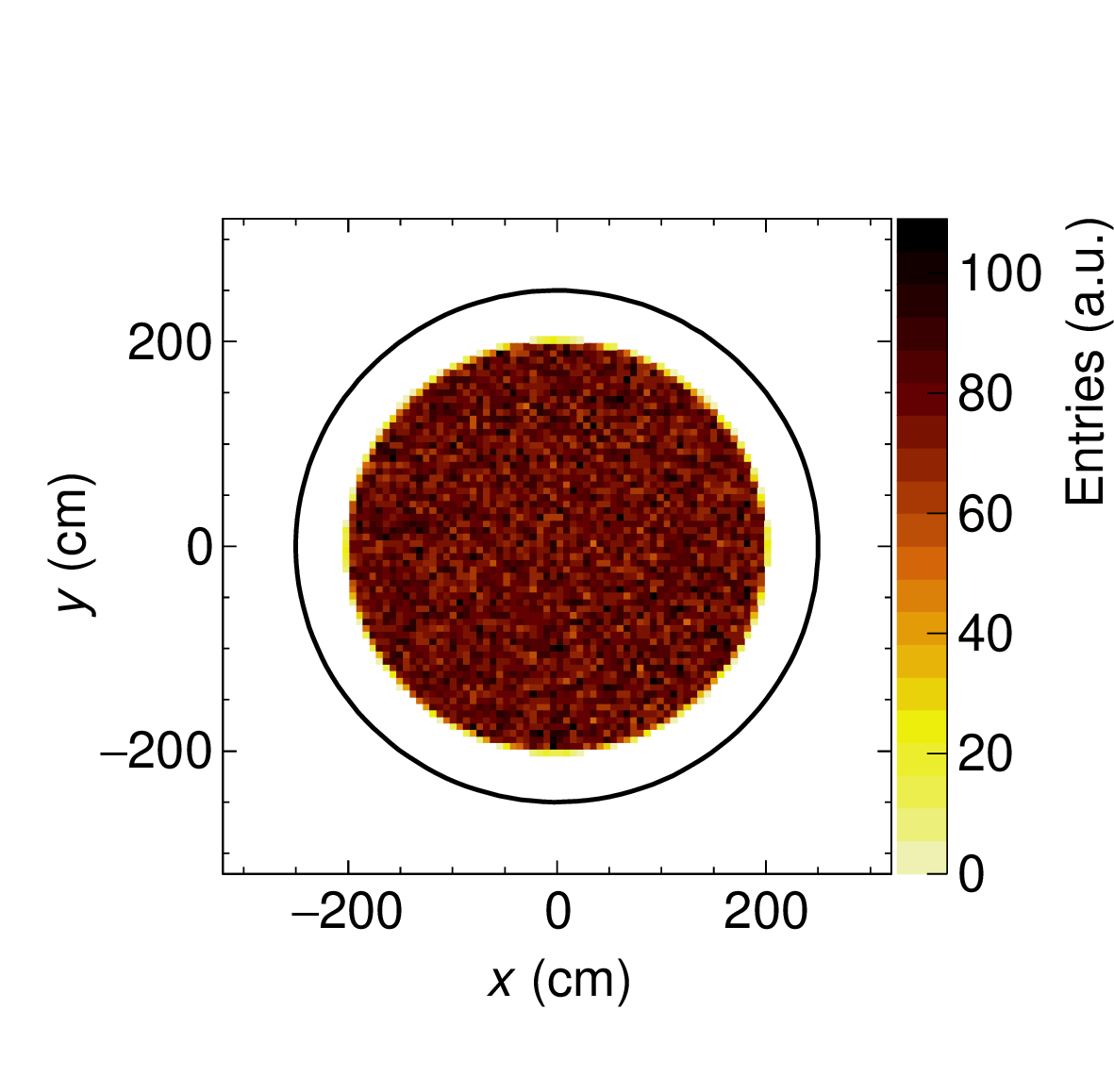}
         \caption{}
         \label{fig:XYViewGAr}
     \end{subfigure}
        \caption{Starting positions for secondary particles in the PS sample. The primaries are not shown, as all of their starting positions are in $(x,y,z)=(0,0,0)$. The left plot shows the distribution in the $zy$ plane, while the plot on the right shows the distribution in the $xy$ plane. The edges of the TPC are drawn on top. } \label{fig:ViewGAr}
\end{figure}

The first sample, which includes a total of $5\times10^5$ tracks, contains a wide variety of detector properties, particle types and energies and was used to validate the \texttt{CKF} algorithm and its \texttt{Seed} in as wide a parameter space as possible. We will refer to this sample as the parameter scan sample or PS sample. The PS sample is composed of two equally large sub-samples with different starting position distributions: a sample of primaries---emulated as in a collider event geometry---starting from the center of the detector $(x,y,z)=(0,0,0)\ \text{cm}$ and a sample of secondaries with randomized starting positions within a fiducial cylinder of radius $r=200 \ \text{cm}$ and length $l = 400 \ \text{cm}$. The initial spatial distribution of the secondaries in the sample is shown in Fig.~\ref{fig:ViewGAr}.

\begin{figure}[t]
     \centering
     \begin{subfigure}[b]{0.32\textwidth}
         \centering
         \includegraphics[width=\textwidth]{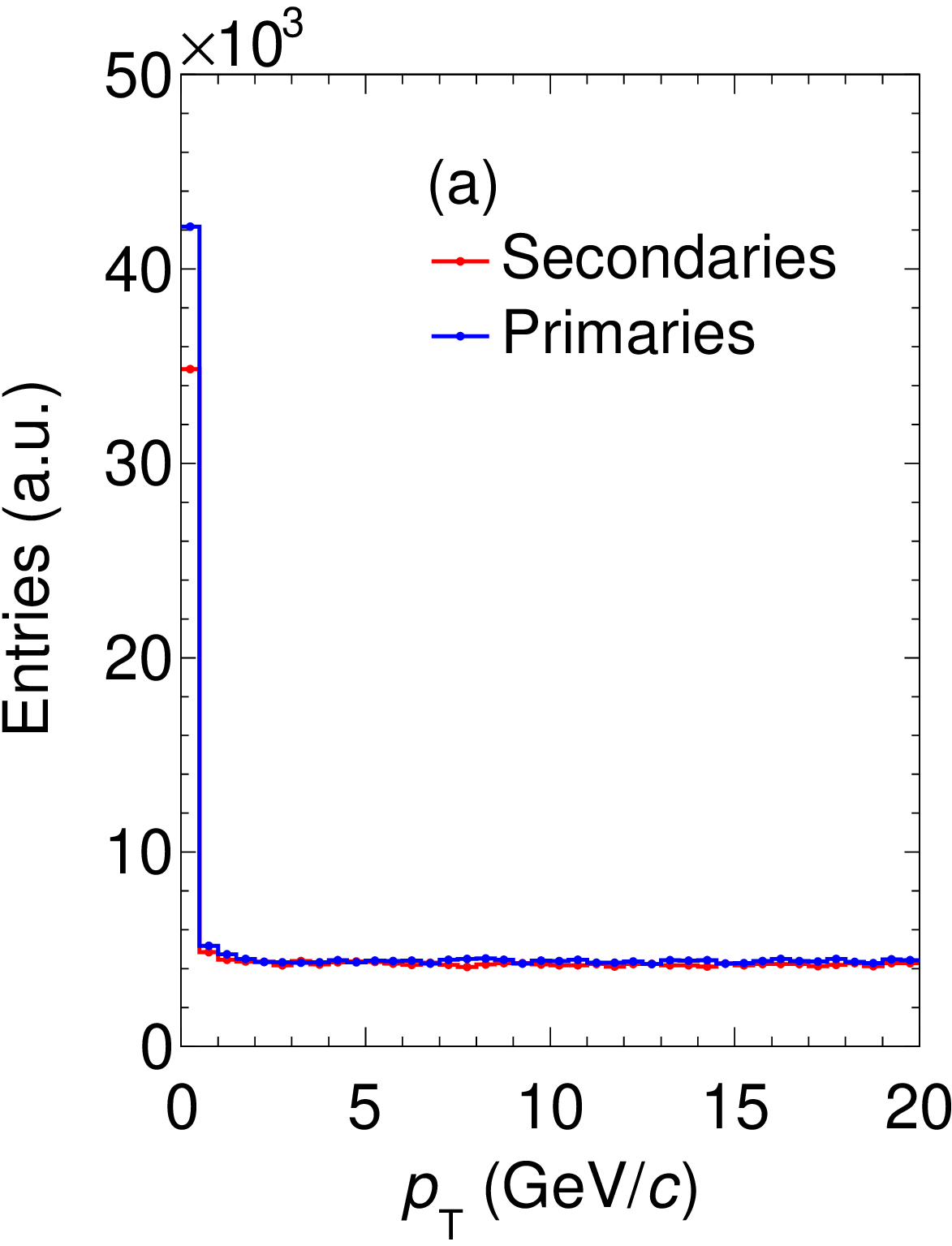}
         \caption{}
         \label{fig:ptTPC}
     \end{subfigure}
     \begin{subfigure}[b]{0.32\textwidth}
         \centering
         \includegraphics[width=\textwidth]{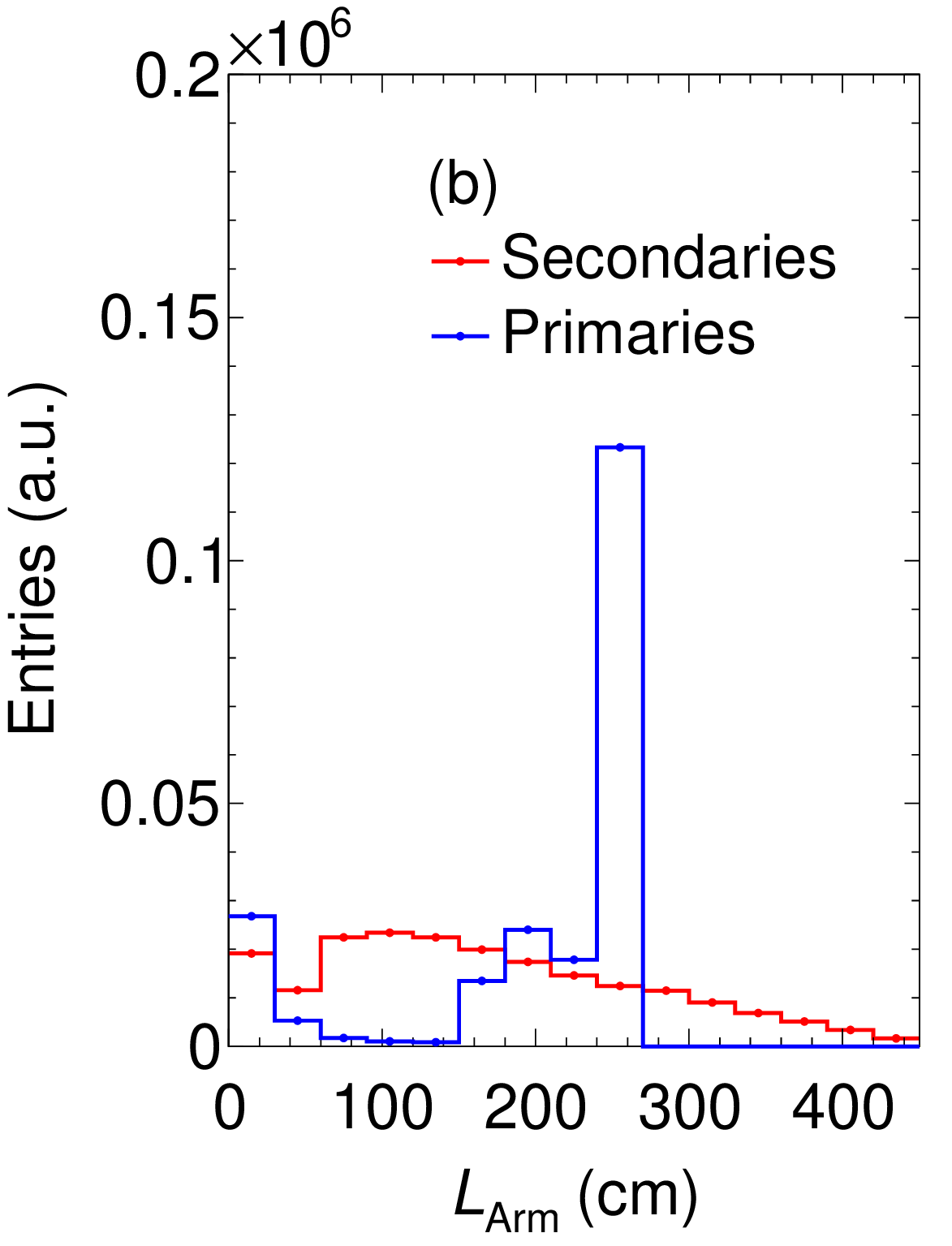}
         \caption{}
         \label{fig:LTPC}
     \end{subfigure}
          \begin{subfigure}[b]{0.32\textwidth}
         \centering
         \includegraphics[width=\textwidth]{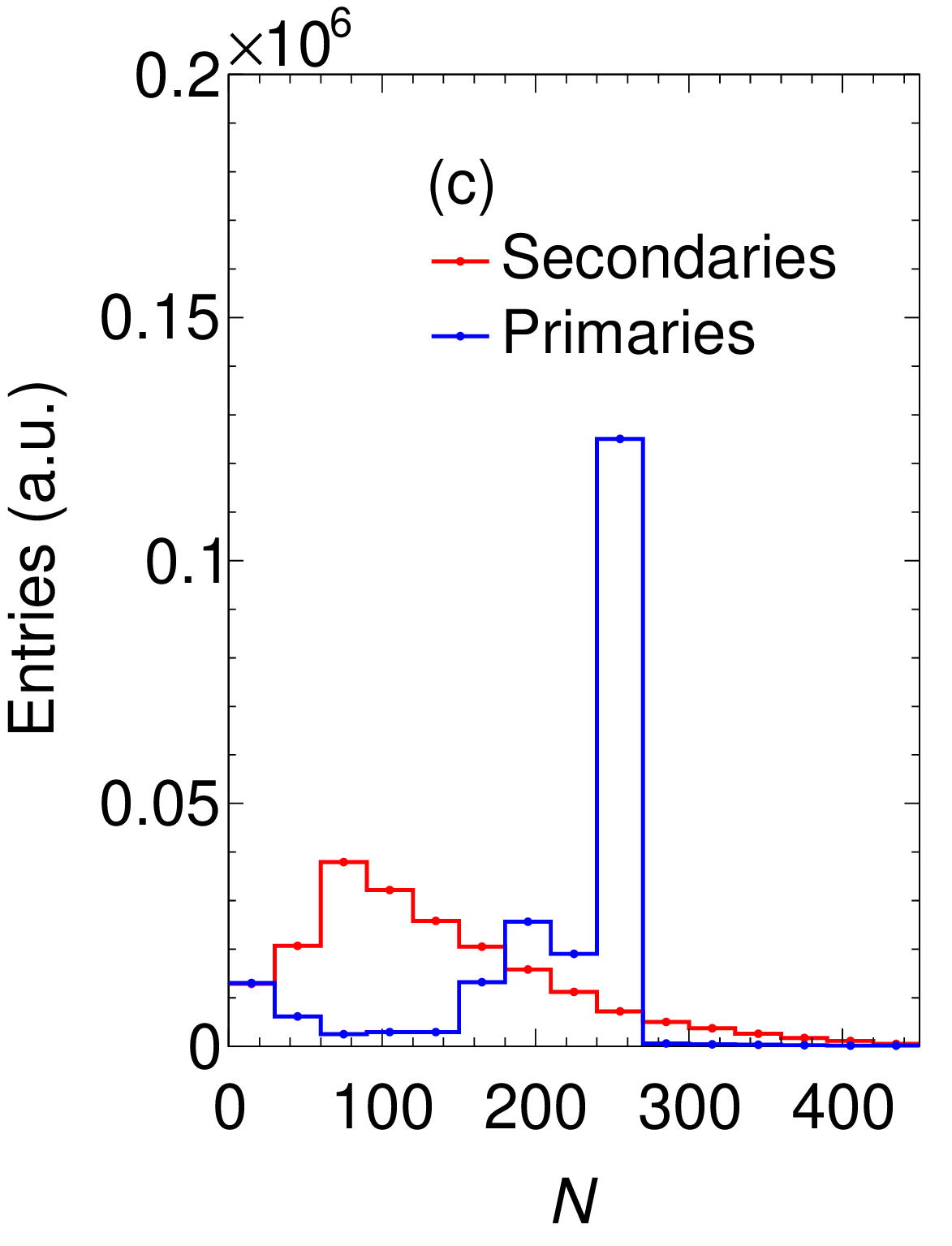}
         \caption{}
         \label{fig:NTPC}
     \end{subfigure}
        \caption{Distributions of (a) transverse momentum $p_{\textrm{T}}$,  (b) lever arm $L_{\textrm{Arm}}$ and (c) number of points per track $N$ in the PS sample. In all the plots the distributions for primary and secondary particles are shown separately. The low $p_{\textrm{T}}$ portion of the spectrum 
 is produced via Eq.~\ref{eq:LowpT}. The spikes around $N=250$ and $L_{\textrm{Arm}}=250$ in the primary sample are explained by the simulated geometry having 250 radial pad layers and the particles starting from the center of the detector. } \label{fig:TPCProperties}
\end{figure}

The tracking pad response as well as the gas properties of the detector are sampled in each simulated event: the resolutions $\sigma_{r\phi}=\sigma_z$ are uniformly distributed between 0.1 cm and 0.5 cm and the pressure $P_\textrm{gas}$ was randomized between 0.1 atm and 10 atm. The gas composition was taken to be the Ne/CO2/N2 (90/10/5) gas mixture used by the ALICE experiment during Run-1~\cite{ALICE:2008ngc}. The radiation length and density of the gas at atmospheric pressure are $X_0=1.2763\times10^4 \ \text{cm}$ and $\rho = 0.0016265 \ \text{g/cm}^3$. The particles produced are equally divided in electrons, muons, pions, kaons and protons, corresponding to the ALICE’s convention for particle types, $t_{\textrm{ID}}$, 0, 1, 2, 3, and 4, respectively. The angles $\phi$ and $\lambda$ are fully randomized. The initial $p_{\textrm{T}}$ is sampled from a two-component distribution: a high-$p_{\text{T}}$ component uniformly distributed in $[0,20]$ GeV/c, which covers 70\% of the total, and a low-$p_{\text{T}}$ component flat in $1/p_\textrm{T}$:
\begin{equation} \label{eq:LowpT}
   p_{\textrm{T}}=\frac{p_{\textrm{T}_{\textrm{min}}}}{p_{\textrm{T}_{\textrm{min}}}/p_{\textrm{T}_{\textrm{max}}}+j},
\end{equation}
where $p_{\textrm{T}_{\textrm{min}}}=0.01$ GeV$/c$, $p_{\textrm{T}_{\textrm{max}}}=20 \ \text{GeV}/c$ and $j$ is a random variable uniformly distributed between 0 and 1. Some key properties of the tracks composing the PS sample are plotted in Fig.~\ref{fig:TPCProperties}. These include the $p_{\textrm{T}}$ spectrum, the lever arm $L_{\textrm{Arm}}$ and the number of points per track $N$ separated between primaries and secondaries. The lever arm is defined as the distance in the $xy$ plane between the first and last point in the track. All the tracks included in this and future plots have been successfully reconstructed, unless stated otherwise.

Primaries and secondaries have analogous $p_{\textrm{T}}$, with the only variations arising from the fact that primaries and secondaries have different reconstruction efficiencies. In particular the reconstruction for low momentum secondaries is more likely to fail than for primaries, resulting in a harder spectrum. This arises from the combination of two factors: the low momentum secondaries have a higher risk of having the start of their track at a $\sin\phi$ angle that is close to the edges of its range, making the application of the mirroring algorithm difficult since not many points are remaining; additionally secondaries are more likely to produce short tracks overall, due to the randomness of their starting points. This is a difficulty which could be at least partially solved, by placing the rotation center in front of each secondary track, to mimic the geometrical construction of the primaries. This solution was however not tested for this study and goes beyond the scope of this work. More significant differences appear in the $L_{\textrm{Arm}}$ and $N$ distributions. Since the primaries all start at the center of the detector, most tracks will cross the detector exiting from the cylinder's barrel, producing a track with as many points as the 250 pad layers. Alternatively the track can exit from the sides of the detector, producing tracks with a slightly smaller $N$ or be stopped inside the detector having $N<50$. For the secondaries the spread is much more homogeneous and the chance of producing tracks with $N>250$ is more significant.

\begin{figure}[!ht]
     \centering
     \begin{subfigure}[b]{0.32\textwidth}
         \centering
         \includegraphics[width=\textwidth]{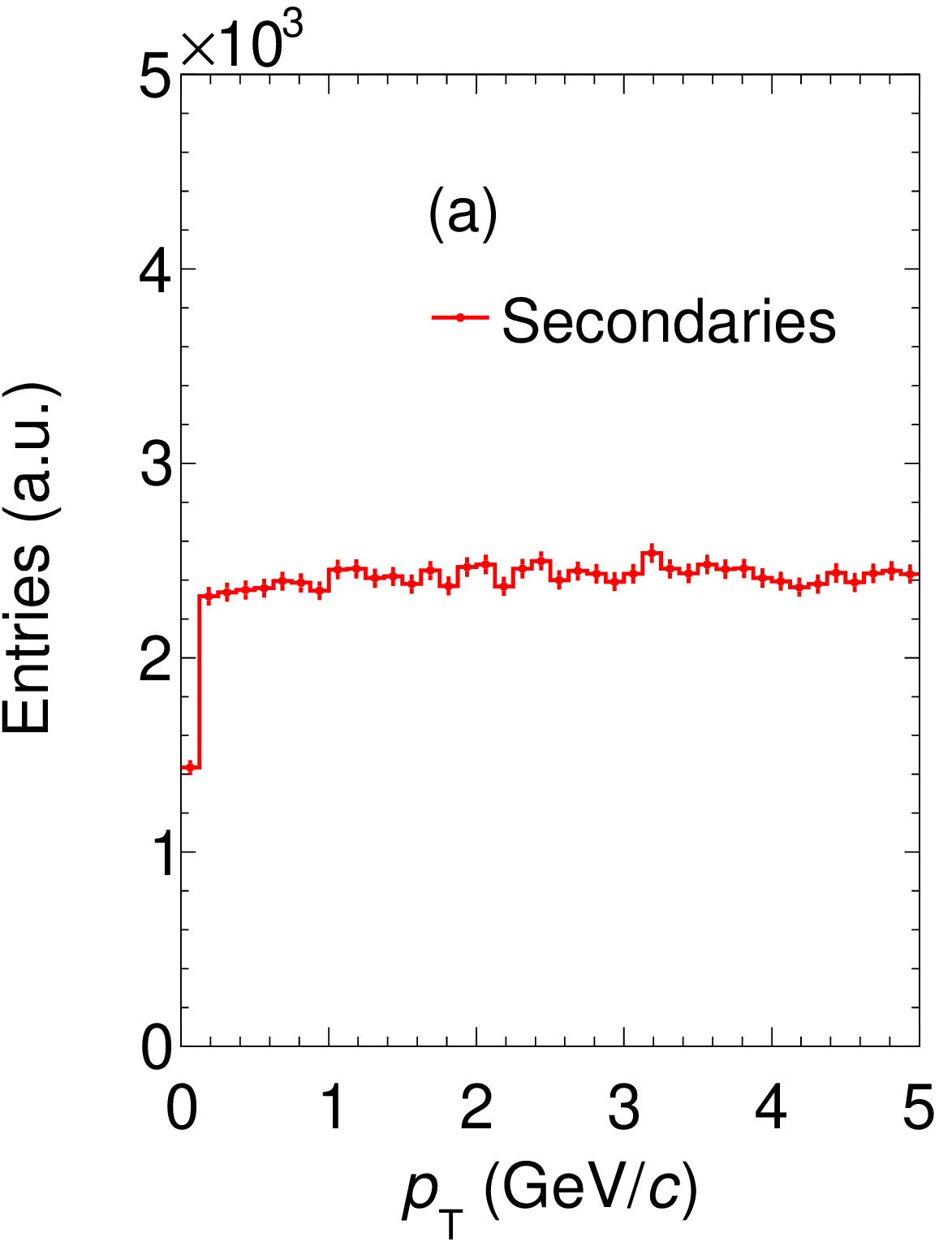}
         \caption{}
         \label{fig:ptGAr}
     \end{subfigure}
     \begin{subfigure}[b]{0.32\textwidth}
         \centering
         \includegraphics[width=\textwidth]{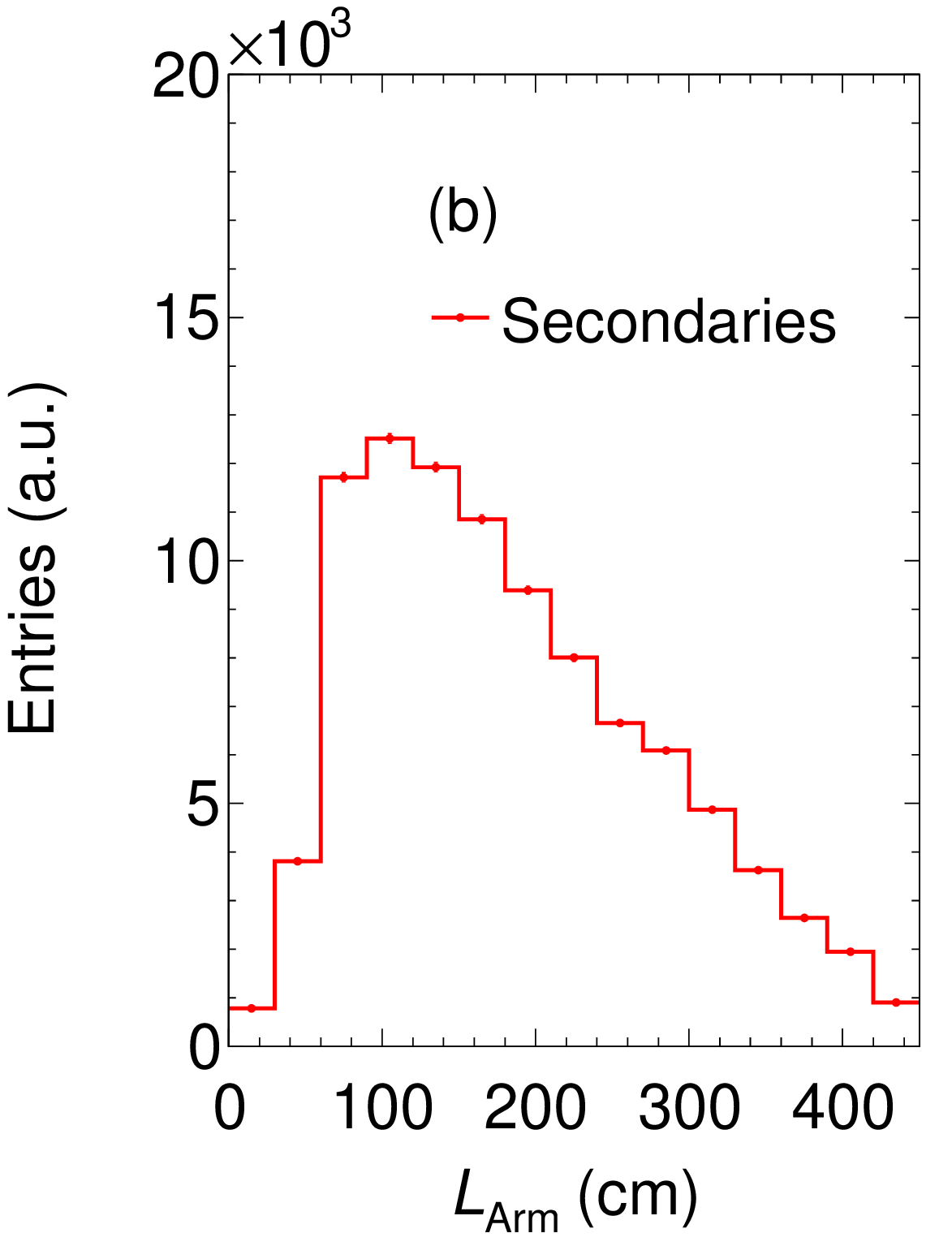}
         \caption{}
         \label{fig:LGAr}
     \end{subfigure}
          \begin{subfigure}[b]{0.32\textwidth}
         \centering
         \includegraphics[width=\textwidth]{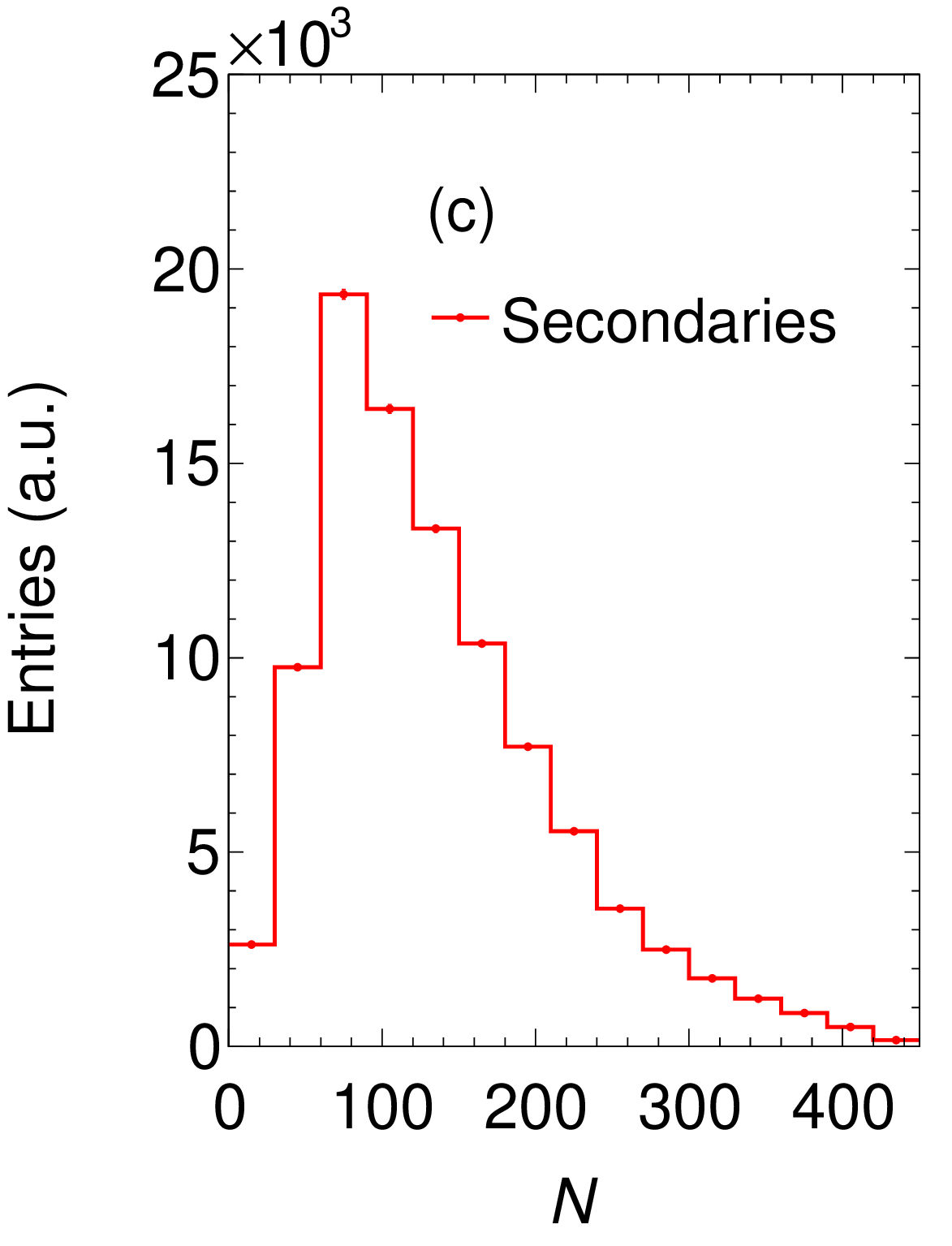}
         \caption{}
         \label{fig:NGAr}
     \end{subfigure}
        \caption{Distributions of (a) transverse momentum $p_{\textrm{T}}$,  (b) lever arm $L_{\textrm{Arm}}$ and (c) number of points per track $N$ in the HP sample. For the HP sample only secondaries are produced, to emulate particles produced in neutrino interactions inside the detector. } \label{fig:GArProperties}
\end{figure}

A second sample containing a total of $10^5$ particle tracks was produced to recreate conditions analogous to the ones that would be experienced by a HPgTPC in a accelerator neutrino experiment, such as the ND-GAr detector. The goal for this second sample is to explore the potential performance of such a detector, using realistic particle spectra and spatial resolutions. This sample will be referred to as the high-pressure sample or HP sample. The HP sample is produced with randomized starting positions in the same manner as the ones applied to the secondaries in the PS sample. This is done to emulate the randomness of particle track formation in a neutrino experiment. The detector characteristics are fixed, having the same cylinder dimensions and pad distribution of the previous sample. The point resolutions are taken as $\sigma_{r\phi}=\sigma_z=0.1 \ \text{cm}$, comparable to what is quoted by ALICE~\cite{LIPPMANN2012}. This figure is also used as the benchmark to estimate the point resolution for ND-GAr, given that the MWPC used by the two detector are going to be the same. The gas is a mixture of argon and methane at a 90 to 10 ratio at 10 atm of pressure, which is the nominal gas suggested for the ND-GAr detector in the DUNE Near Detector~\cite{DUNE:2021tad}. This composition corresponds to a $X_0=1.193 \times 10^3 \ \text{cm}$ and a density of $\rho = 0.01677 \ \text{g/cm}^3$. Only three particle types are considered in this case: muons, pions and protons. These were chosen because they are the key particles produced in $\nu_\mu$ charged-current interactions that are the most relevant in an accelerator neutrino experiment such as DUNE. The initial transverse momenta are randomized to be uniformly distributed between 0.01 GeV$/c$ and 5 GeV$/c$ and the angles are randomized  over the whole spectrum. The $p_{\text{T}}$, $L_\textrm{Arm}$ and $N$ distributions for the sample are shown in Fig.~\ref{fig:GArProperties}. 

\subsection{Tests and results: parameter scan sample}
\label{sec:Validation}

The study performed on the PS sample focuses on the validation of the \texttt{Seed} and \texttt{CKF} algorithms as well as on evaluating the improvement in performance produced by the mirroring technique introduced in Sec.~\ref{sec:Loop}. The first test performed on the PS sample was a so-called pull test. A pull, $\Pi$, is defined as the difference between the true value and the reconstructed value of one of the state vector parameters $s=(y,z,\sin{\phi},\tan{\lambda},$ $ q/p_{\text{T}})=(s_0,s_1,s_2,s_3,s_4)$, normalized by the square root of the correspondent diagonal element of the covariance matrix $C_{ii}$:
\begin{equation}
    \Pi_i\equiv\frac{s_i-s_i^{\text{true}}}{\sqrt{C_{ii}}}.
\end{equation}
If the covariance matrix is well defined, the distributions of the pulls should be normal, centered in 0  with $\sigma\simeq1$.

\begin{figure}[!ht]
     \centering
     \begin{subfigure}{0.32\textwidth}
         \centering
         \includegraphics[width=\textwidth]{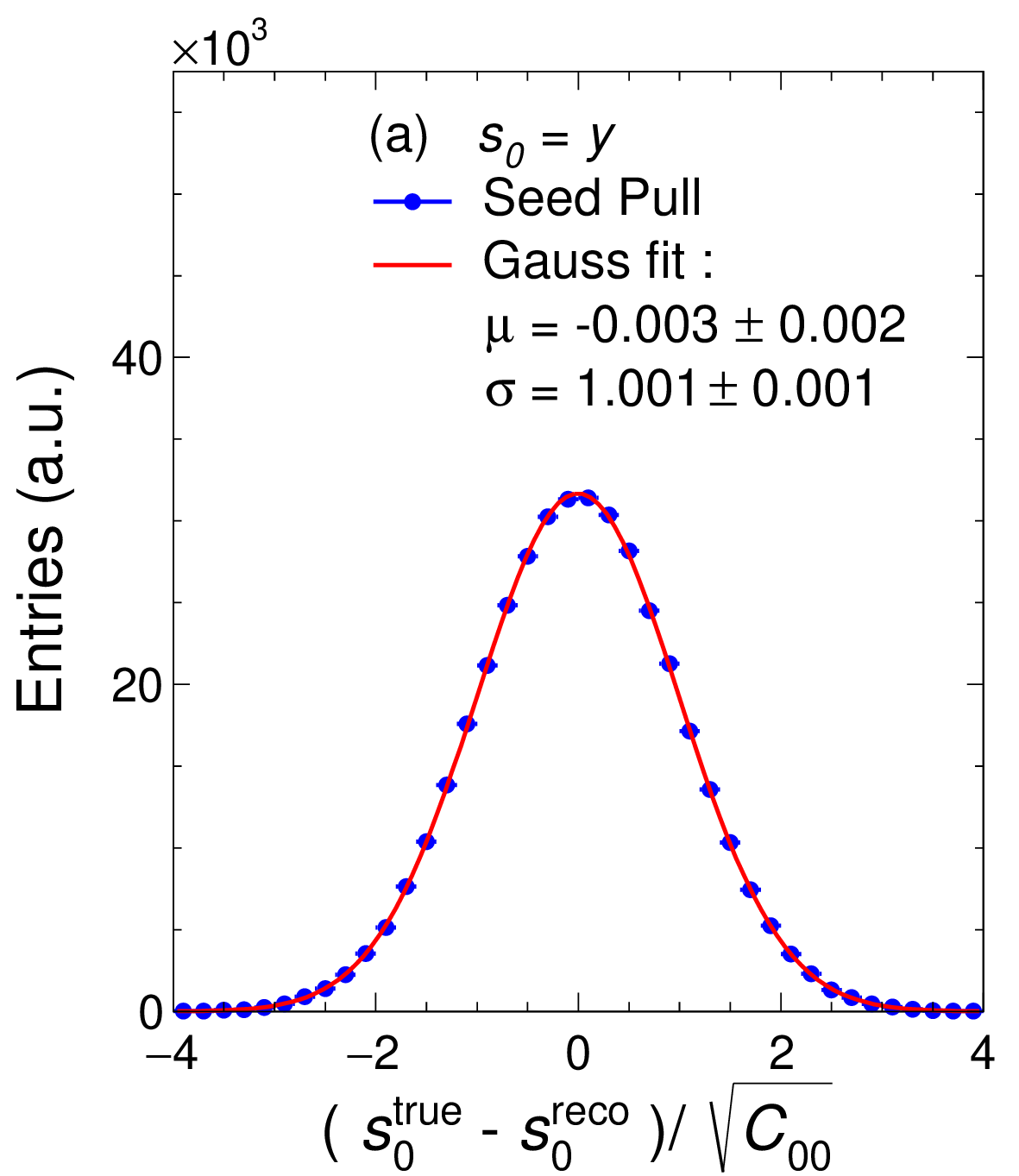}
         \caption{}
         \label{fig:resp0SeedGAr}
     \end{subfigure}
     \begin{subfigure}{0.32\textwidth}
         \centering
         \includegraphics[width=\textwidth]{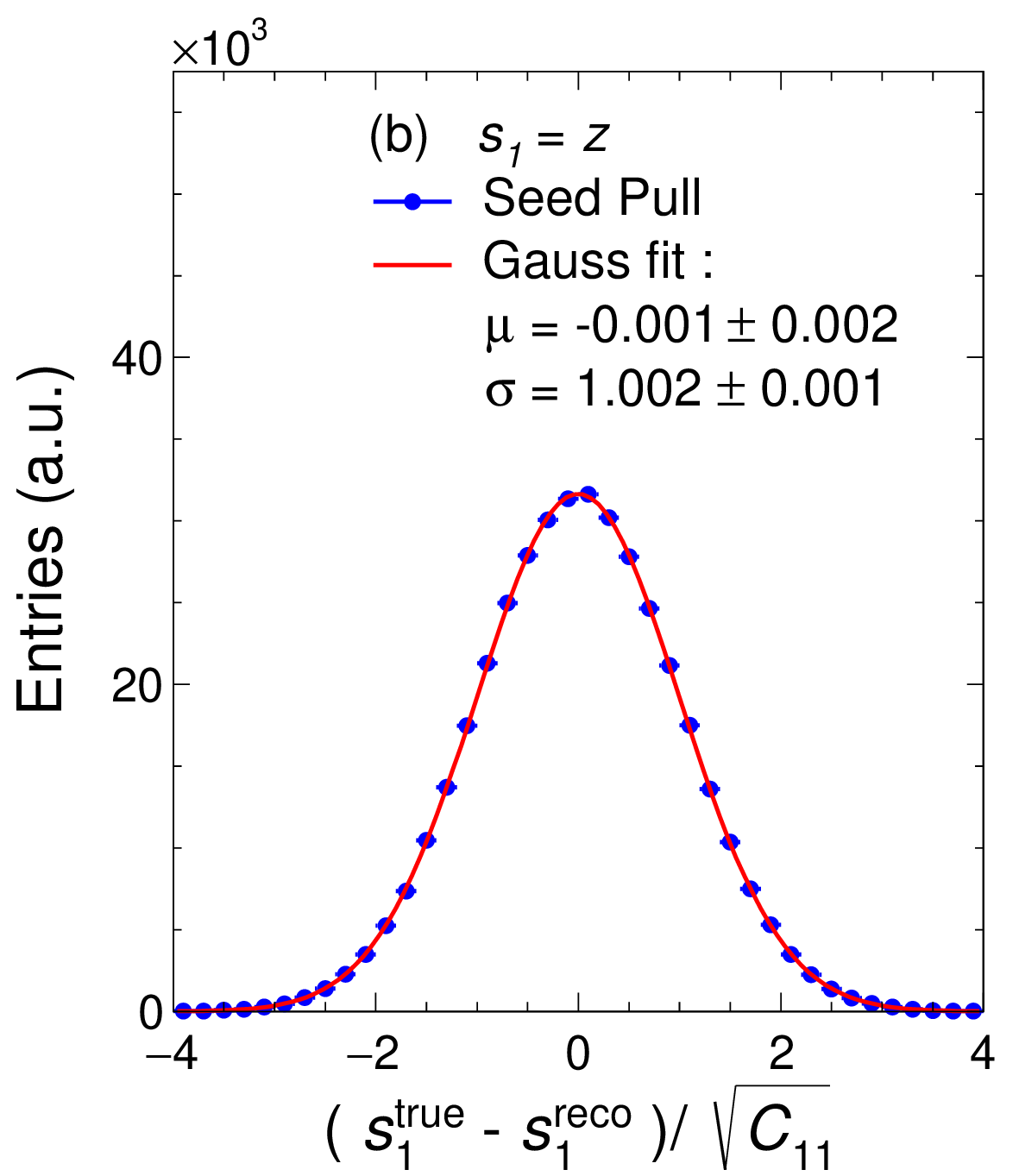}
         \caption{}
         \label{fig:resp1SeedGAr}
     \end{subfigure}
    \begin{subfigure}{0.32\textwidth}
         \centering
         \includegraphics[width=\textwidth]{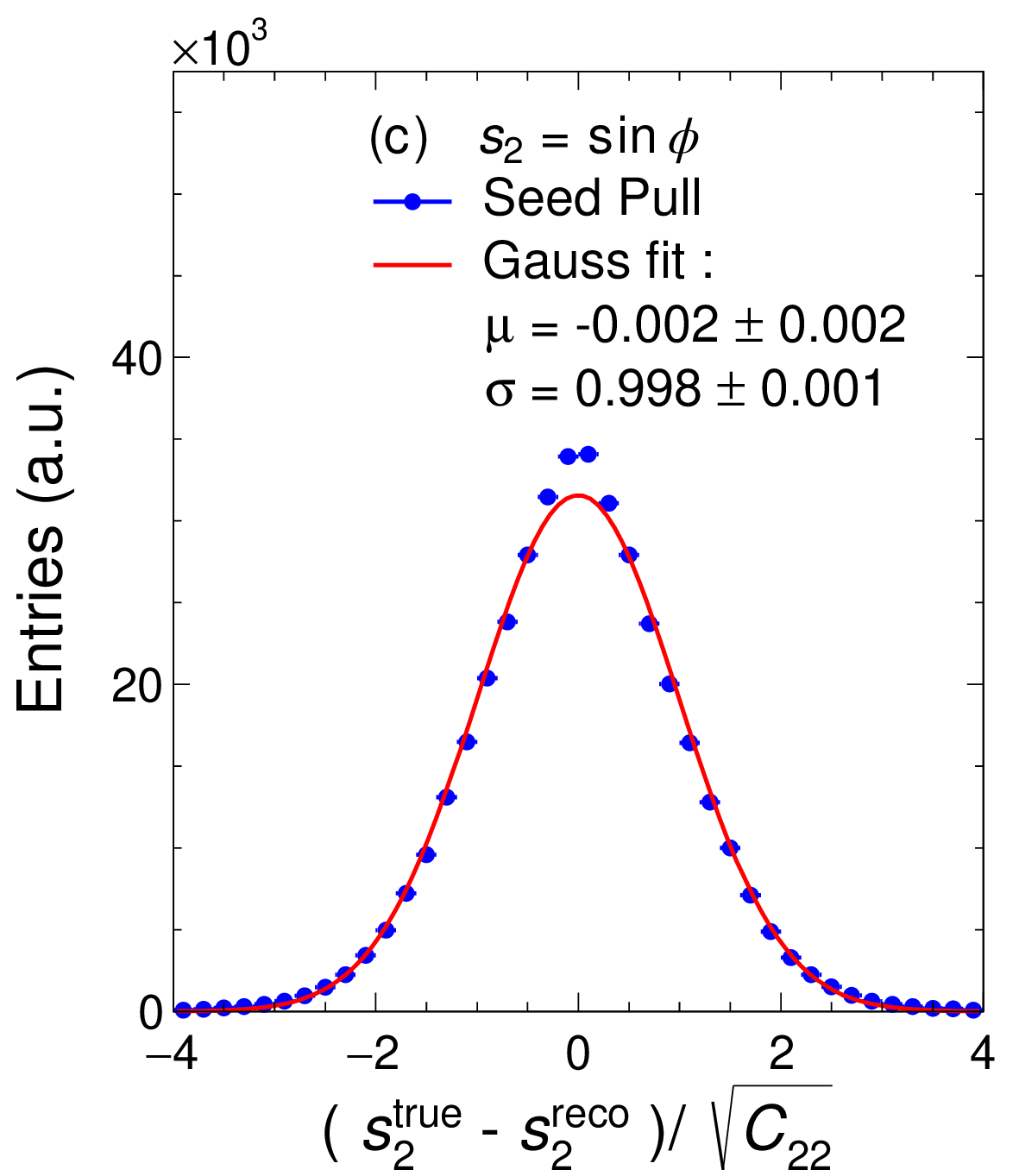}
         \caption{}
         \label{fig:resp2SeedGAr}
     \end{subfigure}
          \begin{subfigure}{0.32\textwidth}
         \centering
         \includegraphics[width=\textwidth]{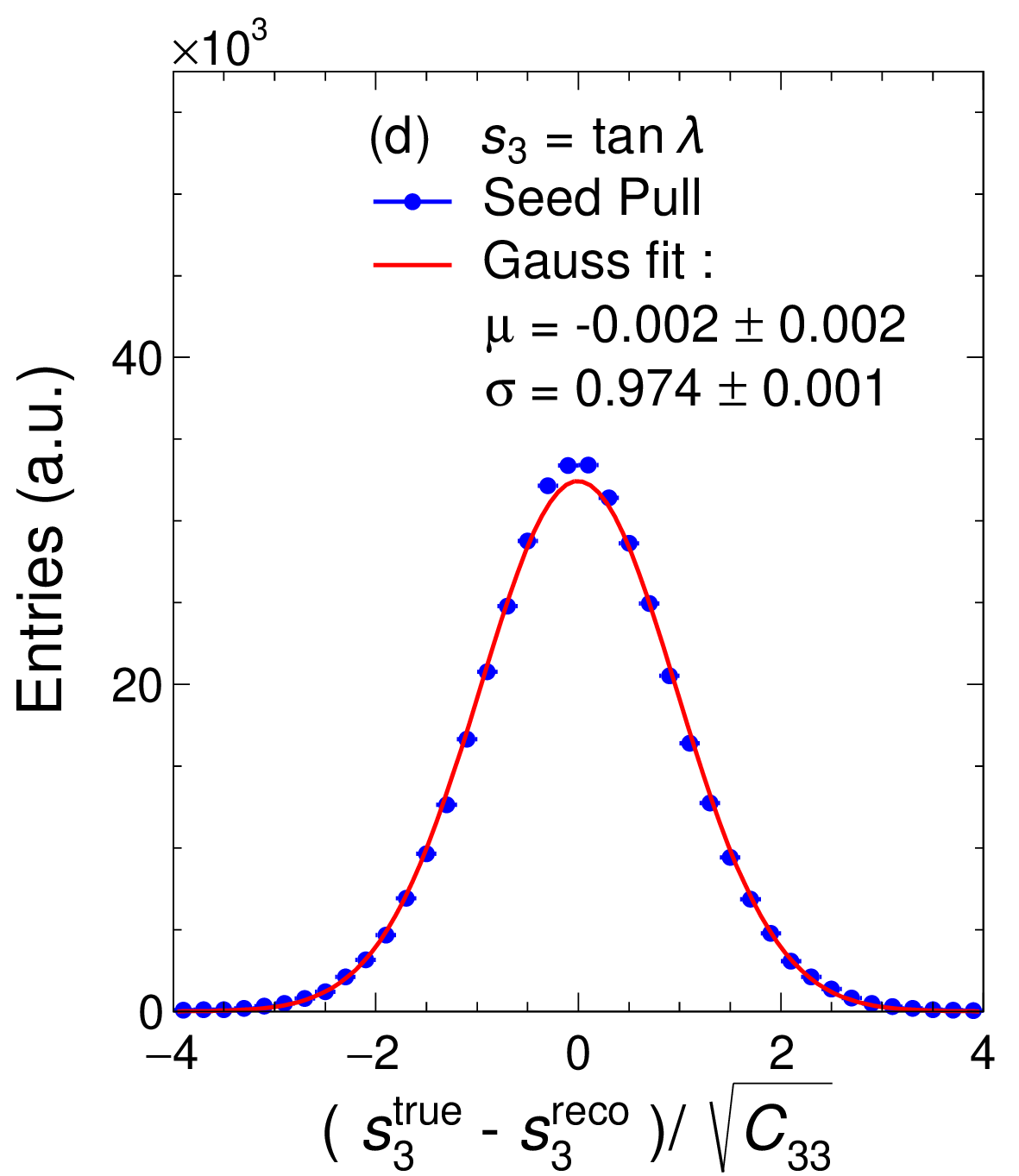}
         \caption{}
         \label{fig:resp3SeedGAr}
     \end{subfigure}
     \begin{subfigure}{0.32\textwidth}
         \centering
         \includegraphics[width=\textwidth]{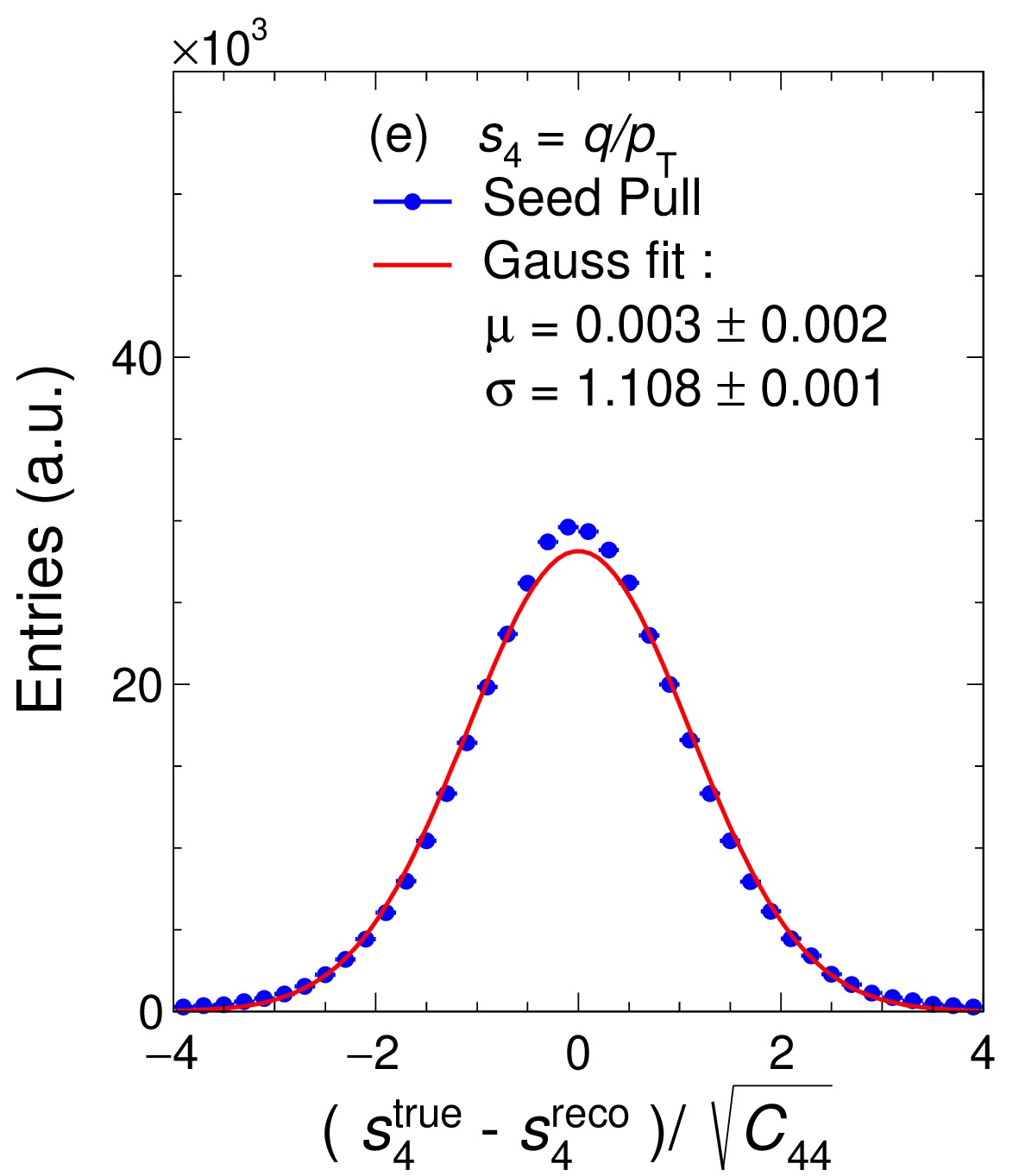}
         \caption{}
         \label{fig:resp4SeedGAr}
     \end{subfigure}
        \caption{Pull distributions for the \texttt{Seed} algorithm over the whole PS sample. All distributions were fitted to a Gaussian function. Results for parameters $s_0$ to $s_4$ (i.e. $y$, $x$, $\sin\phi$, $\tan\lambda$ and $q/p_{\text{T}}$) are shown from left to right and labeled from (a) to (e) accordingly. }
        \label{fig:UnitGAr}
\end{figure}

\begin{figure}[!ht]
     \centering
     \begin{subfigure}{0.32\textwidth}
         \centering
         \includegraphics[width=\textwidth]{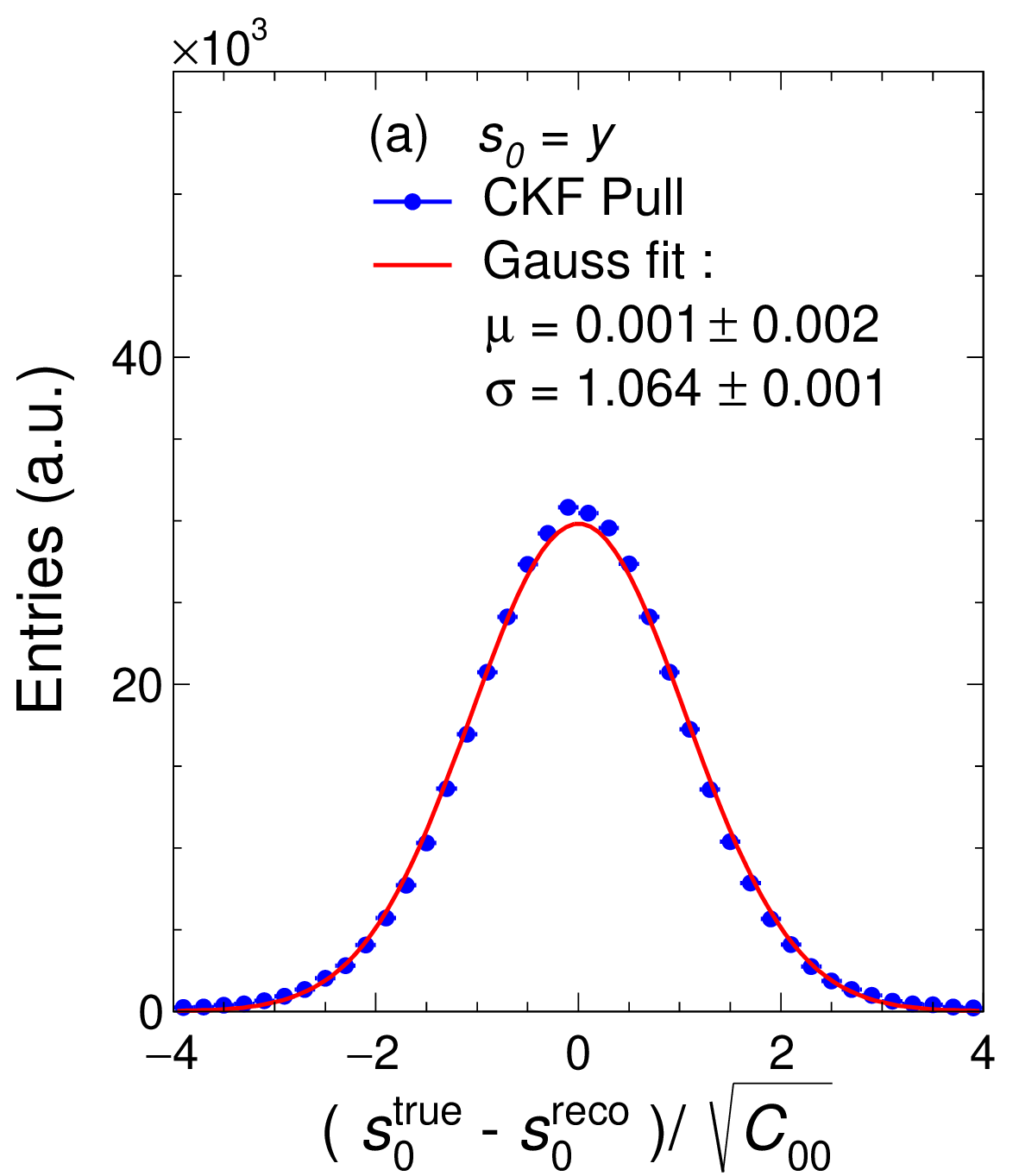}
         \caption{}
         \label{fig:resp0KFGAr}
     \end{subfigure}
     \begin{subfigure}{0.32\textwidth}
         \centering
         \includegraphics[width=\textwidth]{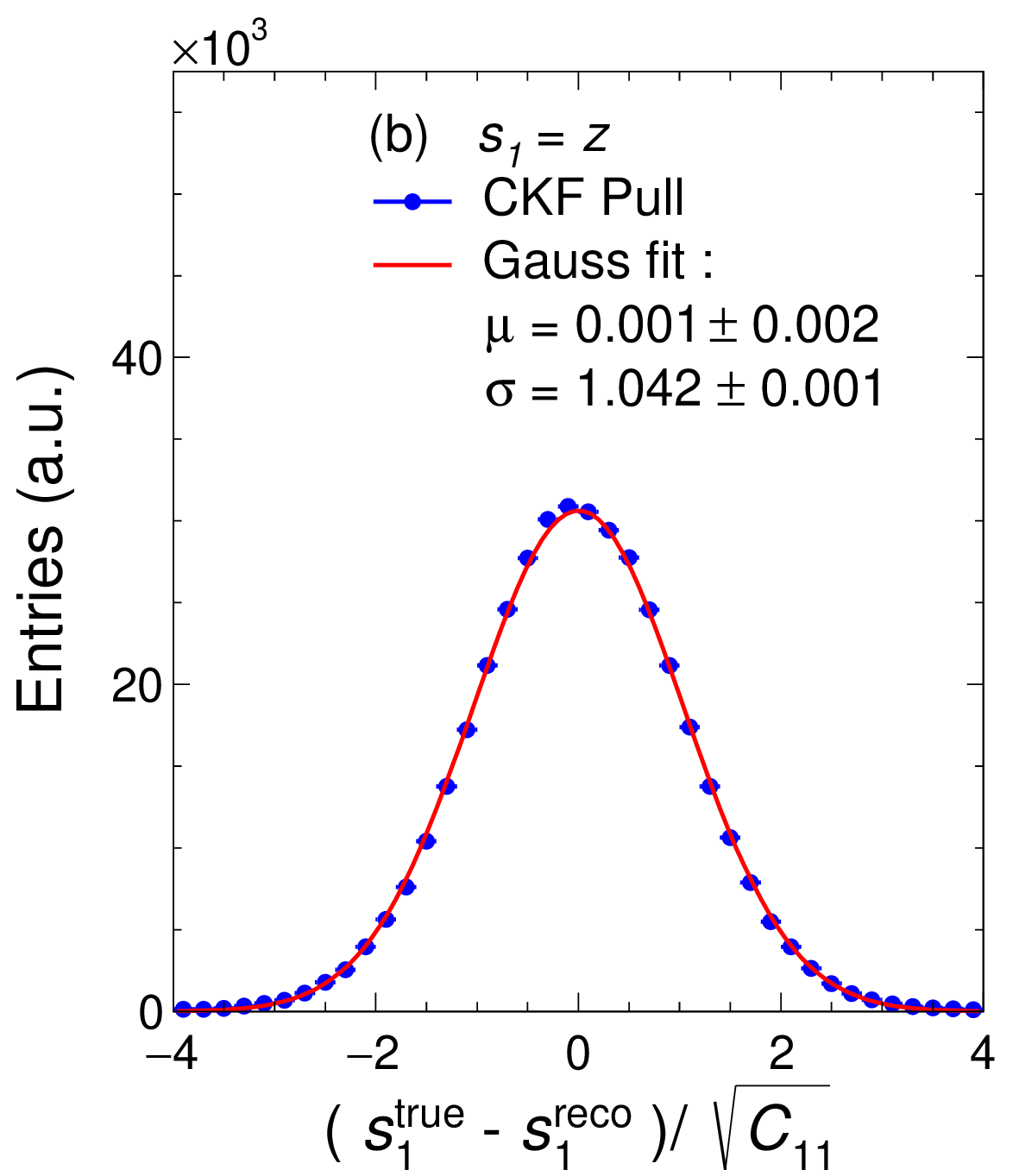}
         \caption{}
         \label{fig:resp1KFGAr}
     \end{subfigure}
    \begin{subfigure}{0.32\textwidth}
         \centering
         \includegraphics[width=\textwidth]{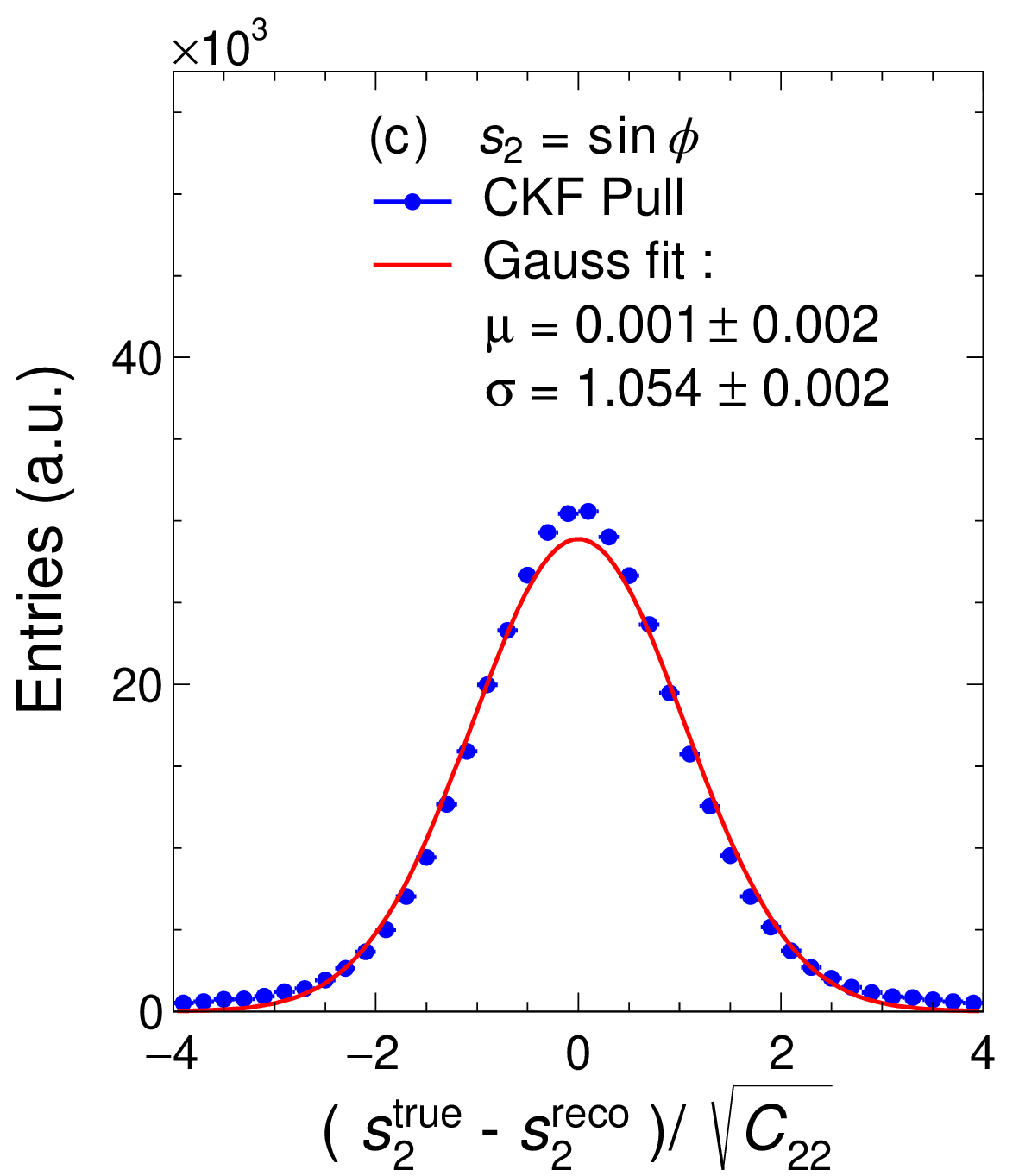}
         \caption{}
         \label{fig:resp2KFGAr}
     \end{subfigure}
          \begin{subfigure}{0.32\textwidth}
         \centering
         \includegraphics[width=\textwidth]{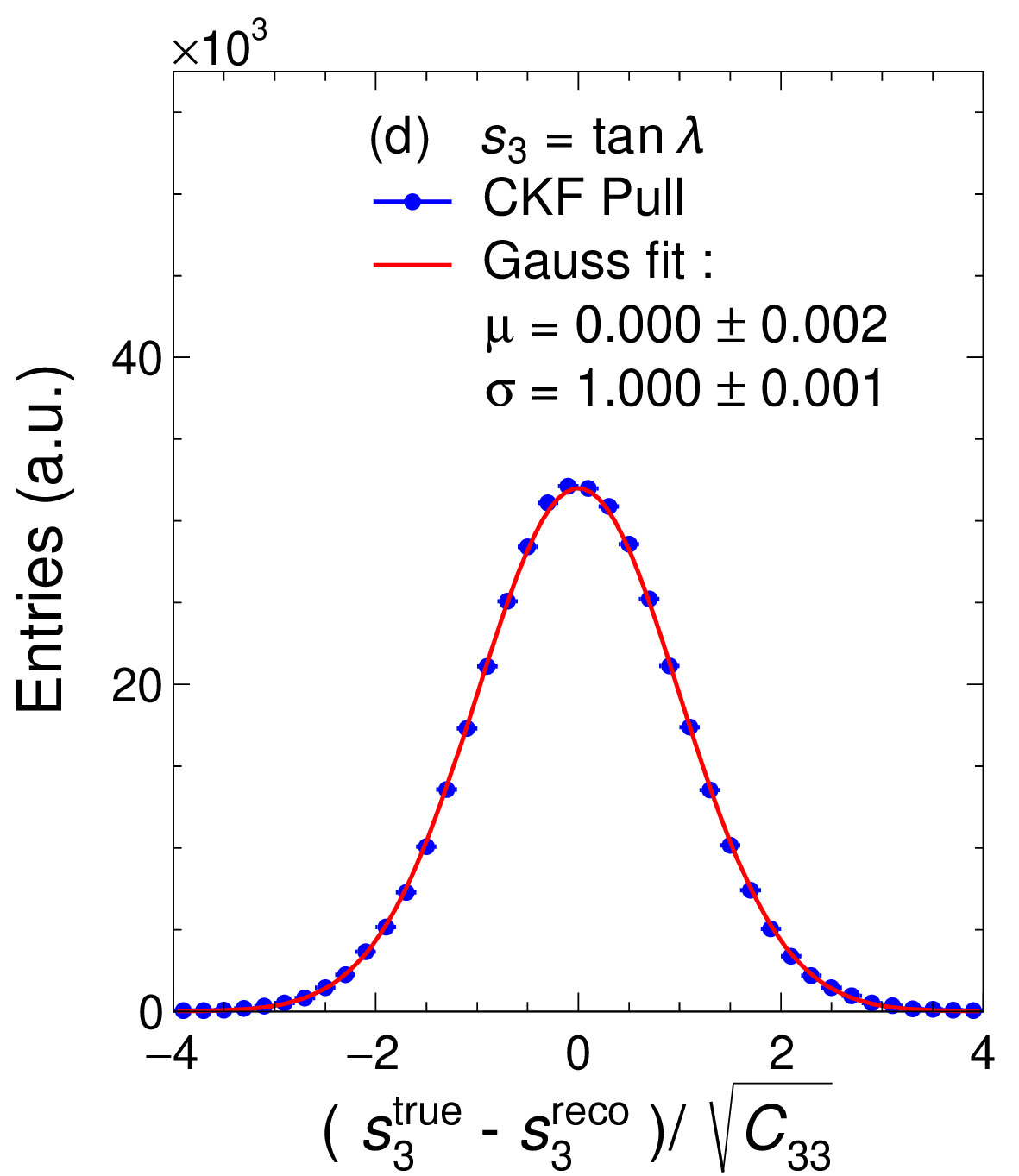}
         \caption{}
         \label{fig:resp3KFGAr}
     \end{subfigure}
     \begin{subfigure}{0.32\textwidth}
         \centering
         \includegraphics[width=\textwidth]{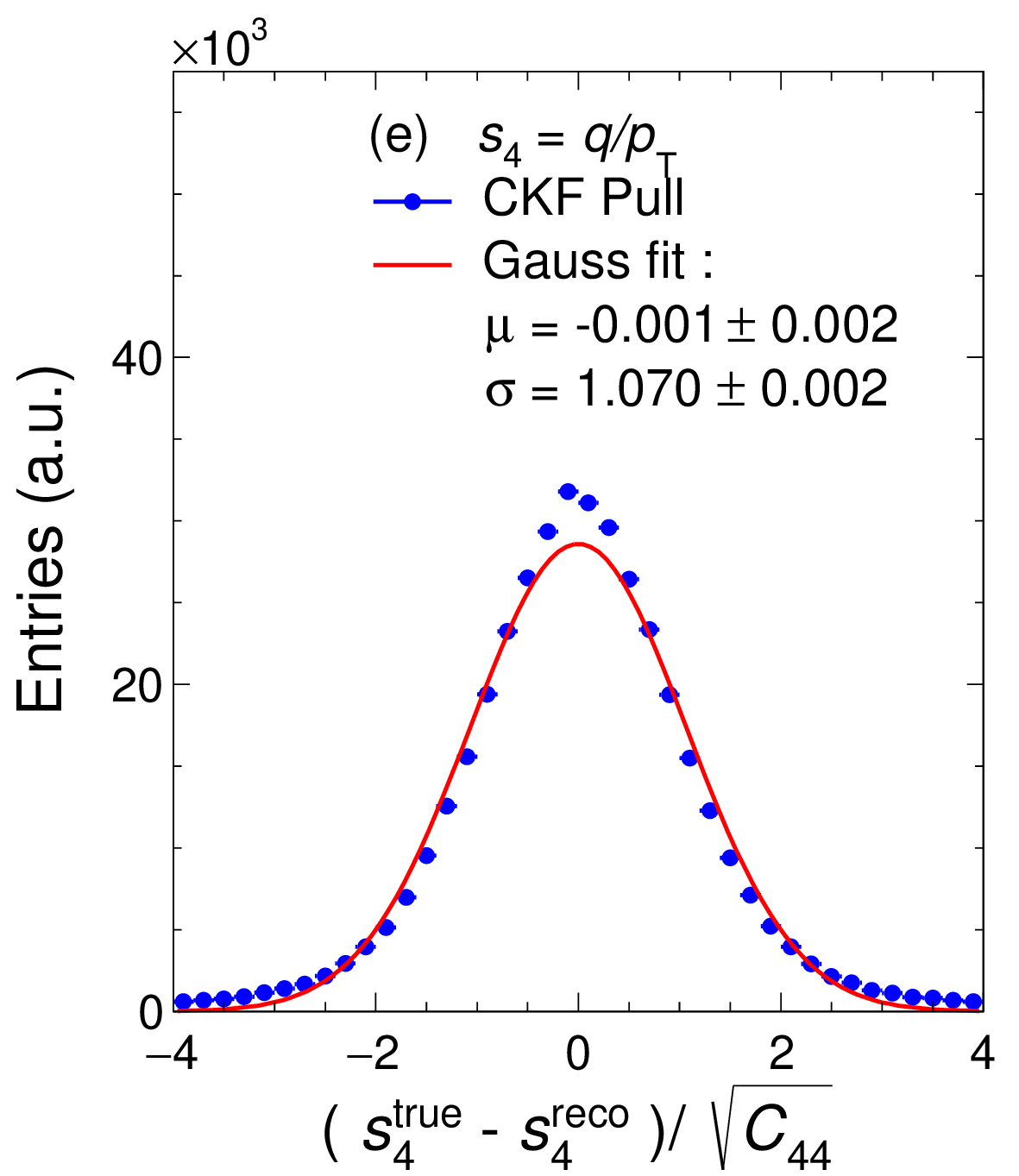}
         \caption{}
         \label{fig:resp4KFGAr}
     \end{subfigure}
        \caption{Pull distributions obtained after the full propagation of the \texttt{CKF} algorithm over the whole PS sample. All distributions were fitted to a Gaussian function. Results for parameters $s_0$ to $s_4$ (i.e. $y$, $x$, $\sin\phi$, $\tan\lambda$ and $q/p_{\text{T}}$) are shown from left to right and labeled from (a) to (e) accordingly.}
        \label{fig:UnitGArKF}
\end{figure}

The pulls were tested for the sample, both for the results of the \texttt{Seed} and for the estimates evaluated at the start of the track, after the full propagation of the \texttt{CKF}: the resulting distributions for all the state vector parameters are shown in Figs.~\ref{fig:UnitGAr} and~\ref{fig:UnitGArKF}, respectively. All the pull distributions were fitted to a standard Gaussian distribution and were found to be centered in 0 and have $\sigma\sim1$; this implies that the diagonal elements of the covariance matrices well describe the uncertainties.The only significant deviations can be seen for $s_4 = q/p_\text{T}$ for which $\sigma\sim 1.1$ and to a lesser extent $s_3 = \tan\lambda$. In both cases the underestimations of the matrix elements are likely due to the approximations made during the energy loss correction step. Due to the fact that the $\text{d}E/\text{d}x$ depends on the momentum of the particle and indirectly its energy, the correction procedure is estimated using a step integration. However this method can only in part compensate for the non-linearity in the energy loss and is not completely reliable over long steps. For this reason the more significant deviations are seen for the \texttt{Seed} algorithm.

\begin{figure}[!ht]
     \centering
     \begin{subfigure}[b]{0.48\textwidth}
         \centering
         \includegraphics[width=\textwidth]{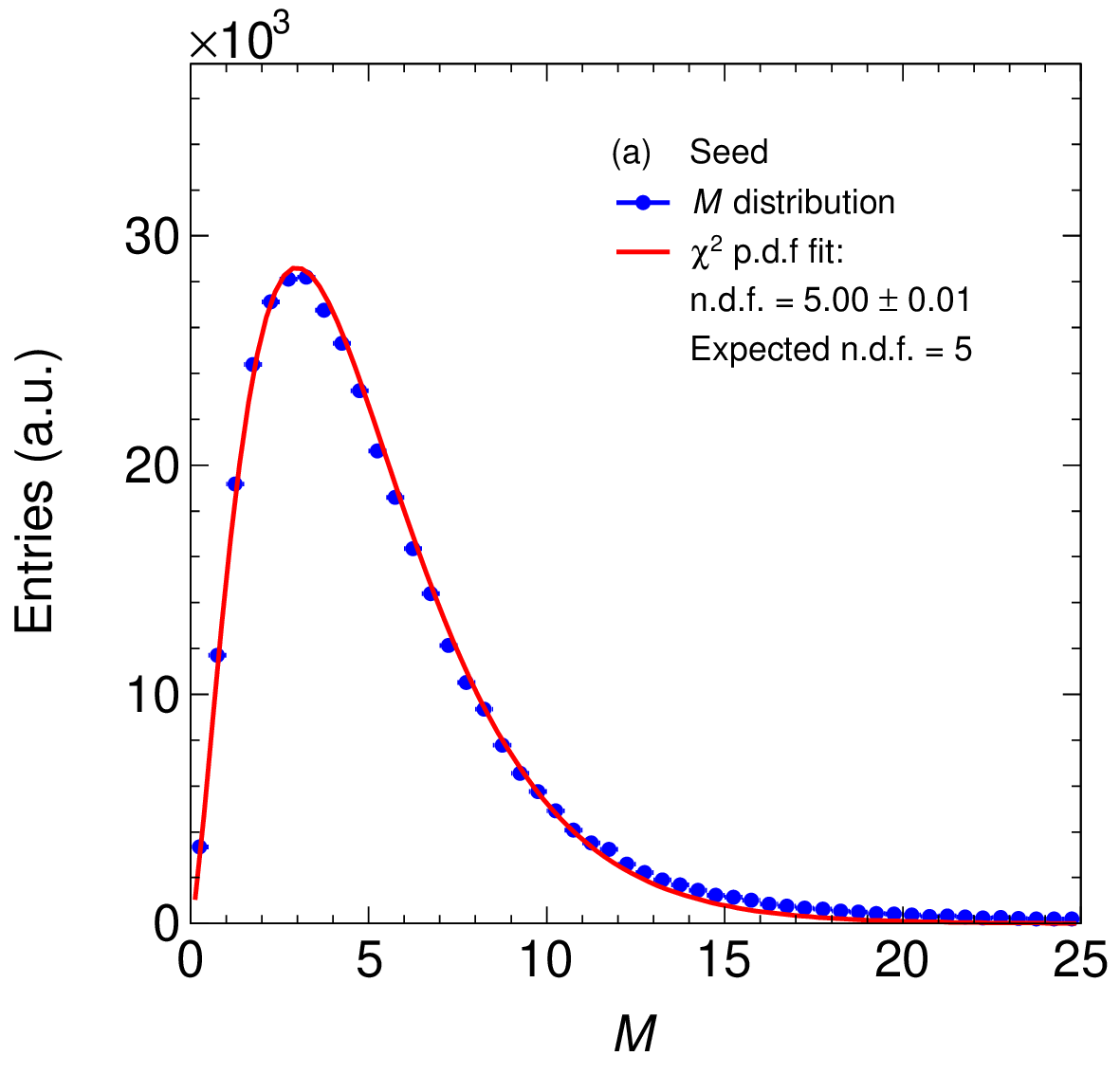}
         \caption{}
         \label{fig:chi2Seed}
     \end{subfigure}
     \begin{subfigure}[b]{0.48\textwidth}
         \centering
         \includegraphics[width=\textwidth]{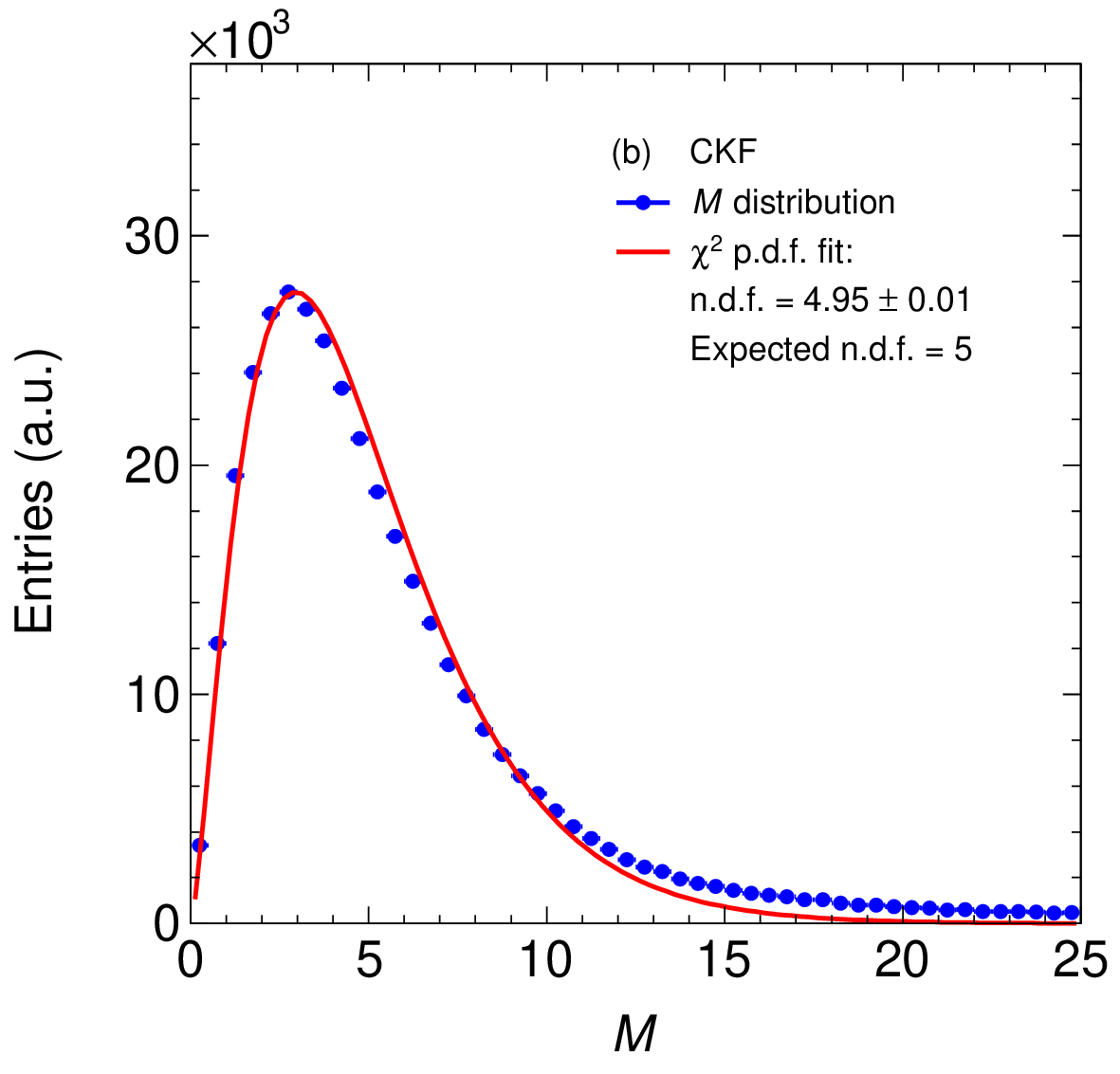}
         \caption{}
         \label{fig:chi2Kalman}
     \end{subfigure}
        \caption{Mahalanobis distance $M$ distribution for the PS Sample fitted by a standard $\chi^2$ p.d.f. showing the results for the $n.d.f.$ parameter. The expected result for a 5-dimensional matrix is $n.d.f. = 5$. The results for the \texttt{Seed} and for the fully propagated \texttt{CKF} are shown in plots (a) and (b) respectively. } \label{fig:chi2}
\end{figure}

The standard pull distributions, while being effective at testing the uncertainties associated with the individual parameters, do not provide any information regarding the off-diagonal correlation terms. In order to test the quality of the estimates for the full covariance matrix, the Mahalanobis distance was used~\cite{M-distance}. Given a probability distribution, $D$, on $\mathbb{R}^n$ with mean $\mu$ and positive-definite covariance matrix, $C$, the Mahalanobis distance, $M$, of a point $s$ from $D$, is defined as:
\begin{equation}
    M = \sqrt{(s-\mu)^TC^{-1}(s-\mu)},
\end{equation}
where in our case, $\mu$ corresponds to the true value of state vector, $s^{\text{true}}$, $s$ and $C$ are the estimates obtained from the reconstruction and $n=5$. The Mahalanobis distance, $M$, of a set of points belonging to the distribution $D$, follows a $\chi^2$ distribution with $n$ degrees of freedom. One can check if $C$ is well defined, by verifying that the corresponding $M$ follow a $\chi^2$ distribution with the correct number of degrees of freedom. In Fig.~\ref{fig:chi2}, we show the results of a $\chi^2$ fit over the $M$ distribution for the whole sample. The plot on the left shows the results obtained from the \texttt{Seed} algorithm, while the one on the right shows the results after the full \texttt{CKF} propagation. In both cases the n.d.f. obtained with the $\chi^2$ fit are very close to $n=5$: both \texttt{Seed} and \texttt{CKF} estimates accurately the covariances of the reconstructed track parameters in the state vector. The only deviations are caused by the same $\text{d}E/\text{d}x$ related issue which was pointed out for the Pull test results. It is also important to note that this test is particularly demanding, since all deviations present in the matrix add up to a single figure of merit.

\begin{figure}[!ht]
     \centering
     \begin{subfigure}[b]{0.99\textwidth}
         \centering
         \includegraphics[width=\textwidth]{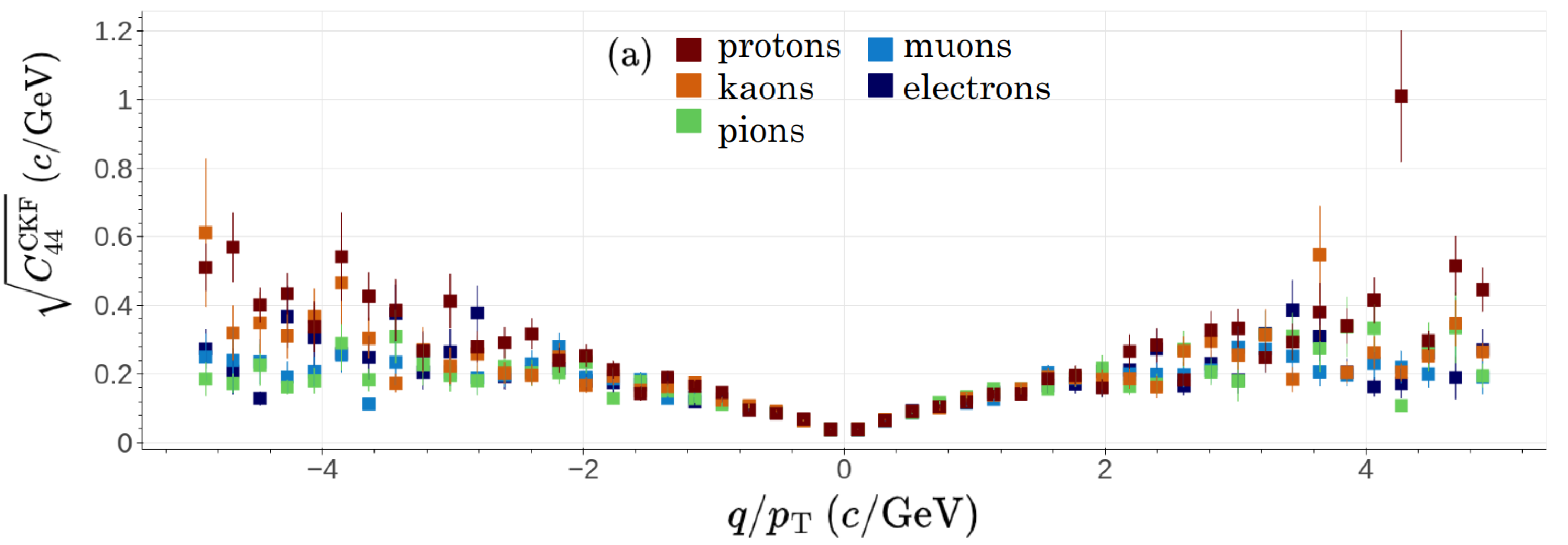}
         \caption{}
         \label{fig:CheckAna_pID_noNorm}
     \end{subfigure}
     \begin{subfigure}[b]{0.99\textwidth}
         \centering
         \includegraphics[width=\textwidth]{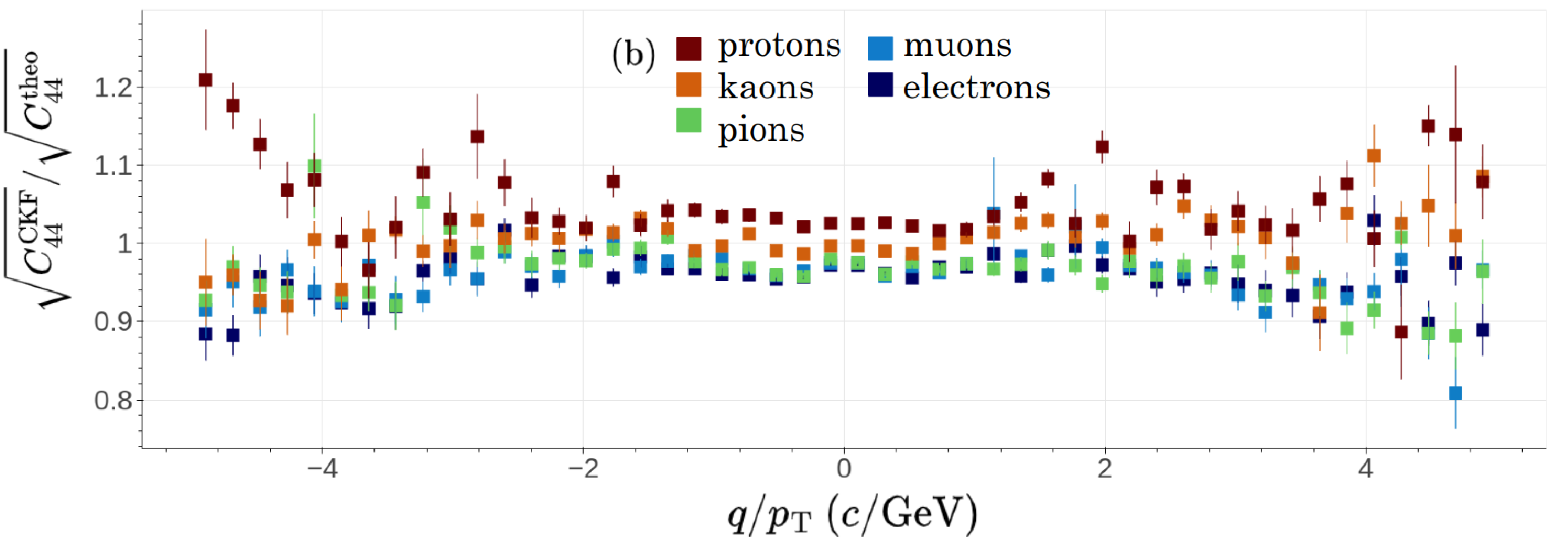}
         \caption{}
         \label{fig:CheckAna_pID_Norm}
     \end{subfigure}
        \caption{ (a) \texttt{CKF} $q/p_{\text{T}}$ resolution $\sigma_{\text{CKF}}(q/p_{\text{T}})=\sqrt{C_{44}^{\textrm{CKF}}}$ as a function of the true $q/p_{\text{T}}$. (b) Ratio of the \texttt{CKF} $q/p_{\text{T}}$ resolution, over the theoretical expectations $\sigma_{\text{theo}}(q/p_\text{T})=\sqrt{C_{44}^{\textrm{theo}}}$, as a function of the true $q/p_\text{T}$. The histograms include all particles in the PS sample and are color-coded according to their particle types $t_\textrm{ID} = (0,1,2,3,4) = (\textrm{e},\mu,\pi,K,\textrm{p})$. Only tracks with a minimum of 10 points are considered. These plots have been produced using the interactive analytical tool \texttt{ROOTInteractive}~\cite{RootInt}. The error bars are statistical. See Fig. \ref{fig:Check_Ana_pID_zoom} in Appendix for a zoom-out version.}
        \label{fig:Check_Ana_pID}
\end{figure}

\begin{figure}[!ht]
     \centering
     \begin{subfigure}[b]{0.99\textwidth}
         \centering
         \includegraphics[width=\textwidth]{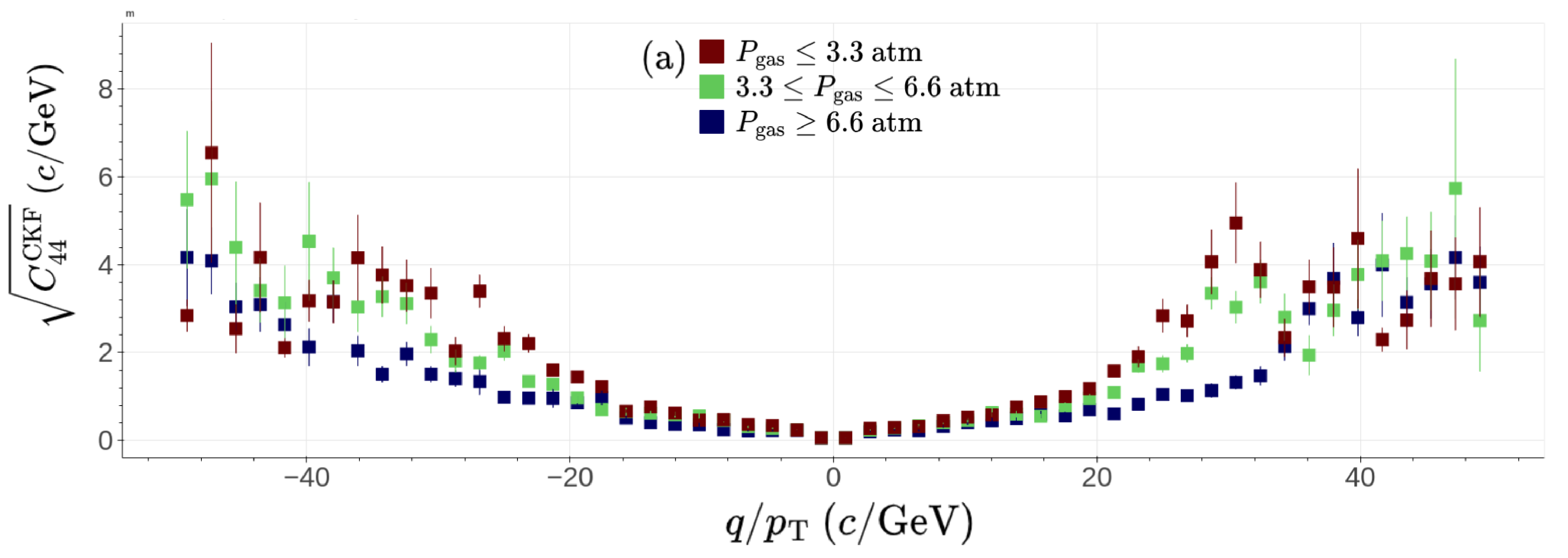}
         \caption{}
         \label{fig:CheckAna_dens_noNorm}
     \end{subfigure}
     \begin{subfigure}[b]{0.99\textwidth}
         \centering
         \includegraphics[width=\textwidth]{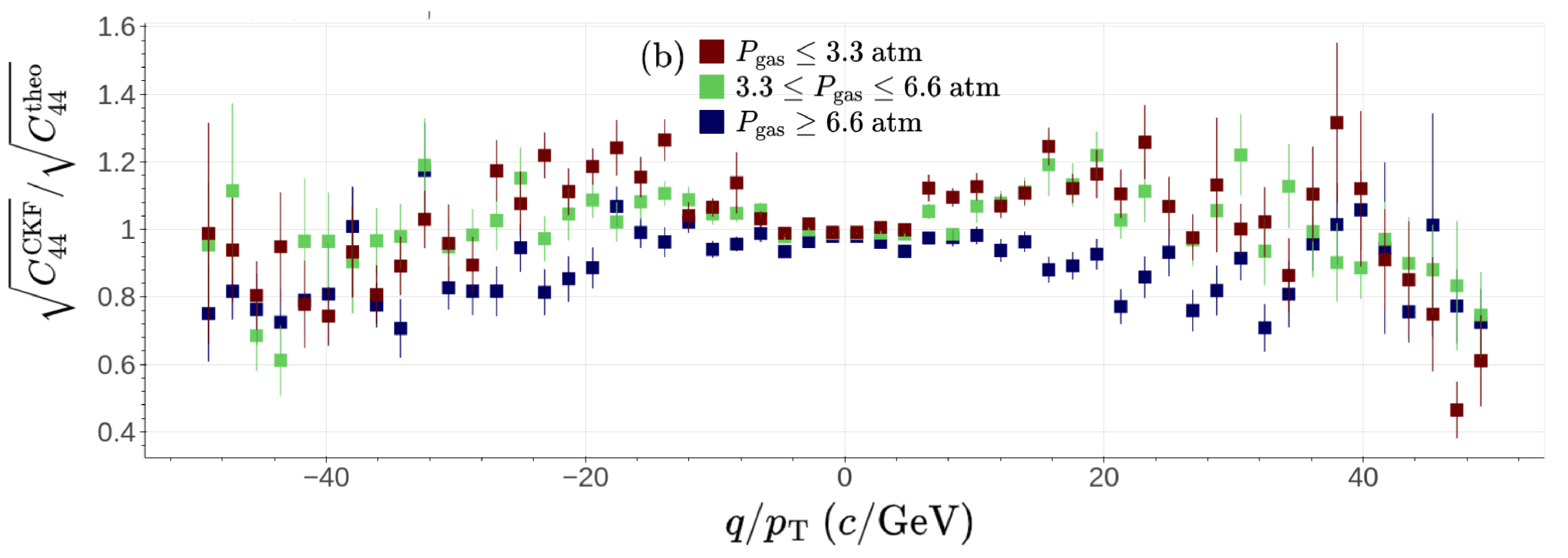}
         \caption{}
         \label{fig:CheckAna_dens_Norm}
     \end{subfigure}
        \caption{ (a) \texttt{CKF} $q/p_{\text{T}}$ resolution $\sigma_{\text{CKF}}(q/p_{\text{T}})=\sqrt{C_{44}^{\textrm{CKF}}}$ as a function of the true $q/p_{\text{T}}$. (b) Ratio of the \texttt{CKF} $q/p_{\text{T}}$ resolution, over the theoretical expectations $\sigma_{\text{theo}}(q/p_\text{T})=\sqrt{C_{44}^{\textrm{theo}}}$, as a function of the true $q/p_\text{T}$. The histograms include all particles in the PS sample and are color-coded according to the gas pressure $P_\text{gas}$ used in the simulation. Only tracks with a minimum of 10 points are considered. These plots have been produced using the interactive analytical tool \texttt{ROOTInteractive}~\cite{RootInt}. The error bars are statistical.}
        \label{fig:CheckAna_dens}
\end{figure}

\begin{figure}[!ht]
     \centering
     \begin{subfigure}[b]{0.99\textwidth}
         \centering
         \includegraphics[width=\textwidth]{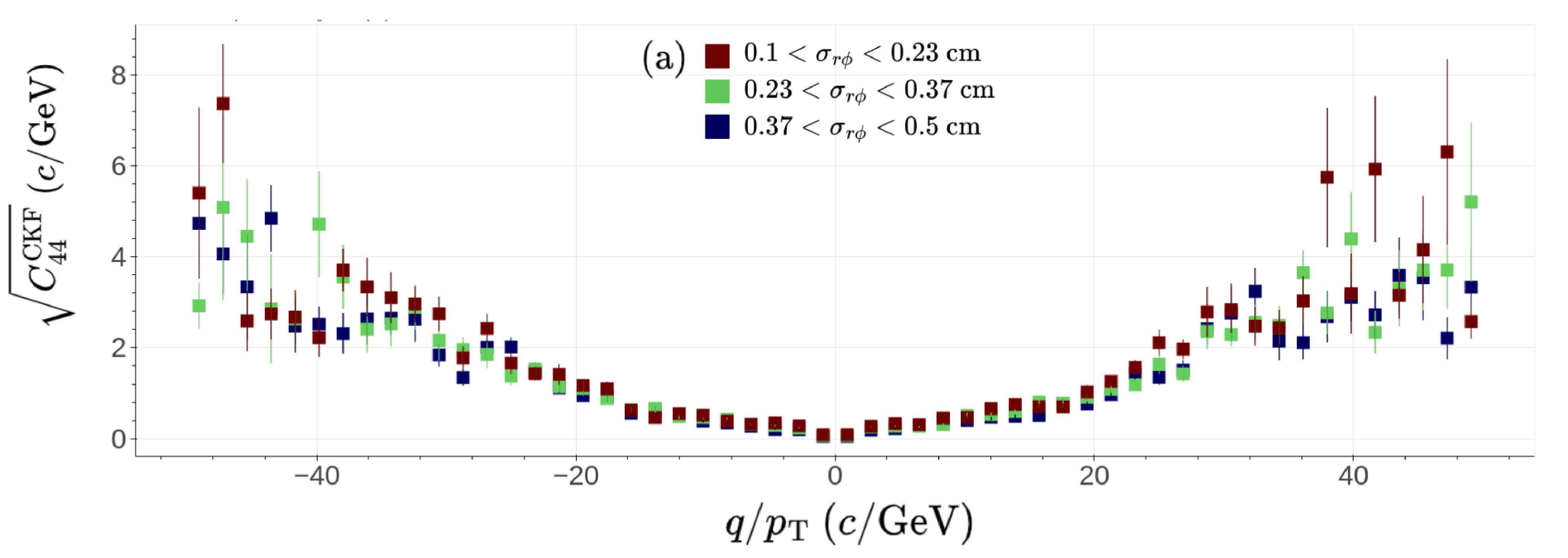}
         \caption{}
         \label{fig:CheckAna_res_noNorm}
     \end{subfigure}
     \begin{subfigure}[b]{0.99\textwidth}
         \centering
         \includegraphics[width=\textwidth]{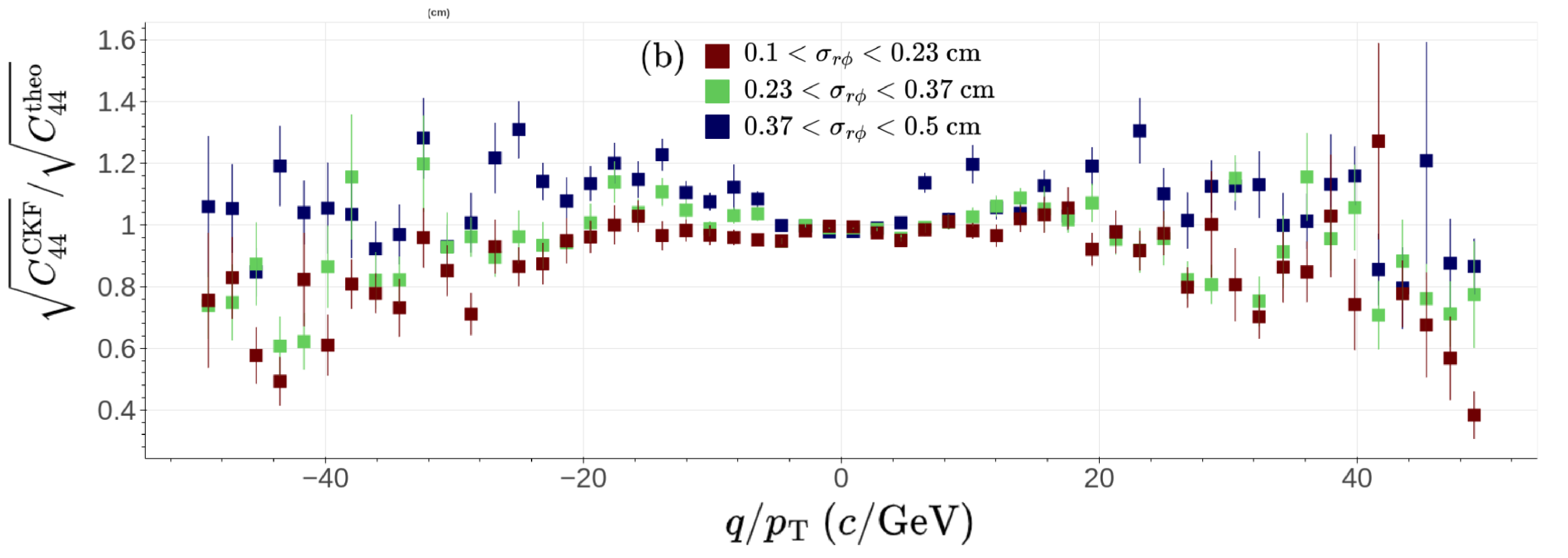}
         \caption{}
         \label{fig:CheckAna_res_Norm}
     \end{subfigure}
        \caption{(a) \texttt{CKF} $q/p_{\text{T}}$ resolution $\sigma_{\text{CKF}}(q/p_{\text{T}})=\sqrt{C_{44}^{\textrm{CKF}}}$ as a function of the true $q/p_{\text{T}}$. (b) Ratio of the \texttt{CKF} $q/p_{\text{T}}$ resolution, over the theoretical expectations $\sigma_{\text{theo}}(q/p_\text{T})=\sqrt{C_{44}^{\textrm{theo}}}$, as a function of the true $q/p_\text{T}$. The histograms include all particles in the PS sample and are color-coded according to the radial resolution $\sigma_{r\phi}$ used in the simulation. Only tracks with a minimum of 10 points are considered. These plots have been produced using the interactive analytical tool \texttt{ROOTInteractive}~\cite{RootInt}. The error bars are statistical.}
        \label{fig:Check_Ana_res}
\end{figure}

The PS sample was also used to test whether the \texttt{CKF} algorithm produced results that are consistent with the theoretical expectations. The analytical formula for the ideal $q/p_{\text{T}}$ resolution, $\sigma_{\text{theo}}(q/p_{\text{T}})=\sqrt{C_{44}^{\textrm{theo}}}$, obtainable using a curvature measurement in a TPC, can be written as~\cite{PDG:34,Gluckstern:1963}:
\begin{equation}
   \sigma_{\text{theo}}(1/p_{\text{T}}) = \sqrt{C_{44}^{\textrm{theo}}} = \sqrt{\sigma_{\text{H}}^2+\sigma_{\text{MS}}^2}.
\end{equation}
The $\sigma_{\text{H}}$ component is determined by the point resolution and can be written as:
\begin{equation}\label{eq:sigmaN}
    \sigma_{\text{H}}(1/p_{\text{T}})=\frac{\sigma_{r\phi}}{0.3BL_{\textrm{Arm}}^2}\sqrt{\frac{720}{N+4}}.
\end{equation} 
The multiple scattering component can be written as:
\begin{equation}\label{eq:sigmaMS}
    \sigma_{\text{MS}}(1/p_{\text{T}})=\bigg\langle\frac{1}{\beta p_{\text{T}}}\bigg\rangle\frac{0.016 \ (\textrm{GeV}/c)}{0.3 B l\cos\lambda}\sqrt{\frac{l}{X_0}},
\end{equation}
where $l$ is the length of the track in the $xy$-plane. 
Note that the value of $1/\left(\beta p_{\text{T}}\right)$ is averaged along the trajectory to take into account energy loss. In Figs.~\ref{fig:Check_Ana_pID},~\ref{fig:CheckAna_dens} and~\ref{fig:Check_Ana_res},  the upper plots show the \texttt{CKF} covariance estimates, $\sqrt{C_{44}^{\textrm{CKF}}}$, while the bottom plots show their ratios to the theoretical expectations, $\sqrt{C_{44}^{\textrm{theo}}}$. The upper plots are shown to demonstrate how widely the values of $\sqrt{C_{44}^{\textrm{CKF}}}$ vary and to highlight, in comparison, how stable the ratio with the theoretical expectation is.  The points analyzed are randomly taken along the reconstructed tracks and down-sampled to $10\%$ of the total to avoid correlations.


In Fig.~\ref{fig:Check_Ana_pID},the histograms are color-scaled based on particle type. Conversely, in Figs.~\ref{fig:CheckAna_dens} and ~\ref{fig:Check_Ana_res}, the color scaling corresponds to gas pressure, $P_{\textrm{gas}}$, and point resolution $\sigma_{r\phi}=\sigma_z$ respectively. The \texttt{CKF} results show overall good agreement with the theoretical expectation, with ratios $\sim 1$ for momenta down to 20 MeV/$c$ within statistical uncertainties. In Fig. \ref{fig:Check_Ana_pID_zoom} in the Appendix we present an alternative version of Fig.~\ref{fig:Check_Ana_pID} focusing on the central region $q/p_\text{T}\in[-5,5] \ c/\text{GeV}$ which is more statistically significant.

\begin{figure}[!ht]
     \centering
     \begin{subfigure}[b]{0.32\textwidth}
         \centering
         \includegraphics[width=\textwidth]{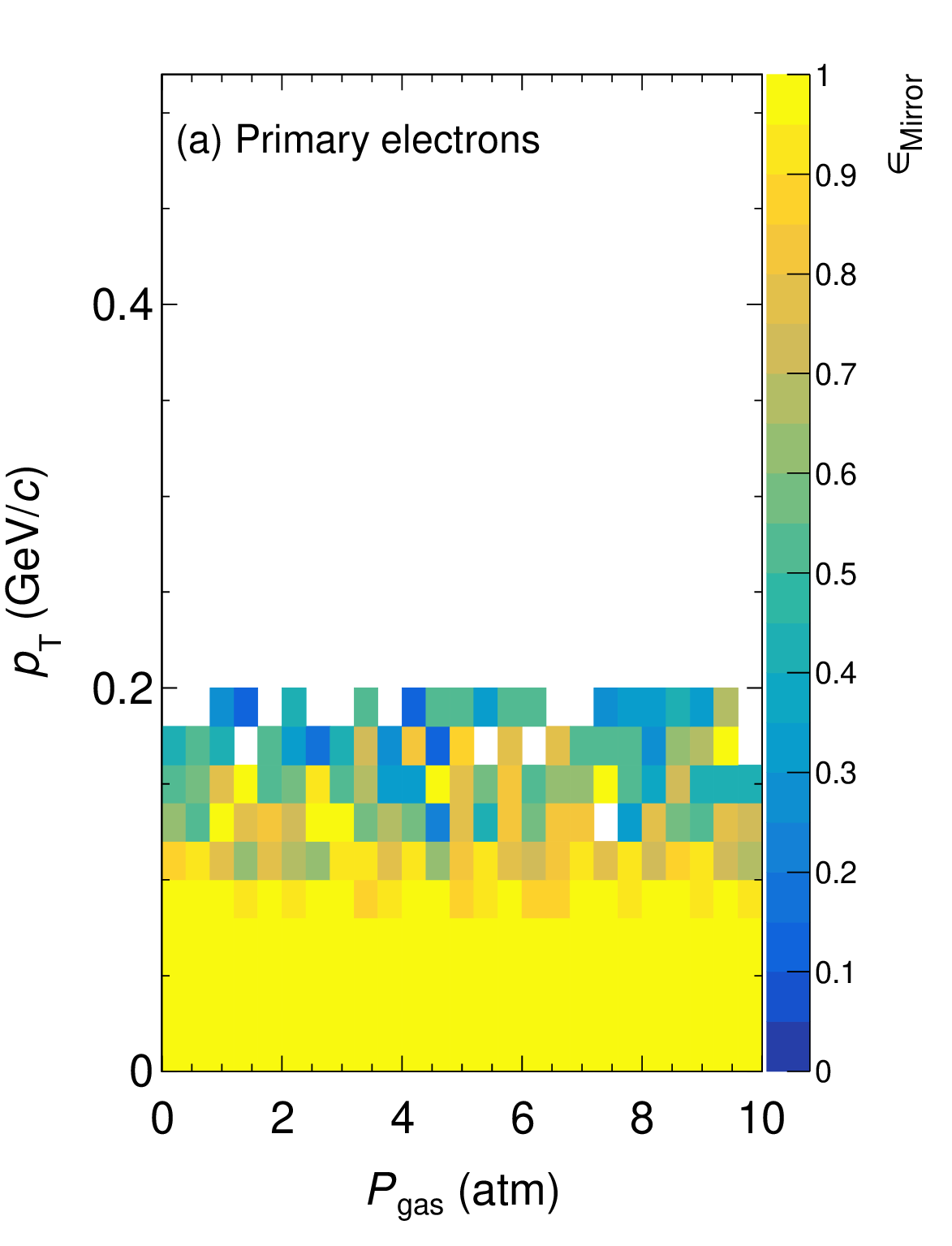}
         \caption{}
         \label{fig:MirrorRatiop_Prim_e}
     \end{subfigure}
     \begin{subfigure}[b]{0.32\textwidth}
         \centering
         \includegraphics[width=\textwidth]{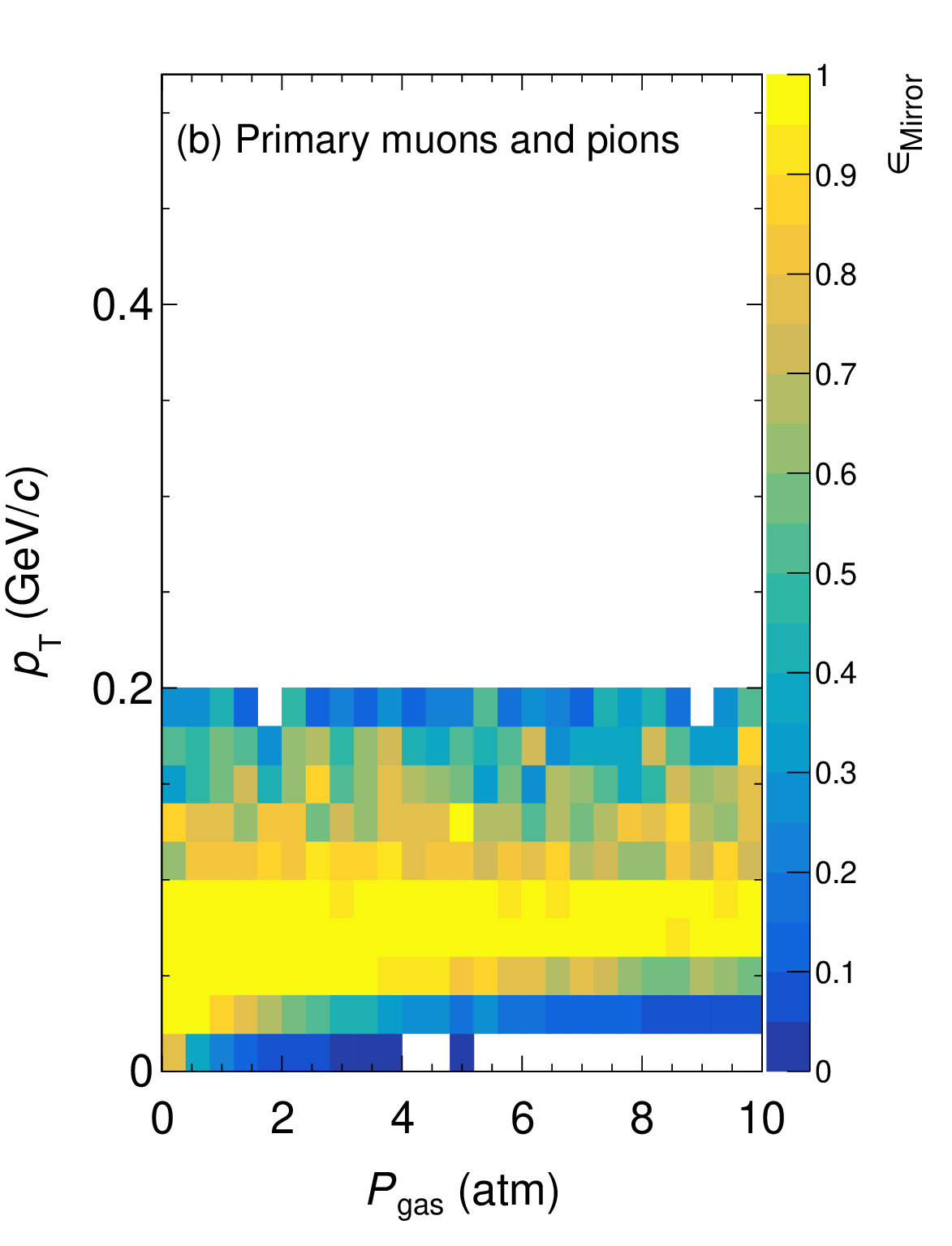}
         \caption{}
         \label{fig:MirrorRatio_Prim_mu}
     \end{subfigure}
          \begin{subfigure}[b]{0.32\textwidth}
         \centering
         \includegraphics[width=\textwidth]{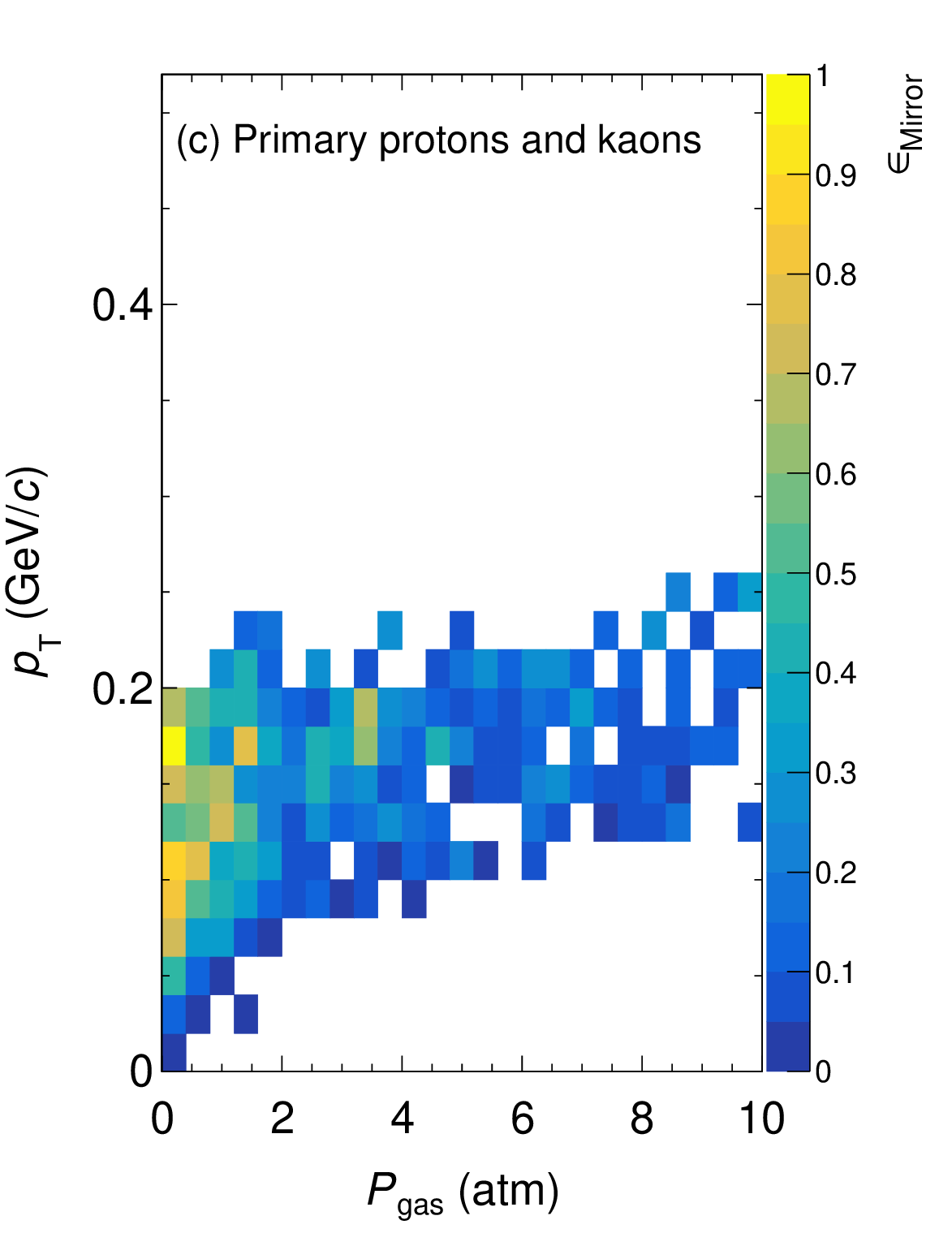}
         \caption{}
         \label{fig:MirrorRation_Prim_k}
     \end{subfigure}
     \begin{subfigure}[b]{0.32\textwidth}
         \centering
         \includegraphics[width=\textwidth]{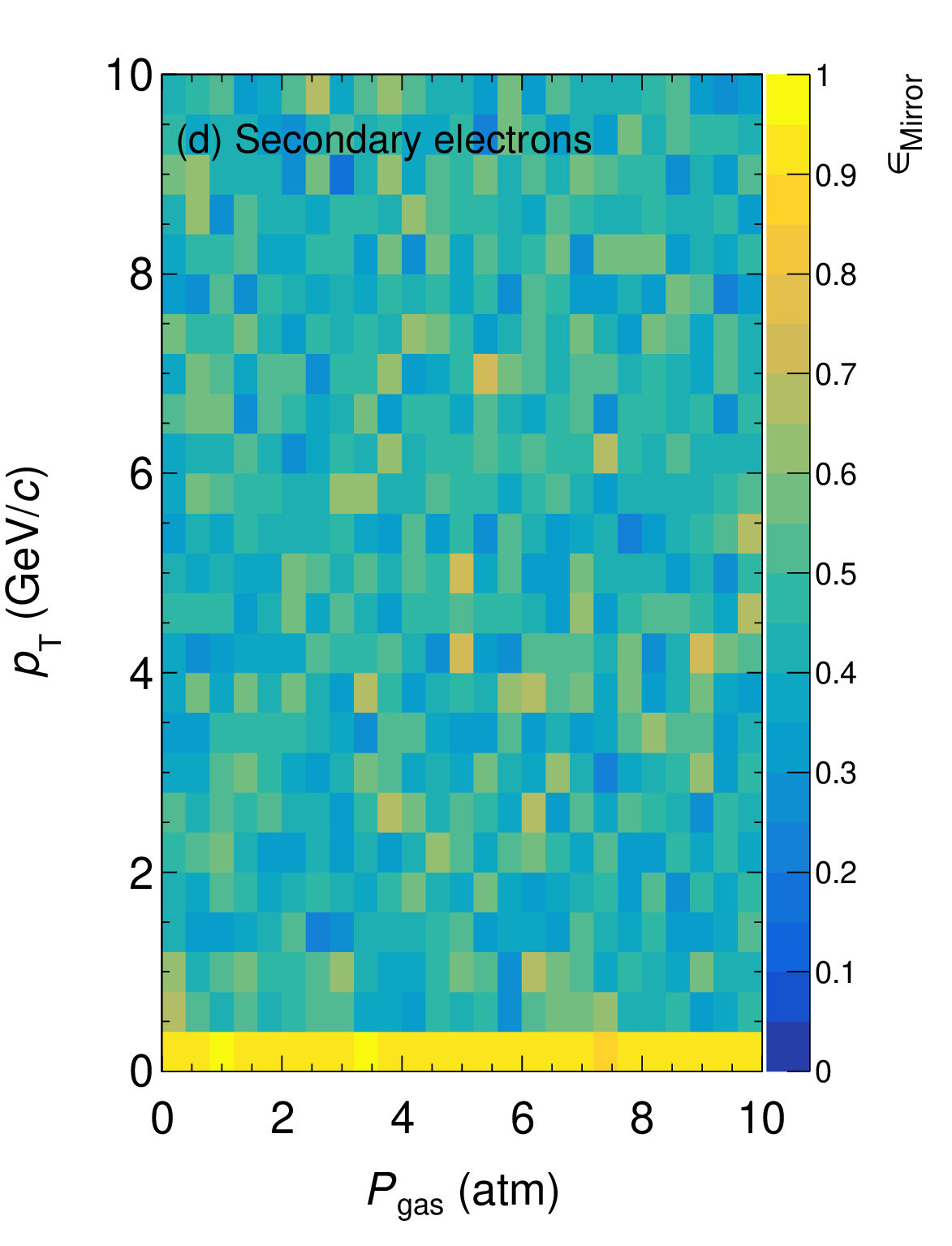}
         \caption{}
         \label{fig:MirrorRatiop_Sec_e}
     \end{subfigure}
     \begin{subfigure}[b]{0.32\textwidth}
         \centering
         \includegraphics[width=\textwidth]{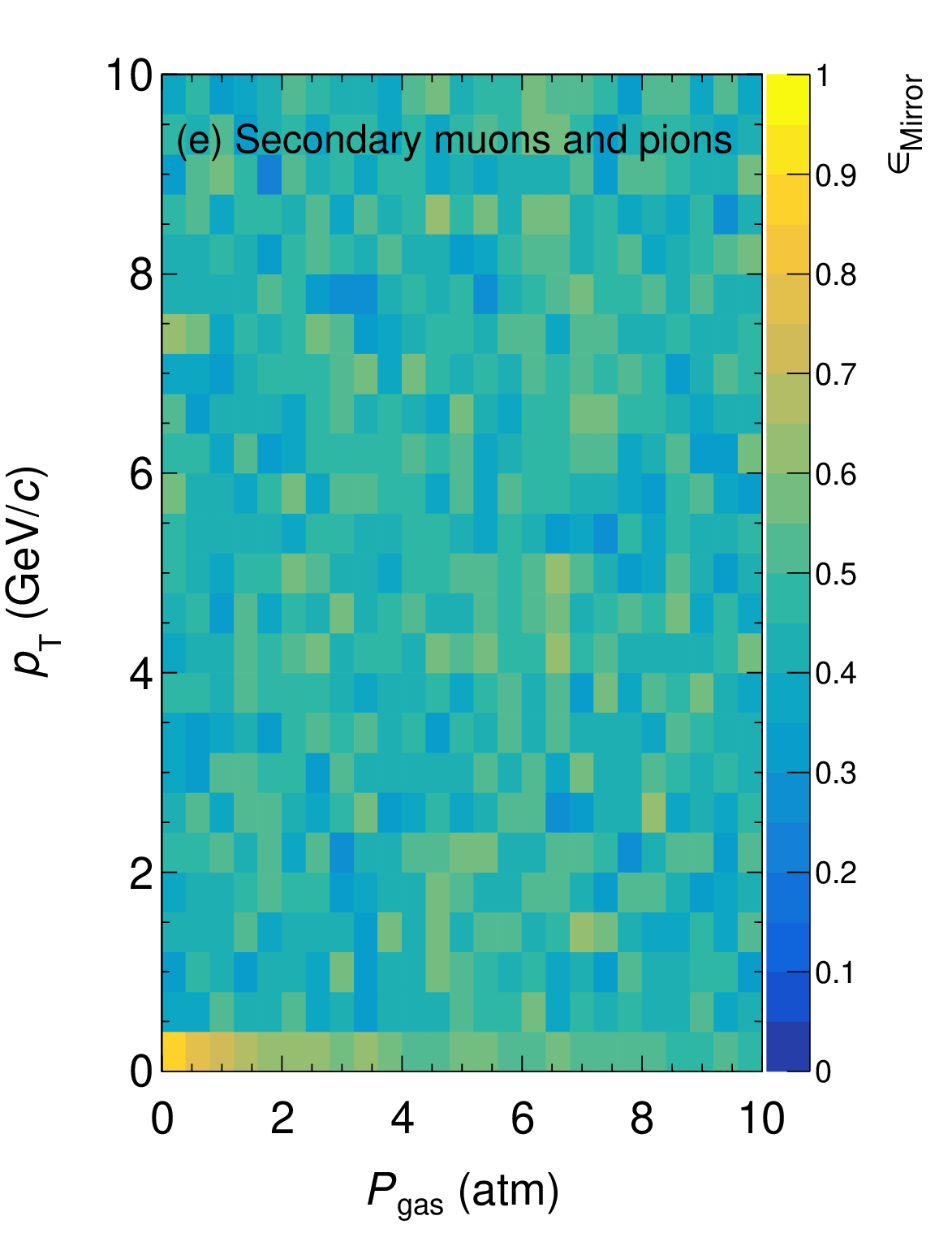}
         \caption{}
         \label{fig:MirrorRatio_Sec_mu}
     \end{subfigure}
          \begin{subfigure}[b]{0.32\textwidth}
         \centering
         \includegraphics[width=\textwidth]{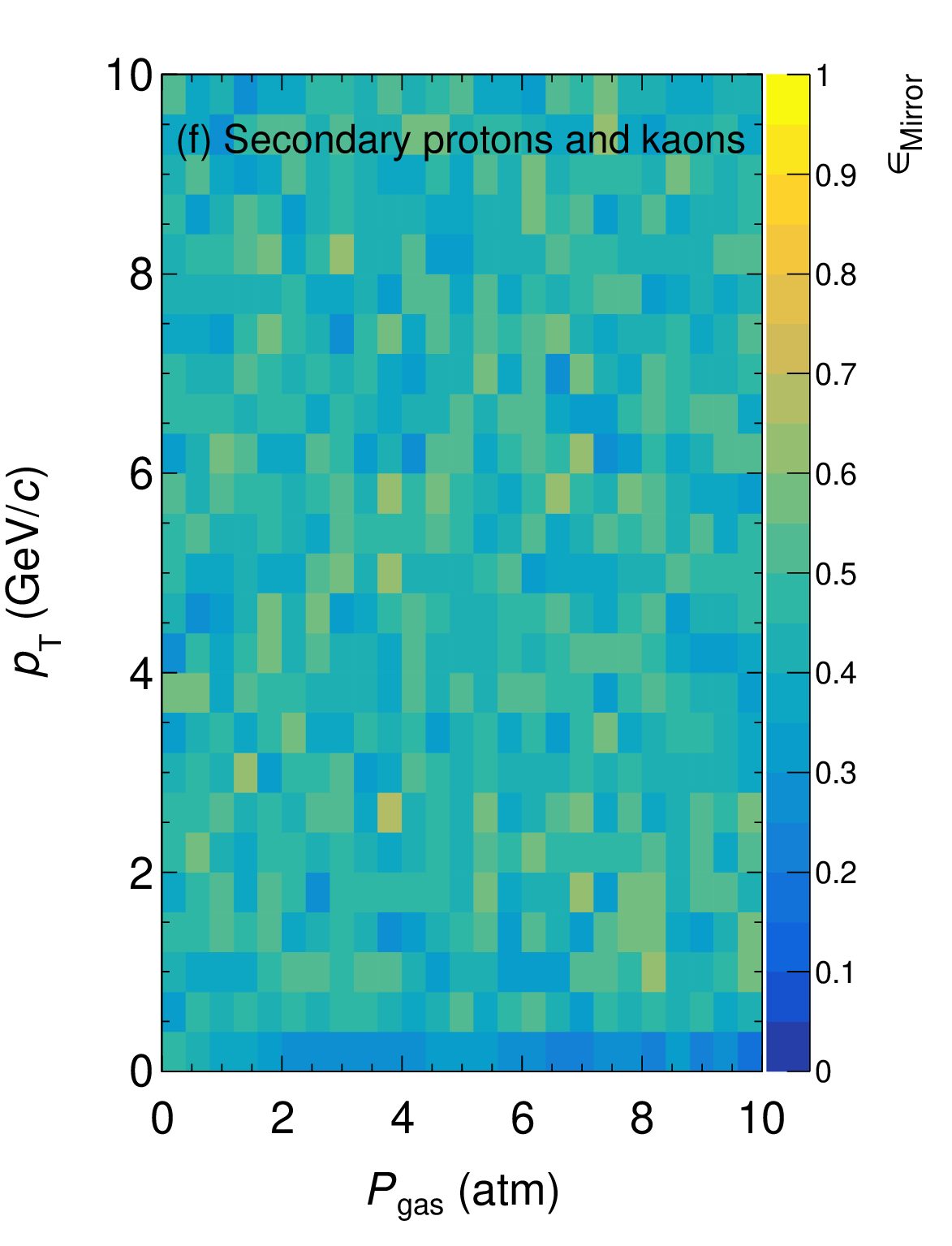}
         \caption{}
         \label{fig:MirrorRation_Sec_k}
     \end{subfigure}
        \caption{Portion of tracks in the PS sample for which the mirror rotation was applied, $\epsilon_\textrm{Mirror}$, as a function of the gas pressure $P_\textrm{gas}$ and the initial true transverse momentum $p_\textrm{T}$ of the particle. The primaries are shown in the upper row, while the secondaries are shown in the lower row. The particle types are separated based on their mass: (a) and (d) contain only electrons, (b) and (e) muons and pions, (c) and (f) kaons and protons.} \label{fig:MirrorRatio}
\end{figure}

The PS sample was further used to test the improvement in $q/p_{\text{T}}$ resolution brought by the introduction of the \enquote{mirror rotation} technique in the \texttt{CKF}, compared to \texttt{BKF}. The fraction of the total tracks for which the mirroring technique was used $\epsilon_\textrm{Mirror}$ is shown in Fig.~\ref{fig:MirrorRatio} as a function of initial true $p_\textrm{T}$ and $P_\textrm{gas}$. We show the fraction for the primaries in the first row and for the secondaries in the second. The particle types are divided in order of mass between electrons in Figs. \ref{fig:MirrorRatiop_Prim_e} and \ref{fig:MirrorRatiop_Sec_e}, muons and pions in Figs. \ref{fig:MirrorRatio_Prim_mu} and \ref{fig:MirrorRatio_Sec_mu}, kaons and protons in Figs. \ref{fig:MirrorRation_Prim_k} and \ref{fig:MirrorRation_Sec_k}. We can see from the upper row plots and the low $p_\textrm{T}$ component of the lower row plots that the likelihood of the particles producing looping trajectory drops significantly with mass, due to the higher $\textrm{d}E/\textrm{d}x$. At low momenta the pressure of the gas also becomes important. This is especially true for the higher mass particles such as protons and kaons, which at higher $P_\textrm{gas}$ are stopped in the detector before producing any looping trajectory (see Fig. \ref{fig:MirrorRation_Prim_k}). From Figs. \ref{fig:MirrorRatiop_Prim_e}, \ref{fig:MirrorRatio_Prim_mu} and \ref{fig:MirrorRation_Prim_k} it can also be shown that the only primary particle that necessitate the use of the mirroring operation are those that produce a looping trajectory in the detector, which is only possible at low initial transverse momenta $p_\textrm{T}<0.3 \ \textrm{GeV}/c$.  Secondary particle trajectories, on the other hand, can cover more than a semi-plane even if they don't belong to loopers and can thus be produced at any momenta, as shown in Figs. \ref{fig:MirrorRatiop_Sec_e}, \ref{fig:MirrorRatio_Sec_mu} and \ref{fig:MirrorRation_Sec_k}. At low transverse momenta, $\epsilon_\textrm{Mirror}$ is still significantly higher for secondaries. 

\begin{figure}[!ht]
     \centering
     \begin{subfigure}[b]{0.48\textwidth}
         \centering
         \includegraphics[width=\textwidth]{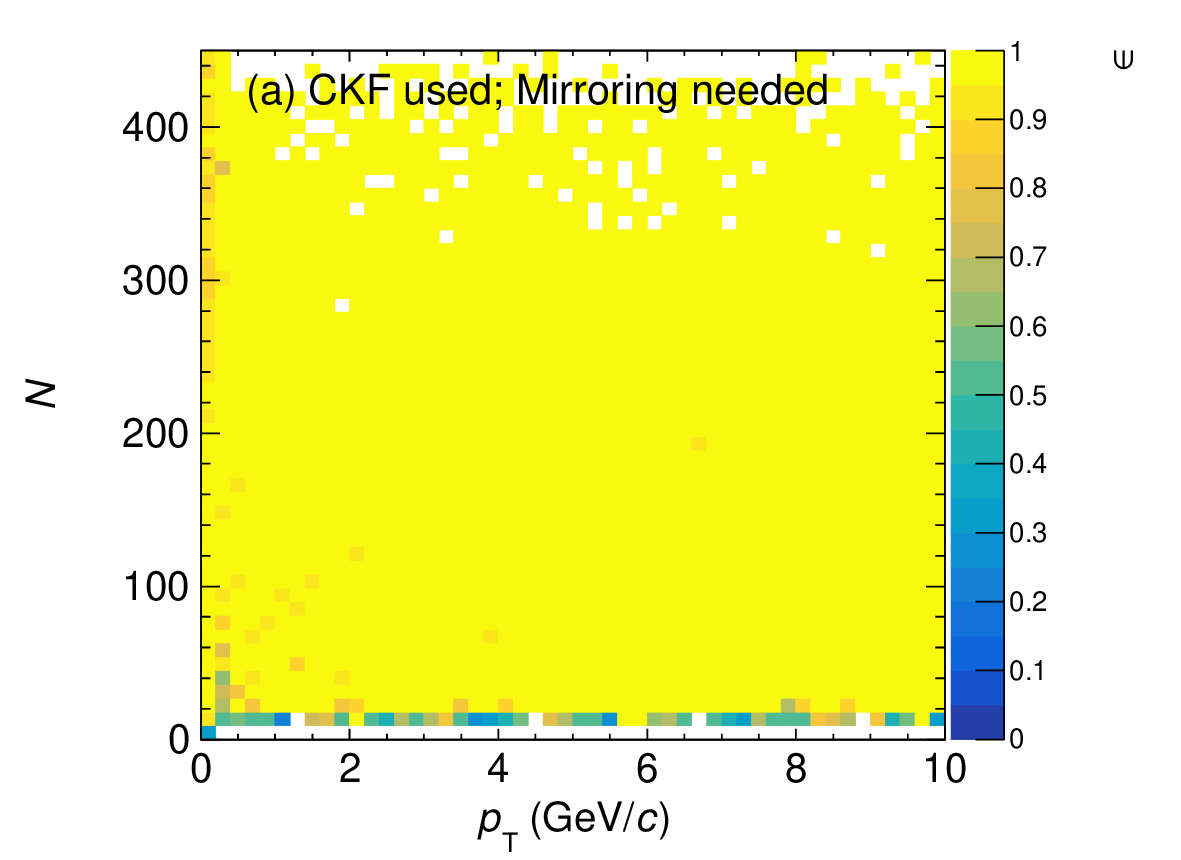}
         \caption{}
         \label{fig:PS_Eff_CKF_Mirror}
     \end{subfigure}
     \begin{subfigure}[b]{0.48\textwidth}
         \centering
         \includegraphics[width=\textwidth]{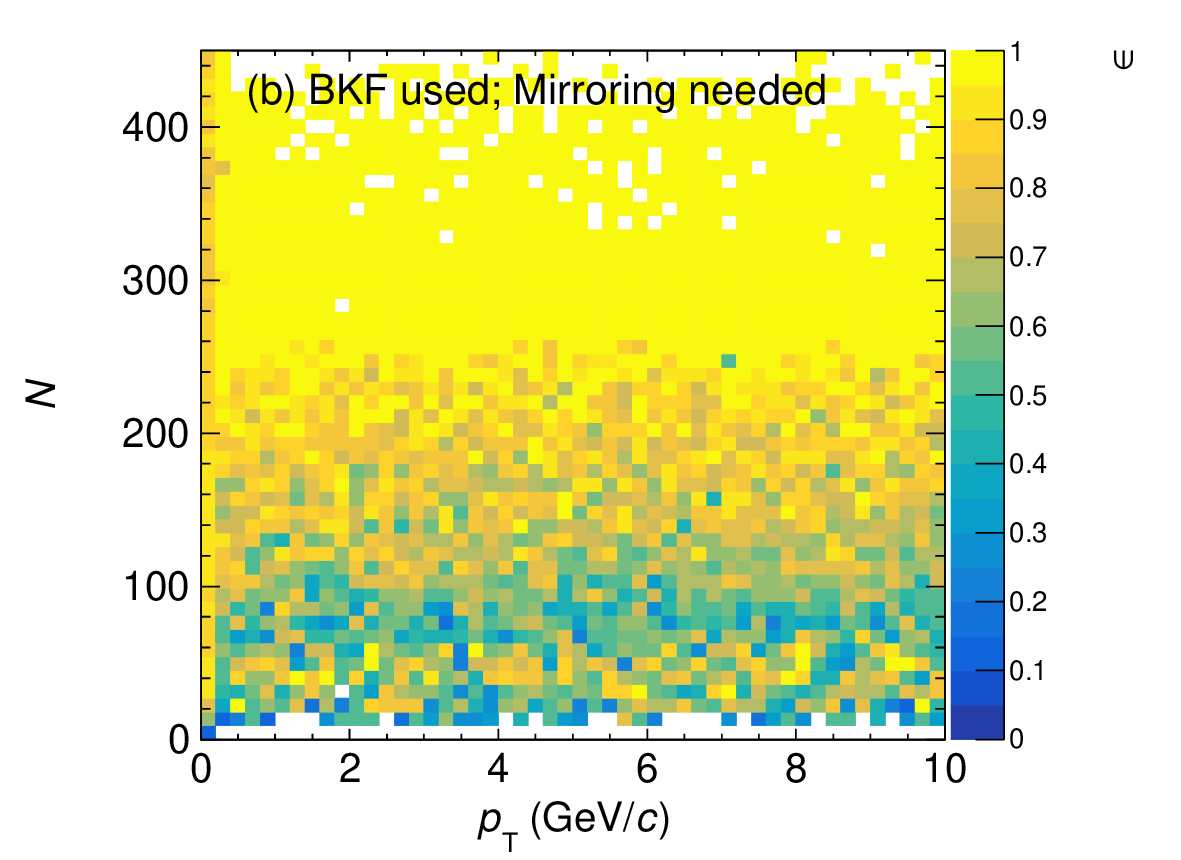}
         \caption{}
         \label{fig:PS_Eff_BKF_Mirror}
     \end{subfigure}
          \begin{subfigure}[b]{0.48\textwidth}
         \centering
         \includegraphics[width=\textwidth]{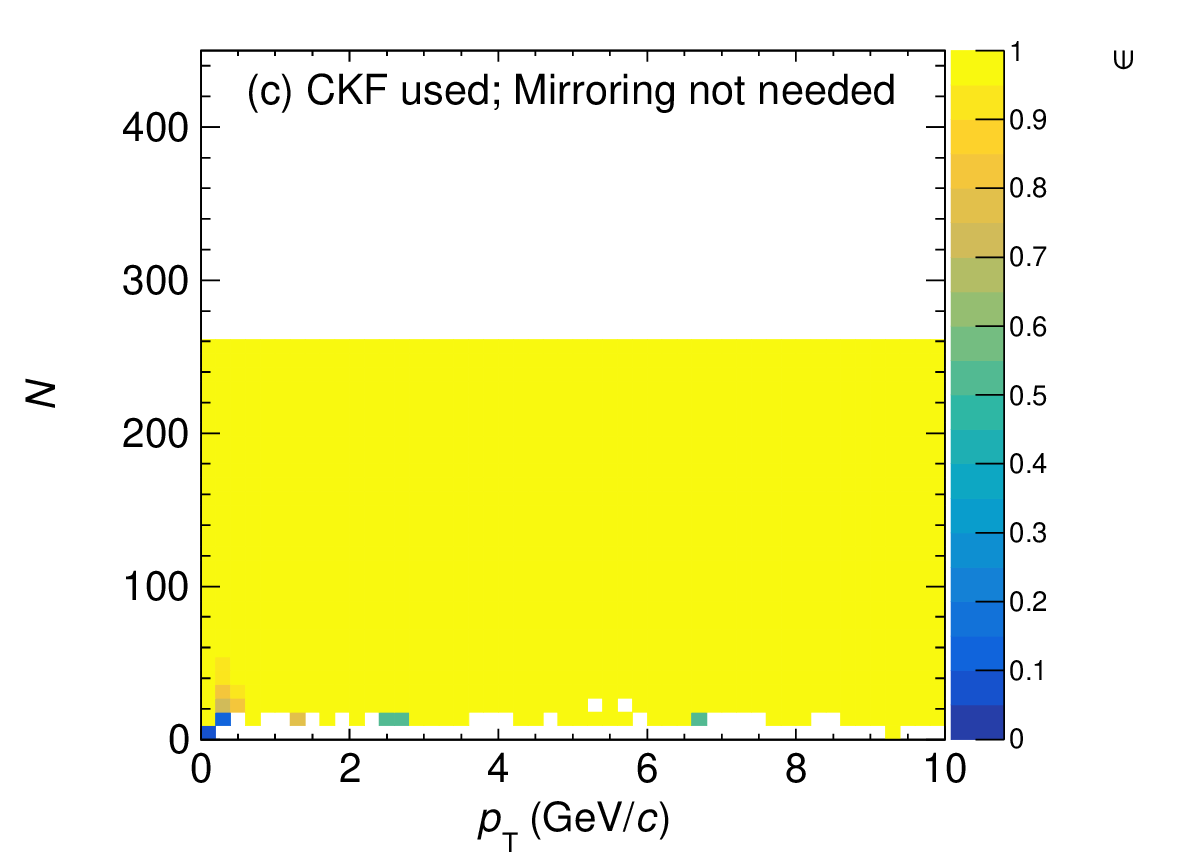}
         \caption{}
         \label{fig:PS_Eff_CKF_NoMirror}
     \end{subfigure}
     \begin{subfigure}[b]{0.48\textwidth}
         \centering
         \includegraphics[width=\textwidth]{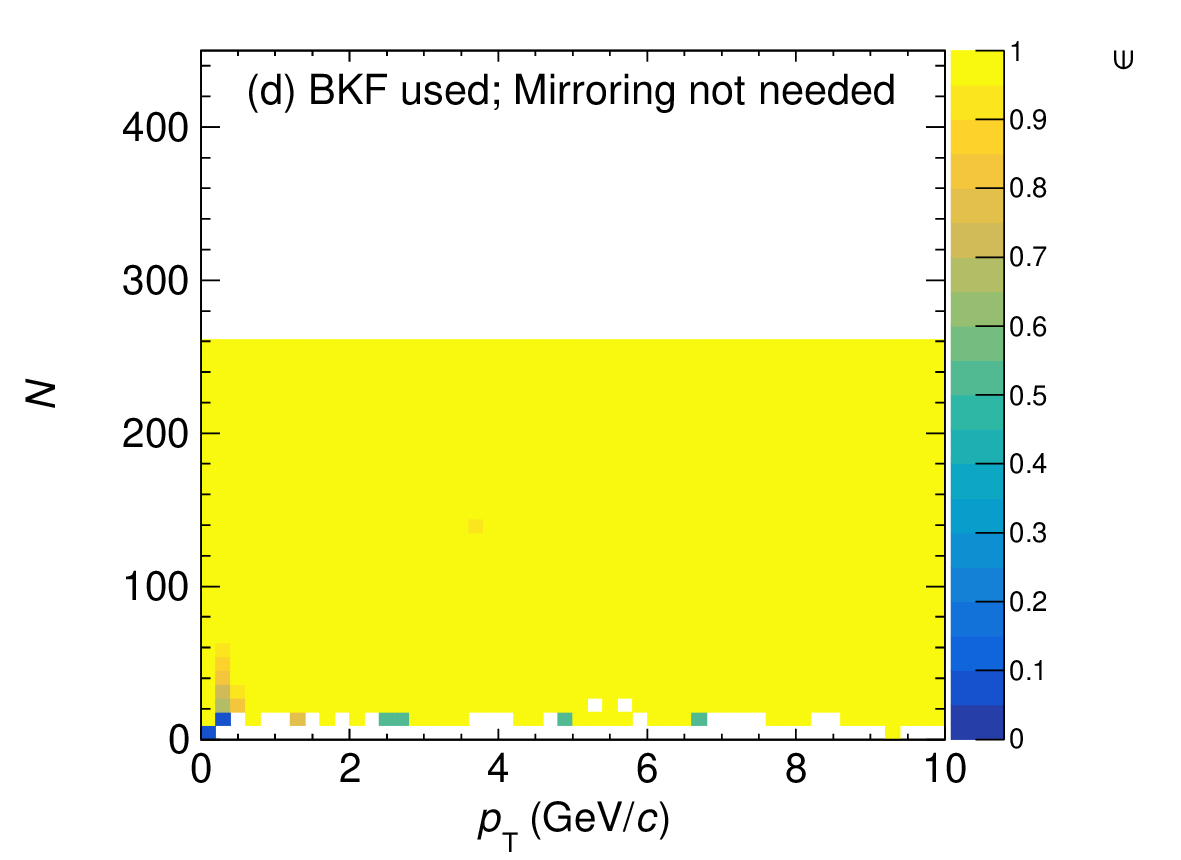}
         \caption{}
         \label{fig:PS_Eff_BKF_NoMirror}
     \end{subfigure}
        \caption{Reconstruction efficiency $\epsilon$ for the PS sample, as a function of the total number of points in the track $N$ and the initial true transverse momentum $p_\textrm{T}$. The results for the tracks for which the mirroring technique is needed to reconstruct all the points, are shown in (a) and (b) for the \texttt{CKF} and \texttt{BKF} respectively. The results for the other tracks are shown in (c) for the \texttt{CKF} and (d) for the \texttt{BKF}.} \label{fig:PS_Eff}
\end{figure}

The reconstruction efficiency $\epsilon$, defined as the fraction of the correctly simulated tracks for which the algorithm is fully propagated, was tested for \texttt{CKF} and \texttt{BKF}. It is shown as a function of the initial true $p_\textrm{T}$ and the $N$ of the total track in the first and second column of Fig \ref{fig:PS_Eff} respectively. In Figs. \ref{fig:PS_Eff_CKF_Mirror} and \ref{fig:PS_Eff_BKF_Mirror} only the tracks for which the mirroring technique is needed to reconstruct all the points are shown, while the other tracks are shown Figs. \ref{fig:PS_Eff_CKF_NoMirror} and \ref{fig:PS_Eff_BKF_NoMirror}. For the tracks that can be fully reconstructed without using the mirroring technique, the efficiency is essentially identical between the \texttt{BKF} and \texttt{CKF} algorithm. It is shown to be very close to 1 except for very low $p_\textrm{T}$ and $N$, for which other approaches  than a Kalman Filter would most likely be used~\cite{RevModPhys.82.1419}. On the other hand $\epsilon$ is significantly improved by the \texttt{CKF} for low $N$ ``mirrored'' tracks, in some cases going from an efficiency of $\sim0.5$ to $\epsilon>0.9$. This is a direct consequence of the increased number of points becoming available through the use of the mirroring technique.

\begin{figure}[!ht]
     \centering
     \begin{subfigure}[h!]{0.9\textwidth}
         \centering
         \includegraphics[width=\textwidth]{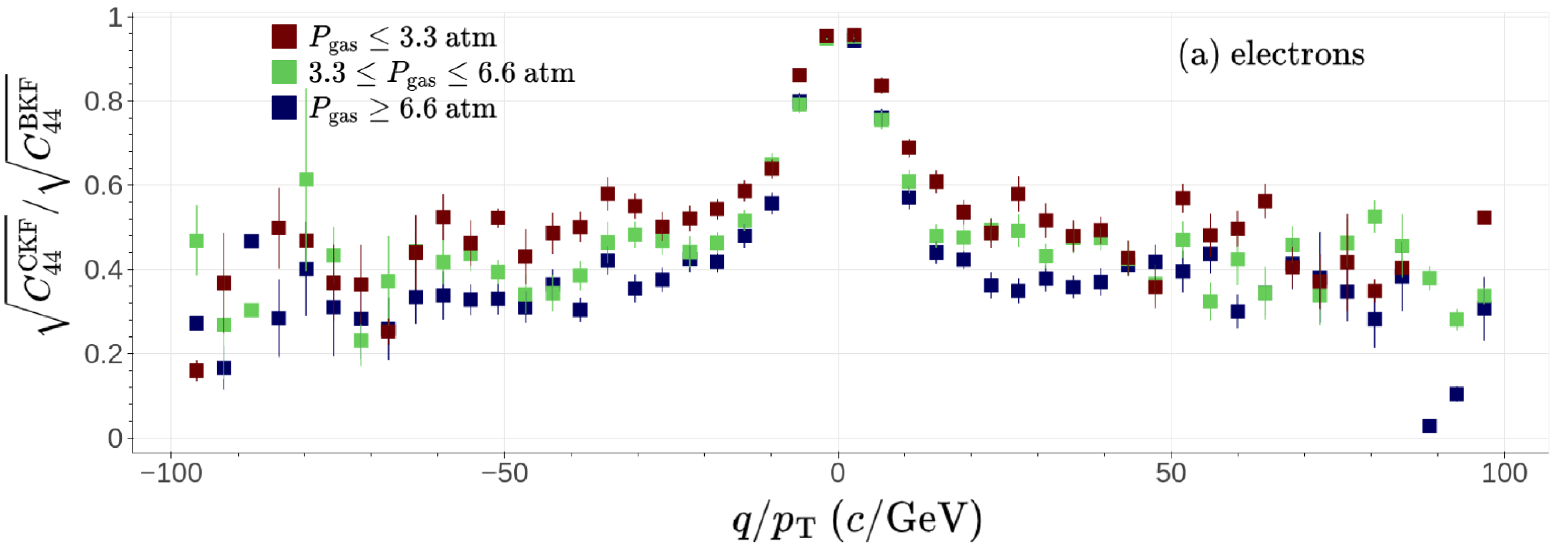}
         \caption{}
         \label{fig:SingleVSLeg_e}
     \end{subfigure}
     \begin{subfigure}[b]{0.9\textwidth}
         \centering
         \includegraphics[width=\textwidth]{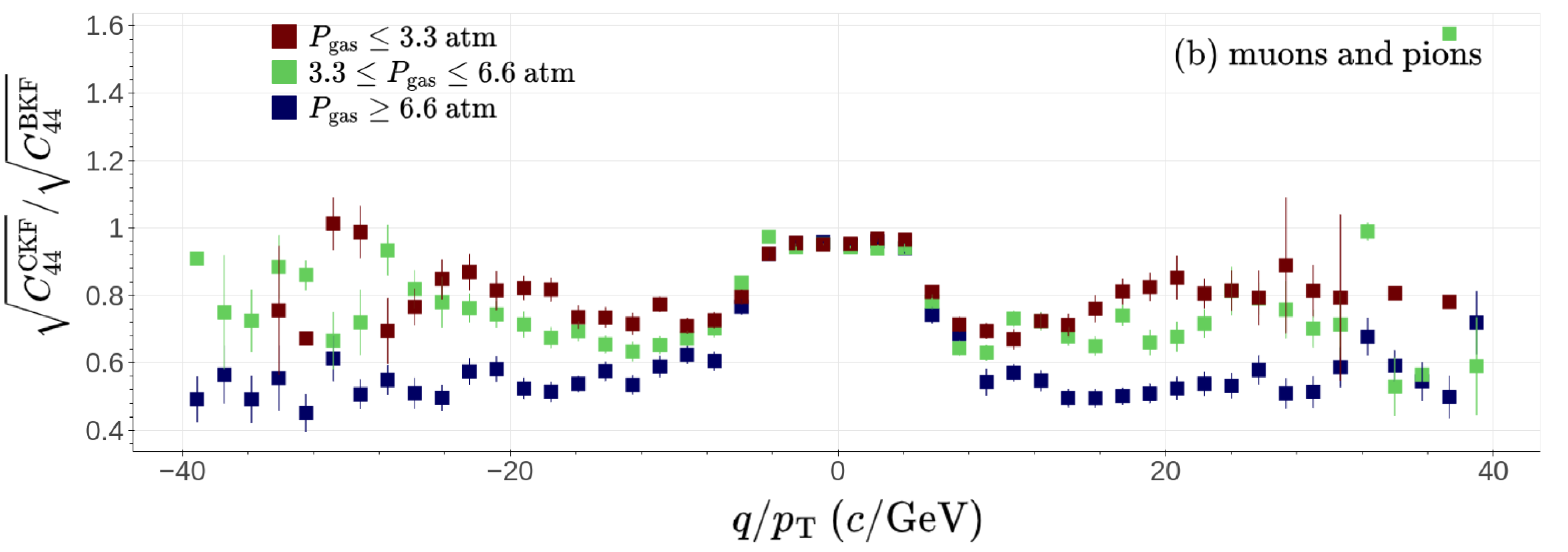}
         \caption{}
         \label{fig:SingleVSLeg_mu_pi}
     \end{subfigure}
          \begin{subfigure}[b]{0.9\textwidth}
         \centering
         \includegraphics[width=\textwidth]{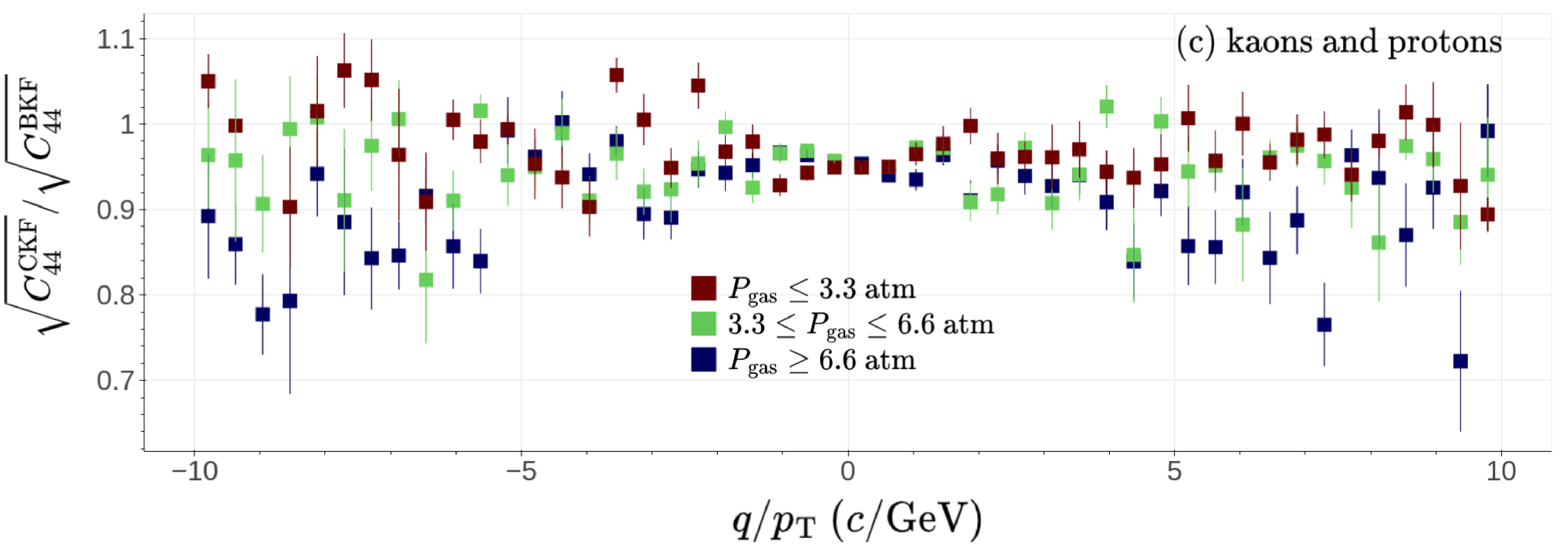}
         \caption{}
         \label{fig:SingleVSLeg_proton_kaon}
     \end{subfigure}
        \caption{Ratios of the $q/p_\text{T}$ resolutions obtained using the full \texttt{CKF} algorithm including the mirror rotation method $\sqrt{C_{44}^{\textrm{CKF}}}$, over the reconstruction without mirror rotation $\sqrt{C_{44}^{\text{BKF}}}$. The plots were produced for the whole PS sample. The histograms are color coded according to the gas pressure $P_{\textrm{gas}}$. Plot (a) contains only electrons, plot (b) contains pions and muons and plot (c) contains kaons and protons. Only tracks with a minimum of 30 points are considered. These plots have been produced using the interactive analytical tool \texttt{ROOTInteractive}~\cite{RootInt}. The error bars are statistical.}
        \label{fig:SingleVSLeg_dens}
\end{figure}

\begin{figure}[!ht]
     \centering
     \begin{subfigure}[h!]{0.99\textwidth}
         \centering
         \includegraphics[width=\textwidth]{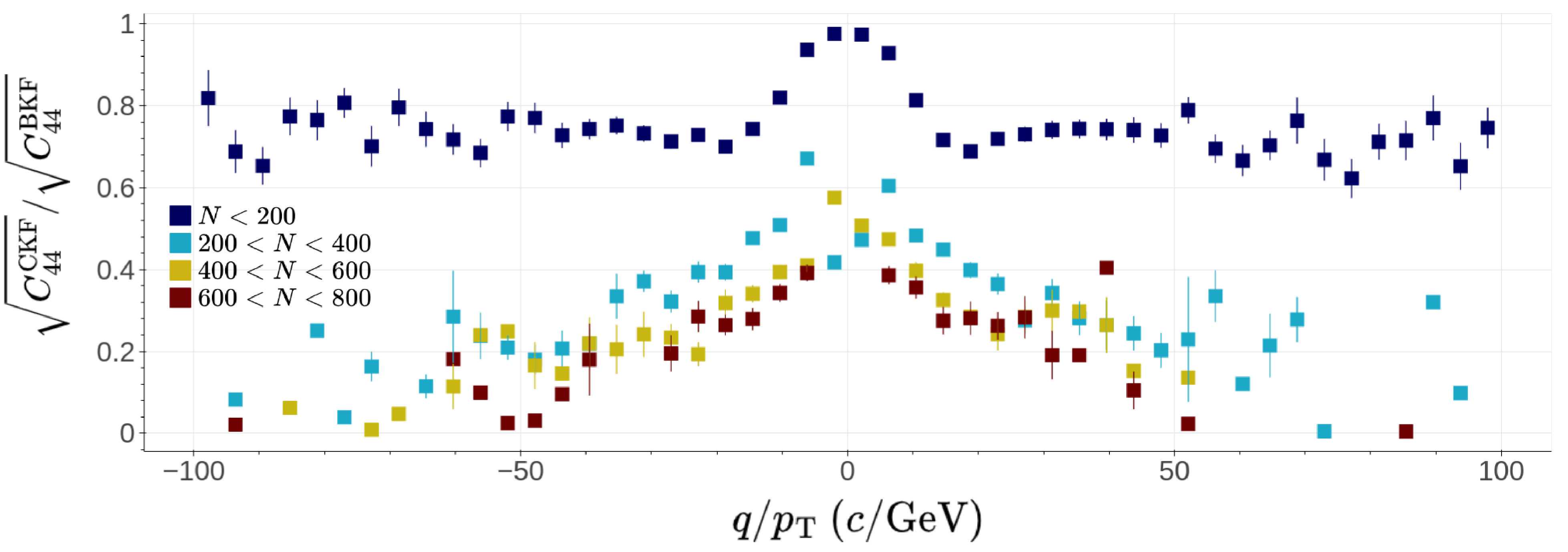}
         \caption{}
         \label{fig:SingleVSLeg_NPoints}
     \end{subfigure}
        \caption{Ratios of the $q/p_\text{T}$ resolutions obtained using the full \texttt{CKF} algorithm including the mirror rotation method $\sqrt{C_{44}^{\textrm{CKF}}}$, over the reconstruction without mirror rotation $\sqrt{C_{44}^{\text{BKF}}}$. The plots were produced for the whole PS sample. The histograms are color coded according to number of points in the tracks $N$. These plots have been produced using the interactive analytical tool \texttt{ROOTInteractive}~\cite{RootInt}. The error bars are statistical.}
        \label{fig:SingleVSLeg_NPointsAll}
\end{figure}

The difference in performance due to the mirror rotation can be quantified using the ratios between the full reconstruction resolution $\sqrt{C_{44}^{\textrm{CKF}}}$ and the $\sqrt{C_{44}^{\text{BKF}}}$ obtained with the basic reconstruction at a given point along the track. Figure~\ref{fig:SingleVSLeg_dens} shows the ratios $\sqrt{C_{44}^{\textrm{CKF}}}/\sqrt{C_{44}^{\text{BKF}}}$ as a function of the true $q/p_{\text{T}}$, color-coded according to the gas pressure $P_{\textrm{gas}}$. Figure~\ref{fig:SingleVSLeg_e} shows the results for a sample of only electrons, while Fig.~\ref{fig:SingleVSLeg_mu_pi} shows the results for a sample of muons and pions and Fig.~\ref{fig:SingleVSLeg_proton_kaon} for a sample of protons and kaons. The points are again randomly taken along the reconstructed tracks and down-sampled to $10\%$ of the total. For the electron sample, an overall relative improvement of $\sim 60\%$ is shown at $p_{\text{T}}<100$ MeV/$c$, with peaks of up to $\sim 80\%$ for the lowest momentum tracks in low pressure environments. This behavior is in agreement with Eqs.~\ref{eq:sigmaN} and~\ref{eq:sigmaMS} which show a dependency of the $1/p_{\text{T}}$ resolution on $1/\sqrt{N}$ and $1/\sqrt{L_\textrm{Arm}}$ respectively: using the mirroring technique, more space points of the tracks are used, resulting in larger $N$ and $L_\textrm{Arm}$. The dependence on the number of points is shown more clearly in Fig.~\ref{fig:SingleVSLeg_NPointsAll}, where the histograms are color coded for total number of points in the track, including those only accessible through the mirroring technique. Tracks containing more than 800 points are excluded for easier legibility of the results. It is shown that for the longest tracks, relative improvements of up to 80\% can be achieved.


The difference between the results is shown to be less dramatic in more pressurized environments, where particles tend to be absorbed sooner and tracks are generally shorter. This trend is confirmed looking at the results obtained for the muons and pions sample in Fig.~\ref{fig:SingleVSLeg_mu_pi}, for which $\textrm{d}E/\textrm{d}x$ will on average be higher due to their higher masses: the improvement in this case is by $\sim 20\%$ for $p_{\text{T}}<150$ MeV/$c$ with peaks of up to $\sim50$\% for the lowest momentum tracks in low pressure environments. No improvements were found for the more massive particles shown in the sample, except for minor ones at lower pressures (Fig.~\ref{fig:SingleVSLeg_proton_kaon}). 

\subsection{Tests and results: high-pressure sample}

The HP sample is used to evaluate the detector performance of a HPgTPC as described in Sec.~\ref{sec:Geometry}. We focus on the total momentum relative resolution and bias, defined as the $\sigma$ and $\mu$ of a standard Gaussian fit applied to the momentum fractional residuals:
\begin{equation}
    R = \frac{p_{\text{reco}}}{p_{\text{true}}} - 1.
\end{equation}
The reconstruction efficiency $\epsilon$ was also tested. The results are shown in the Appendix in Fig. \ref{fig:HP_Eff} and are similar to the PS Sample.

The formulas that we quoted for the expected resolution of the $1/p_{\text{T}}$ factor in Eqs.~\ref{eq:sigmaN} and~\ref{eq:sigmaMS} can be adapted for the relative momentum resolution by applying error propagation. The new formulas can then be written as:
\begin{equation}\label{eq:sigmaNptot}
\frac{\sigma_{\text{H}}(p)}{p}=\frac{\cos\lambda \ p  \ \sigma_{r\phi}}{0.3 BL_\textrm{Arm}^2}\sqrt{\frac{720}{N+4}}.
\end{equation}
\begin{equation}\label{eq:sigmaMSptot}
\frac{\sigma_{\text{MS}}(p)}{p}=\frac{0.016 \ (\textrm{GeV}/c)}{0.3 B l\beta \cos \lambda}\sqrt{\frac{l}{X_0}},
\end{equation}
where $\sigma_{\text{H}}(p)$ is the point resolution component of the total momentum resolution and $\sigma_{\text{MS}}(p)$ is the multiple scattering component.

\begin{figure}[!ht]
     \centering
     \begin{subfigure}[b]{0.49\textwidth}
         \centering
         \includegraphics[width=\textwidth]{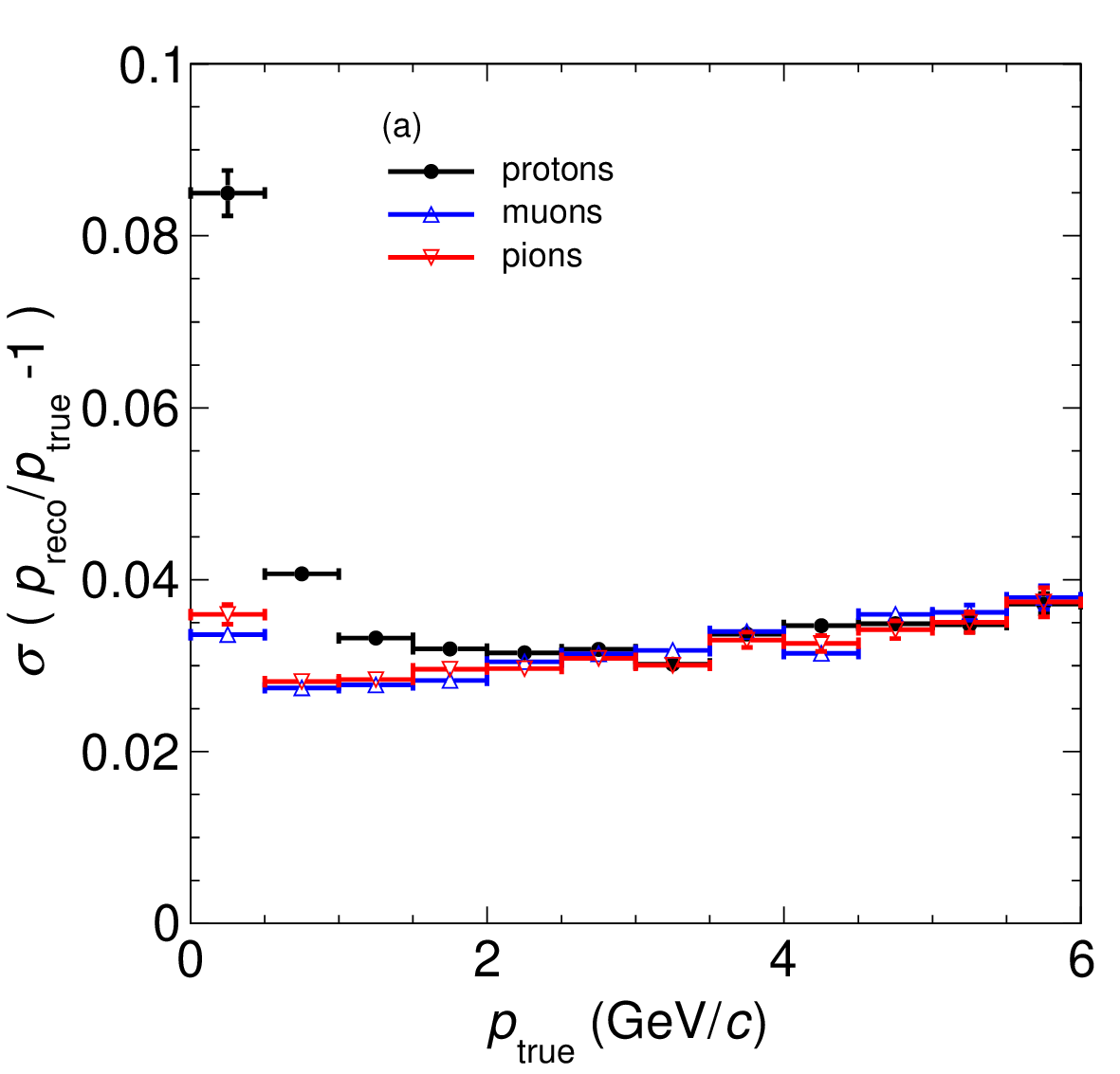}
         \caption{}
         \label{fig:ResND-GArVSp}
     \end{subfigure}
     \begin{subfigure}[b]{0.49\textwidth}
         \centering
         \includegraphics[width=\textwidth]{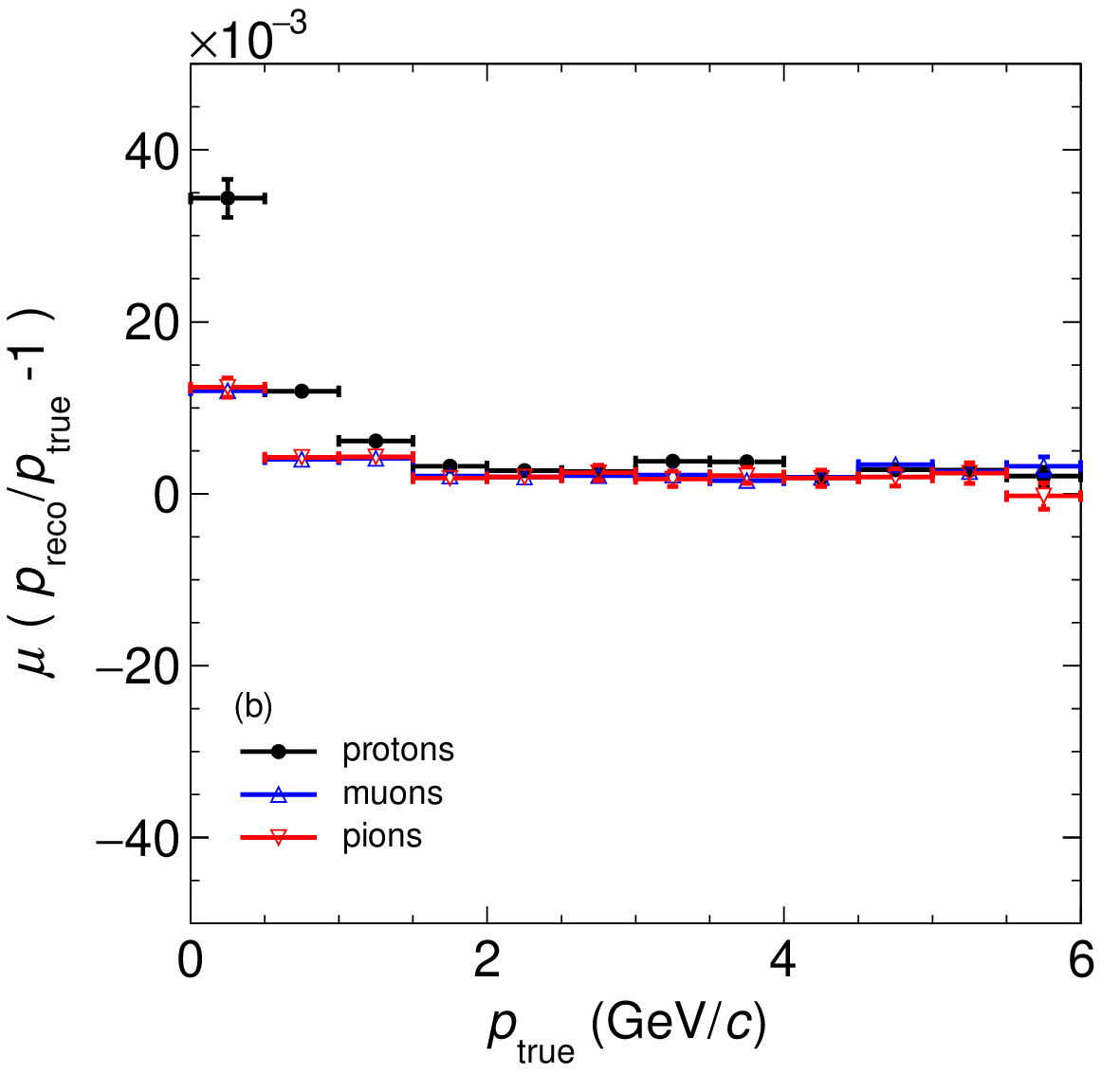}
         \caption{}
         \label{fig:BiasND-GArVSp}
     \end{subfigure}
        \caption{Relative momentum resolution (a) and bias (b) as function of the true momentum, $p_\textrm{true}$, for the HP sample. The two properties are defined as $\mu$ and $\sigma$ of Gaussian fits of the momentum fractional residuals $p_{\text{reco}}/p_{\text{true}}-1$. The three particle types (protons, muons and pions) are drawn separately.}
        \label{fig:ND-GArVSp}
\end{figure}

\begin{figure}[!ht]
     \centering
     \begin{subfigure}[b]{0.49\textwidth}
         \centering
         \includegraphics[width=\textwidth]{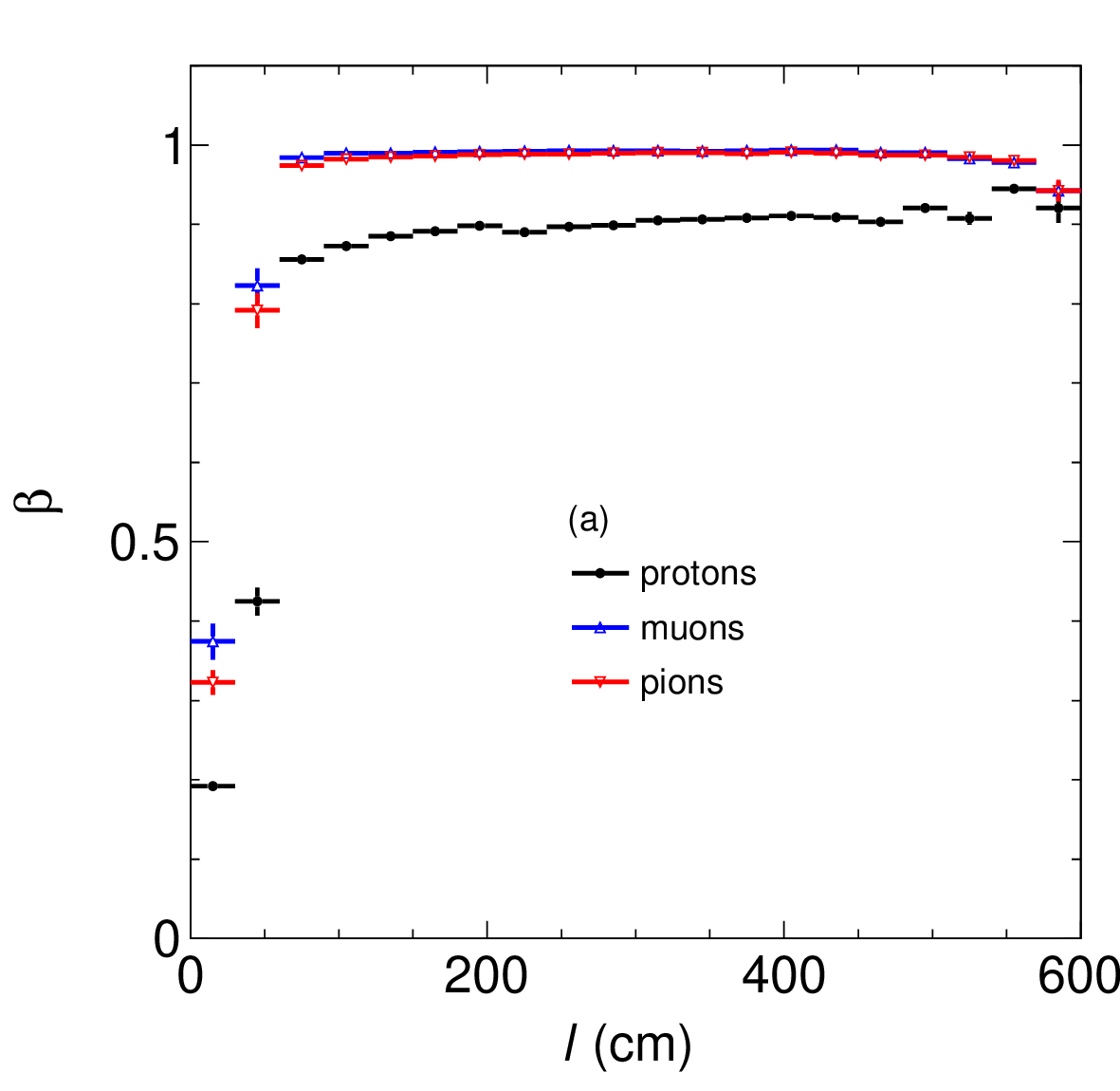}
         \caption{}
         \label{fig:betaVSlength}
     \end{subfigure}
     \begin{subfigure}[b]{0.49\textwidth}
         \centering
         \includegraphics[width=\textwidth]{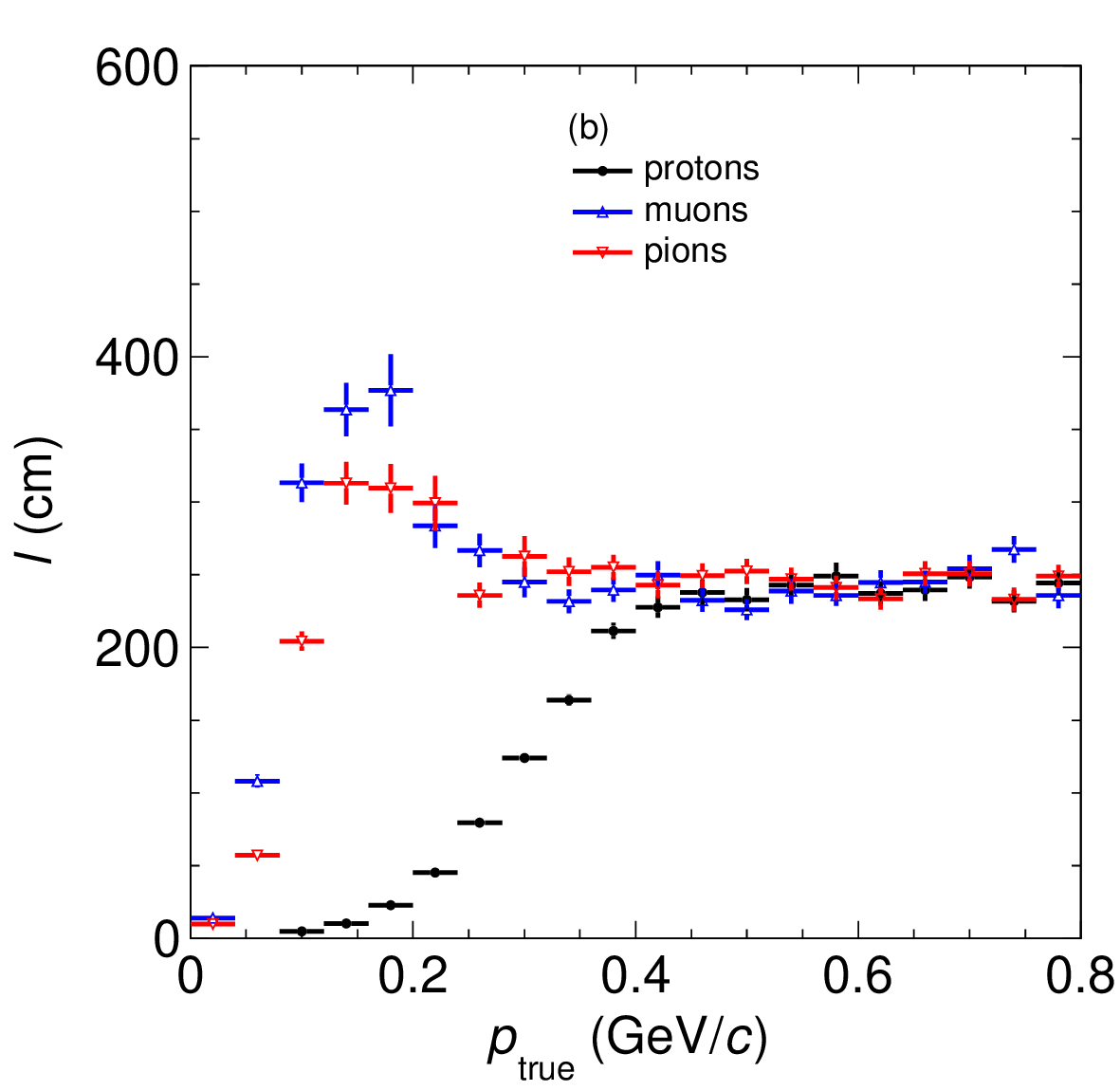}
         \caption{}
         \label{fig:lengthVSp}
     \end{subfigure}
        \caption{(a) Average $\beta$ as a function of the track length $l $ and  (b) average track length as a function of the true momentum $p_\textrm{true}$ for the HP sample.}
        \label{fig:ND-GArextraprops}
\end{figure}

In Fig.~\ref{fig:ND-GArVSp} we show the relative momentum resolution and bias as a function of the true momentum for the three particle types present in the sample. At lower momenta ($p_{\text{true}}<1\text{GeV/}c$) the resolution is close to 2\% for pions and muons while it is closer to 8\% for the protons. In this momentum region the multiple scattering component of the resolution is dominant. This component is inversely proportional to the particles' $\beta$ factor. With a given momentum, the proton, having a higher mass, will always have a smaller $\beta$ and thus a worse expected resolution. Furthermore, while muons and protons at lower energies have the chance to produce longer and even looping tracks inside the detector, protons will tend to lose their energy more quickly, again due to their masses. This is clearly shown in Fig.~\ref{fig:lengthVSp} where the average particle lengths as a function of their true momentum is shown. Somewhat significant biases are also shown at these lower momenta, especially for protons. For momenta $p_{\text{true}}>1~\text{GeV/}c$, the resolution is comparable for the three particle types and increases slowly with the particle momentum. At higher momenta the point resolution component is dominant, so a direct proportionality on the momentum is expected, with no distinction between the particle types. An inverse dependency on the lever arm and number of points and thus indirectly on the length is also expected, but as shown in Fig.~\ref{fig:lengthVSp}, in this momentum range the average length of the track becomes roughly the same for all particle types.

\begin{figure}[!ht]
     \centering
     \begin{subfigure}[b]{0.49\textwidth}
         \centering
         \includegraphics[width=\textwidth]{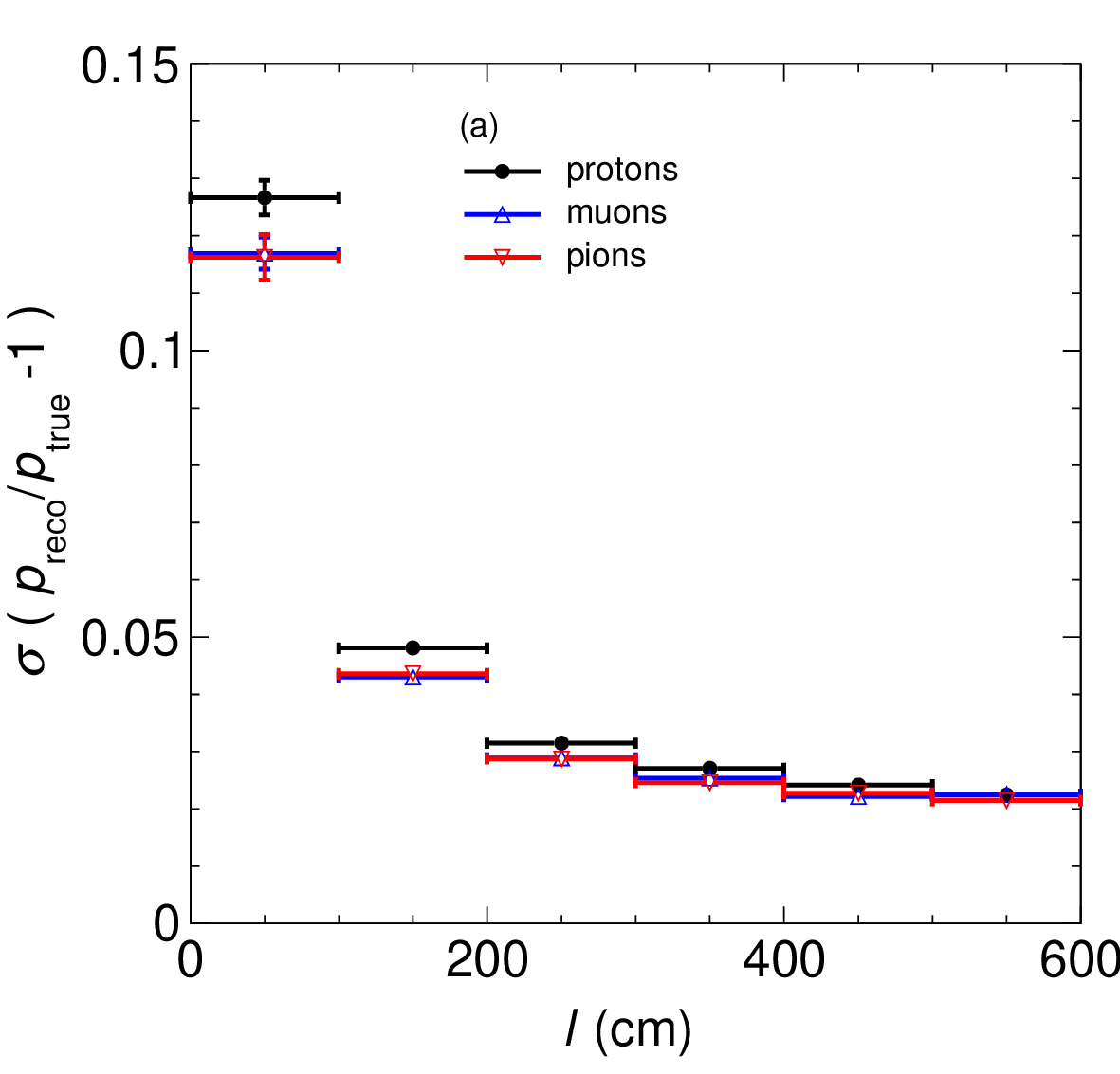}
         \caption{}
         \label{fig:ResND-GArVSlength}
     \end{subfigure}
     \begin{subfigure}[b]{0.49\textwidth}
         \centering
         \includegraphics[width=\textwidth]{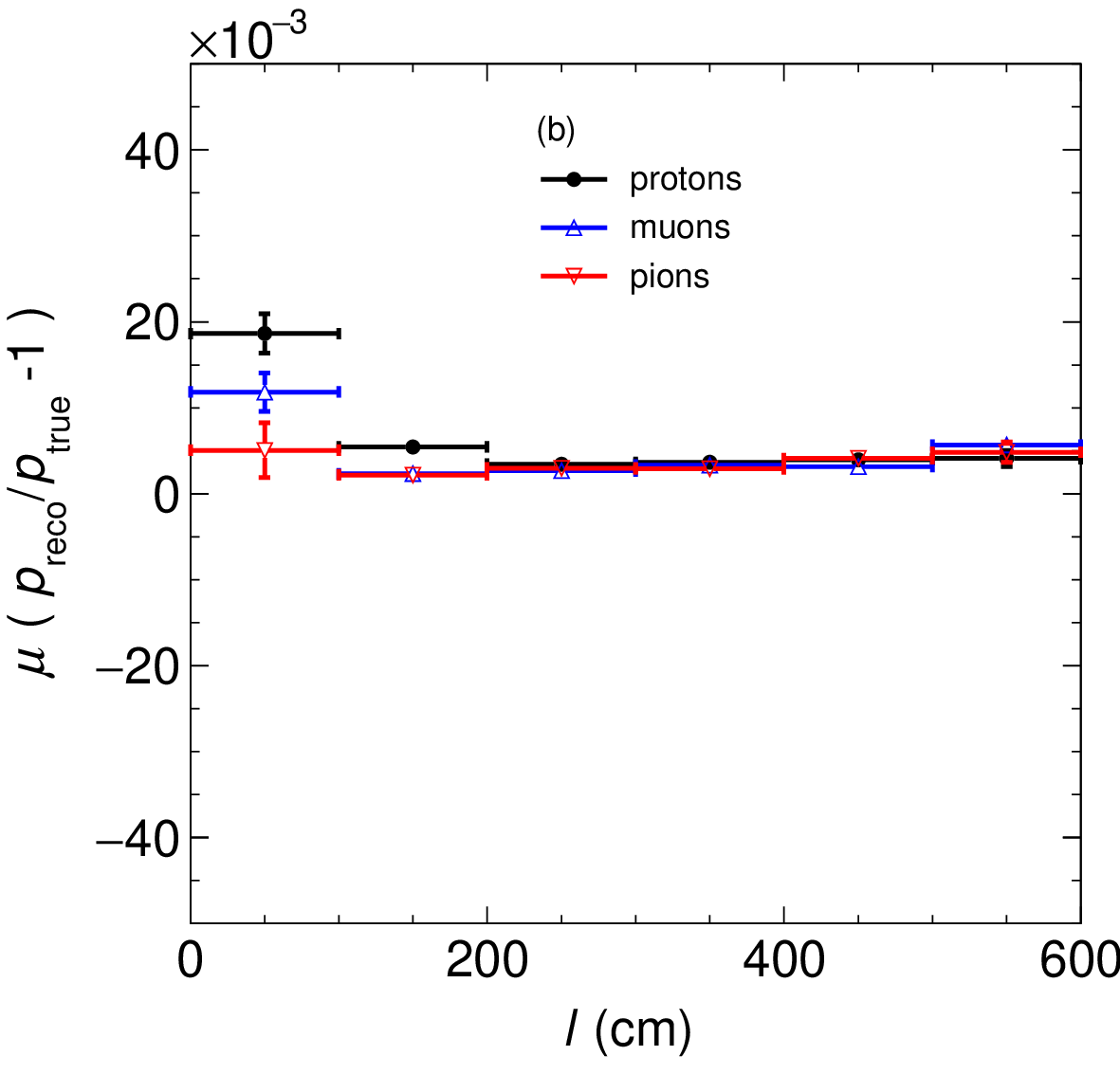}
         \caption{}
         \label{fig:BiasND-GArVSlength}
     \end{subfigure}
        \caption{Relative momentum resolution (a) and bias (b) as function of the true track length, $l$, for the HP sample. The two properties are defined as $\mu$ and $\sigma$ of Gaussian fits of the momentum fractional residuals $p_{\text{reco}}/p_{\text{true}}-1$. The three particle types (protons, muons and pions) are drawn separately.}
        \label{fig:ND-GArVSlength}
\end{figure}

In Fig.~\ref{fig:ND-GArVSlength} we show the momentum resolution and bias as a function of the true track lengths $l$ for the three particle types. As could be predicted from Eq.~\ref{eq:sigmaMSptot}, an inverse proportionality of the relative resolution on $l$ can be observed for all particle types. The worse performance observed for the protons can be explained by the average $\beta$ shown in Fig.~\ref{fig:betaVSlength}, which is always smaller for more massive particles regardless of the length of the tracks. For longer tracks the hit component of the resolution is dominant and the difference in $\beta$ is not as impactful. A somewhat significant bias can be seen at lower lengths. Similar dependencies on lever arm $L_{\textrm{Arm}}$ and number of points in the track $N$, which can be treated as a proxy for $l$ in most cases, are shown in Figs.~\ref{fig:ND-GArVSLArm} and~\ref{fig:ND-GArVSNPoints} in the Appendix. Similar plots have also been produced for the PS sample to allow for direct comparison in Figs. \ref{fig:PSVSp} and \ref{fig:PSVSl} also in the Appendix.

\section{Conclusions}
\label{sec:Conclusions}
In this paper we introduced a Kalman Filter tailored for homogeneous gas TPCs, adapted from the one used in the ALICE experiment. An inherent limitation in the original ALICE approach arises from its suitability only for tracks describing a semi-circle at most in the $xy$ plane, perpendicular to the detector's magnetic field. We discovered that this challenge can be addressed by applying a mirror rotation to the state vector when reaching the semi-circle's boundaries. This adjustment is facilitated by introducing an $xy$ plane rotation during the Kalman Filter's propagation step, converting the longitudinal $x$ coordinate into the radial distance from the rotation center. By implementing this technique, the new algorithm can effectively track trajectories of any length, including multiple circular paths within the detector (loopers). This enhancement has no precedent in the literature and significantly improves upon the ALICE Kalman Filter, making it highly promising for both the ALICE TPC~\cite{Arslandok:2024hhw} and neutrino HPgTPCs like ND-GAr, which, due to the randomness in the production points of charged particles coupled with their relatively low energy, are prone to longer track formations. To evaluate the new algorithm, we developed a toy MC simulation tool named \texttt{fastMCKalman} and generated a sample with diverse detector and particle properties to validate its performance across a wide parameter space. Multiple tests conducted on this sample demonstrated that the algorithm's estimates for parameter covariance effectively describe the sample and align closely with theoretical expectations.This last point is highlighted by Figs. \ref{fig:Check_Ana_pID}, \ref{fig:CheckAna_dens} and \ref{fig:Check_Ana_res} where the theoretically expected resolution was first shown on its own for different particle types, gas densities and resolution and then directly compared with the \texttt{CKF} estimations by showing their ratios. Despite the wide resolution ranges shown for the theoretical expectations, the ratios are uniformly close to 1, demonstrating close agreement. A significant improvement in the reconstruction efficiency for the low $N$ and low $p_\textrm{T}$ mirrored tracks was also shown, in some cases going from $\epsilon\sim 0.5$ to $\epsilon>0.9$. Furthermore, we examined the impact of the mirroring technique by comparing the ratios of $q/p_{\textrm{T}}$ resolutions with and without its application, revealing relative improvements by up to 80\% for low-energy electrons and up to 50\% for muons and pions. Additionally, we evaluated the new Kalman Filter algorithm's performance using a sample of particles propagated in a high material budget environment, simulating conditions akin to a HPgTPC like ND-GAr. Realistic assessments of relative momentum resolution and bias showed behaviors consistent with theoretical expectations, affirming the viability of applying the method to a neutrino gas TPC.

The reconstruction efficiency and resolution improvements brought by the application of the new algorithm to such a detector are very attractive. The improvement in $\mu$ and $\pi$ reconstruction would be extremely valuable when studying nuclear effects using techniques such as TKI, the effectiveness of which is directly dependent on the momentum reconstruction quality. The significant improvement seen for low-momentum electrons will be crucial for DUNE as it allows to more efficiently probe the $\nu_e$ and $\bar{\nu}_e$ fluxes. Additionally, it is likely that the mirroring technique could be applied to track formation. In most track forming algorithms, looping tracks are formed in segments, which are then connected together using various positional arguments. However, this step produces additional combinatorial complexity and makes algorithms more computationally heavy. The mirroring technique could be used to produce looping tracks in a more seamless way, expanding the Kalman Filter approach, which involves adding points one by one using track projection and compatibility $\chi^2$ arguments. This is for example the method currently used by the ALICE experiment~\cite{Ivanov:2003yr, Arslandok:2022dyb}, which could directly benefit from this expansion.    

\section*{Acknowledgments}

We express our gratitude to the ALICE Collaboration for generously providing the open-source code under the BSD License 2.0. Additionally, we extend our thanks to Iuri~Belikov for his essential contributions to the development of the ALICE tracking code, which served as the foundation for the studies presented in this work. 
F.B. would like to thank Thomas Junk and Lukas Koch for the valuable discussion and inputs. 
X.L. is supported by the STFC (UK) Grants No. ST/S003533/1 and ST/S003533/2. 

\bibliographystyle{elsarticle-num} 
\bibliography{reference}
\clearpage
\section*{Appendix}
\begin{figure}[!ht]
     \centering
     \begin{subfigure}[b]{0.48\textwidth}
         \centering
         \includegraphics[width=\textwidth]{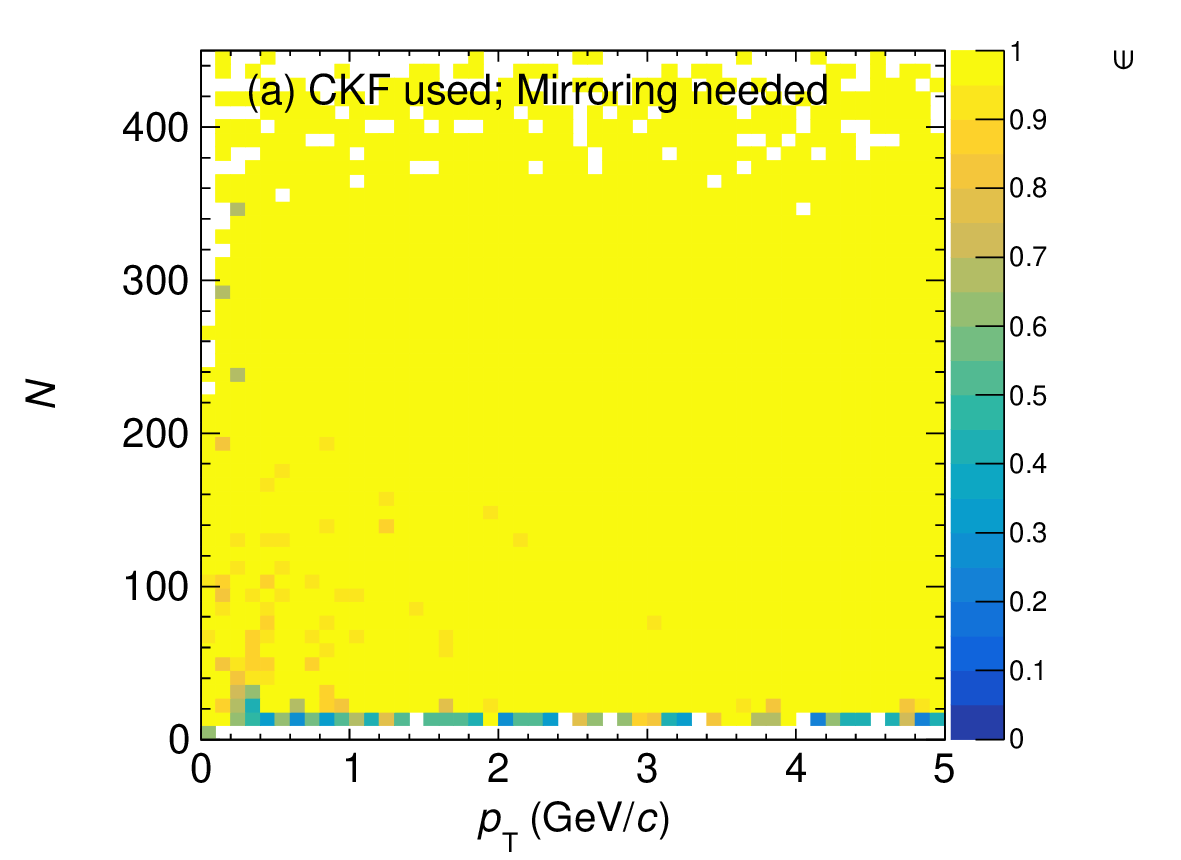}
         \caption{}
         \label{fig:HP_Eff_CKF_Mirror}
     \end{subfigure}
     \begin{subfigure}[b]{0.48\textwidth}
         \centering
         \includegraphics[width=\textwidth]{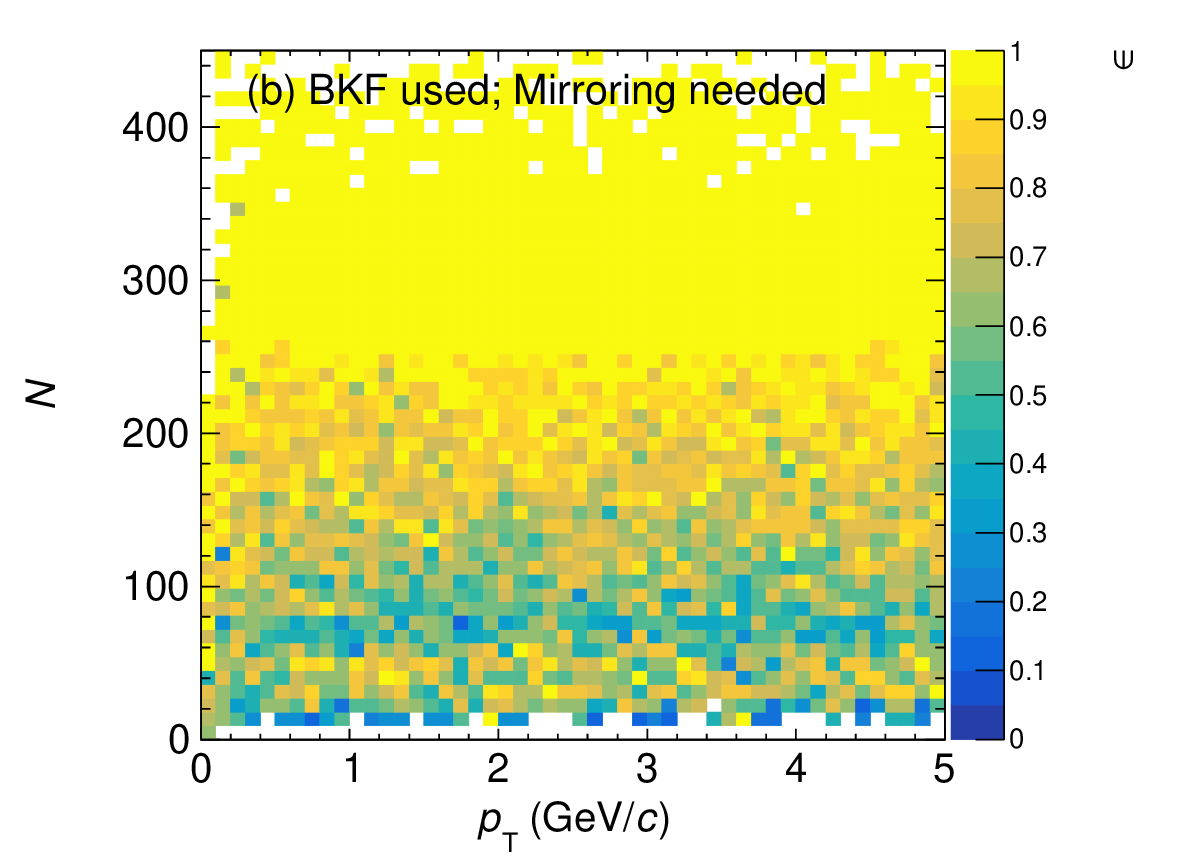}
         \caption{}
         \label{fig:HP_Eff_BKF_Mirror}
     \end{subfigure}
          \begin{subfigure}[b]{0.48\textwidth}
         \centering
         \includegraphics[width=\textwidth]{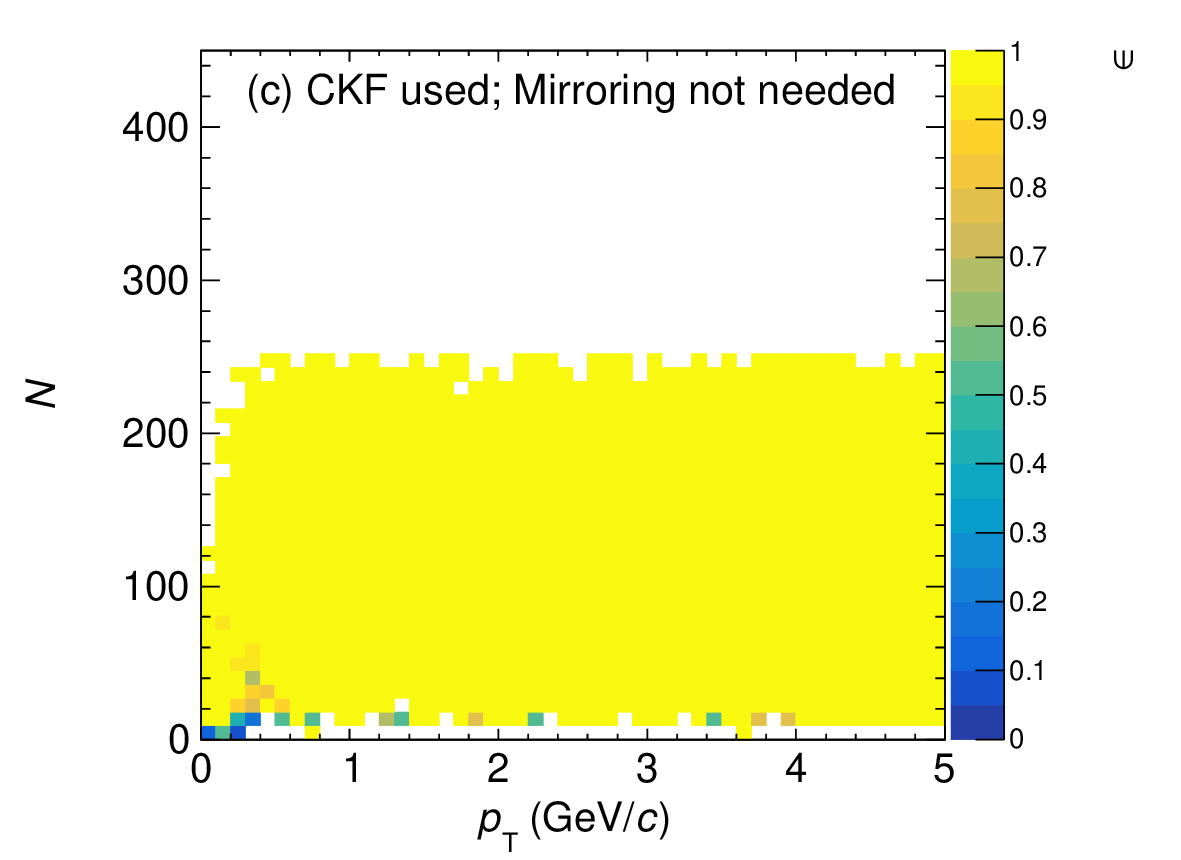}
         \caption{}
         \label{fig:HP_Eff_CKF_NoMirror}
     \end{subfigure}
     \begin{subfigure}[b]{0.48\textwidth}
         \centering
         \includegraphics[width=\textwidth]{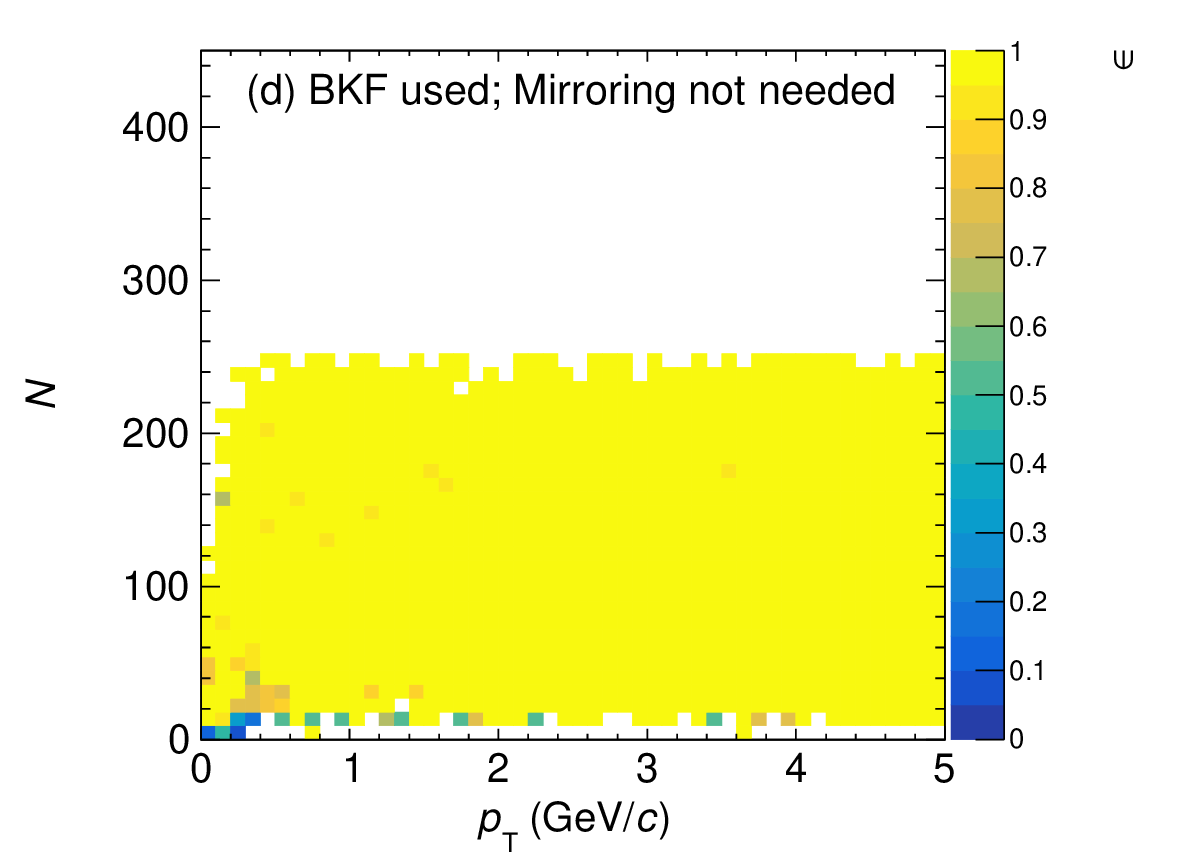}
         \caption{}
         \label{fig:HP_Eff_BKF_NoMirror}
     \end{subfigure}
        \caption{Reconstruction efficiency $\epsilon$ for the HP sample, as a function of the total number of points in the track $N$ and the initial true transverse momentum $p_\textrm{T}$. Analogous to Fig. \ref{fig:PS_Eff}} \label{fig:HP_Eff}
\end{figure}

The reconstruction efficiency for the HP sample is shown in Fig. \ref{fig:HP_Eff}. Analogously to what was shown in Fig. \ref{fig:PS_Eff}, $\epsilon$ is plotted as a function of the initial true $p_\textrm{T}$ and the $N$. The results for the \texttt{CKF} and \texttt{BKF} are shown in the first and second column respectively. In the plots occupying the first row only the tracks for which the mirroring technique is used are shown, while the other tracks are shown in the plots in the second row. The results are analogous to what was shown for the PS sample in Fig. \ref{fig:PS_Eff}.

\label{sec:Appendix}

\clearpage

\begin{figure}[!ht]
     \centering
     \begin{subfigure}[b]{0.99\textwidth}
         \centering
         \includegraphics[width=\textwidth]{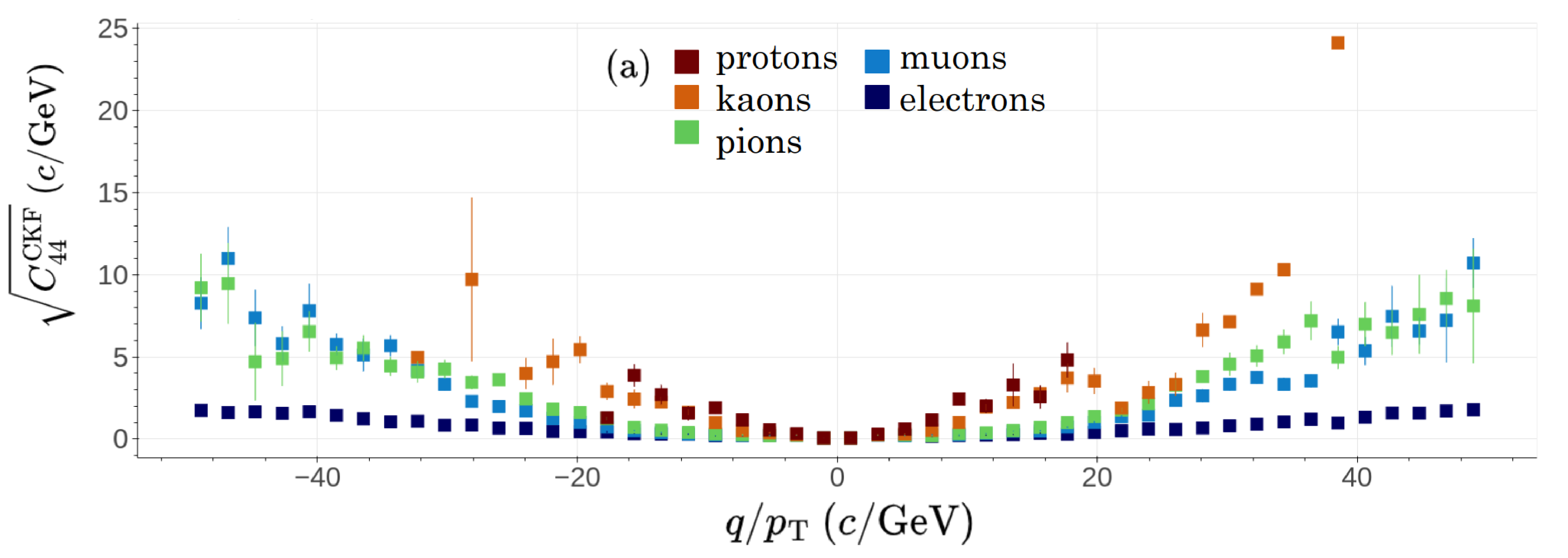}
         \caption{}
         \label{fig:CheckAna_pID_zoom_noNorm}
     \end{subfigure}
     \begin{subfigure}[b]{0.99\textwidth}
         \centering
         \includegraphics[width=\textwidth]{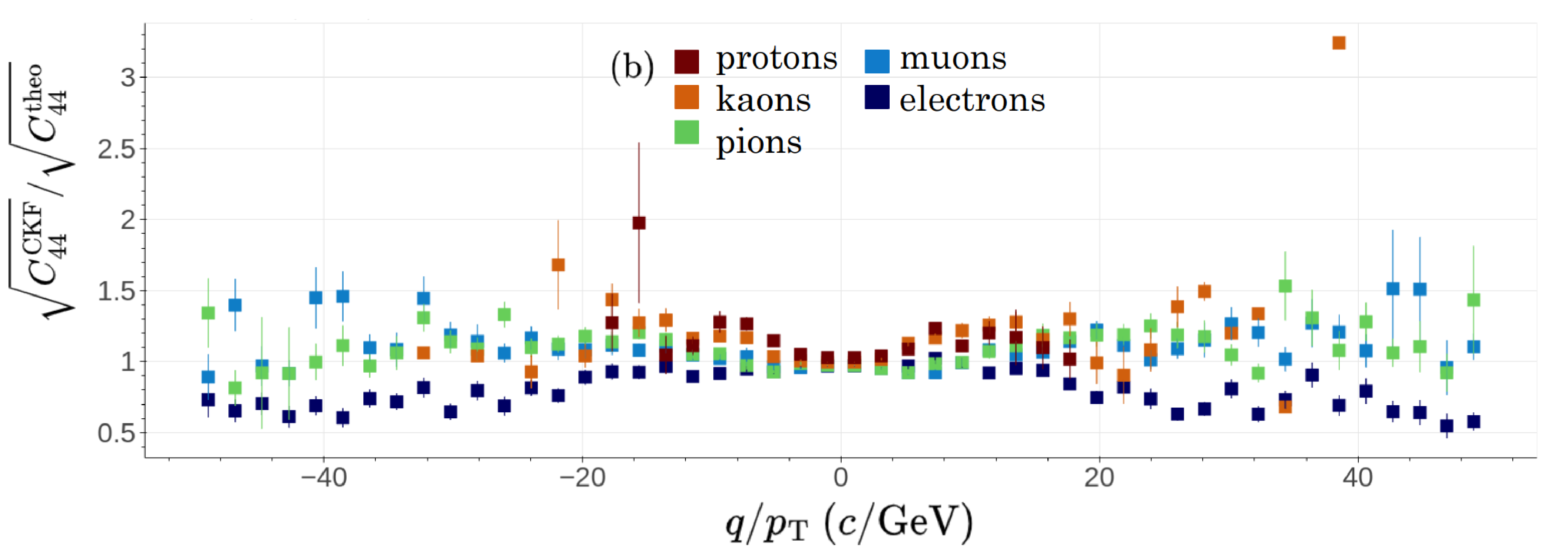}
         \caption{}
         \label{fig:CheckAna_pID_zoom_Norm}
     \end{subfigure}
        \caption{(a) \texttt{CKF} $q/p_{\text{T}}$ resolution $\sigma_{\text{CKF}}(q/p_{\text{T}})=\sqrt{C_{44}^{\textrm{CKF}}}$ as a function of the true $q/p_{\text{T}}$. (b) Ratio of the \texttt{CKF} $q/p_{\text{T}}$ resolution, over the theoretical expectations $\sigma_{\text{theo}}(q/p_\text{T})=\sqrt{C_{44}^{\textrm{theo}}}$, as a function of the true $q/p_\text{T}$. The histograms include all particles in the PS sample and are color-coded according to their particle type. Only tracks with a minimum of 10 points are considered. These plots have been produced using the interactive analytical tool \texttt{ROOTInteractive}~\cite{RootInt}. The error bars are statistical.}
        \label{fig:Check_Ana_pID_zoom}
\end{figure}

In Fig. \ref{fig:Check_Ana_pID_zoom} we show an alternative version of the plot in Fig.\ref{fig:Check_Ana_pID} which focuses on the more statistically rich central region.

\begin{figure}[!ht]
     \centering
     \begin{subfigure}[b]{0.46\textwidth}
         \centering
         \includegraphics[width=\textwidth]{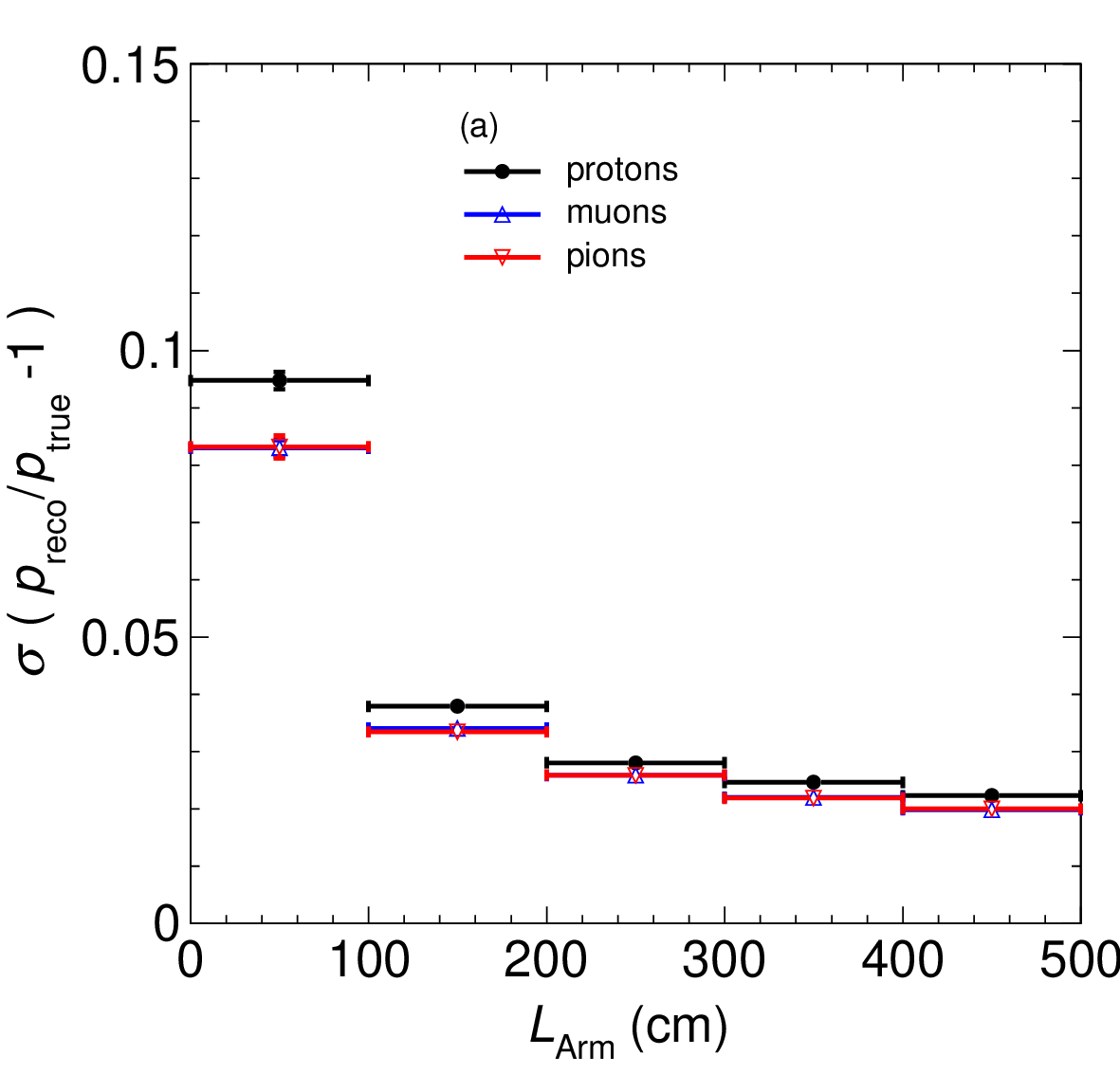}
         \caption{}
         \label{fig:ResND-GArVSLArm}
     \end{subfigure}
     \begin{subfigure}[b]{0.46\textwidth}
         \centering
         \includegraphics[width=\textwidth]{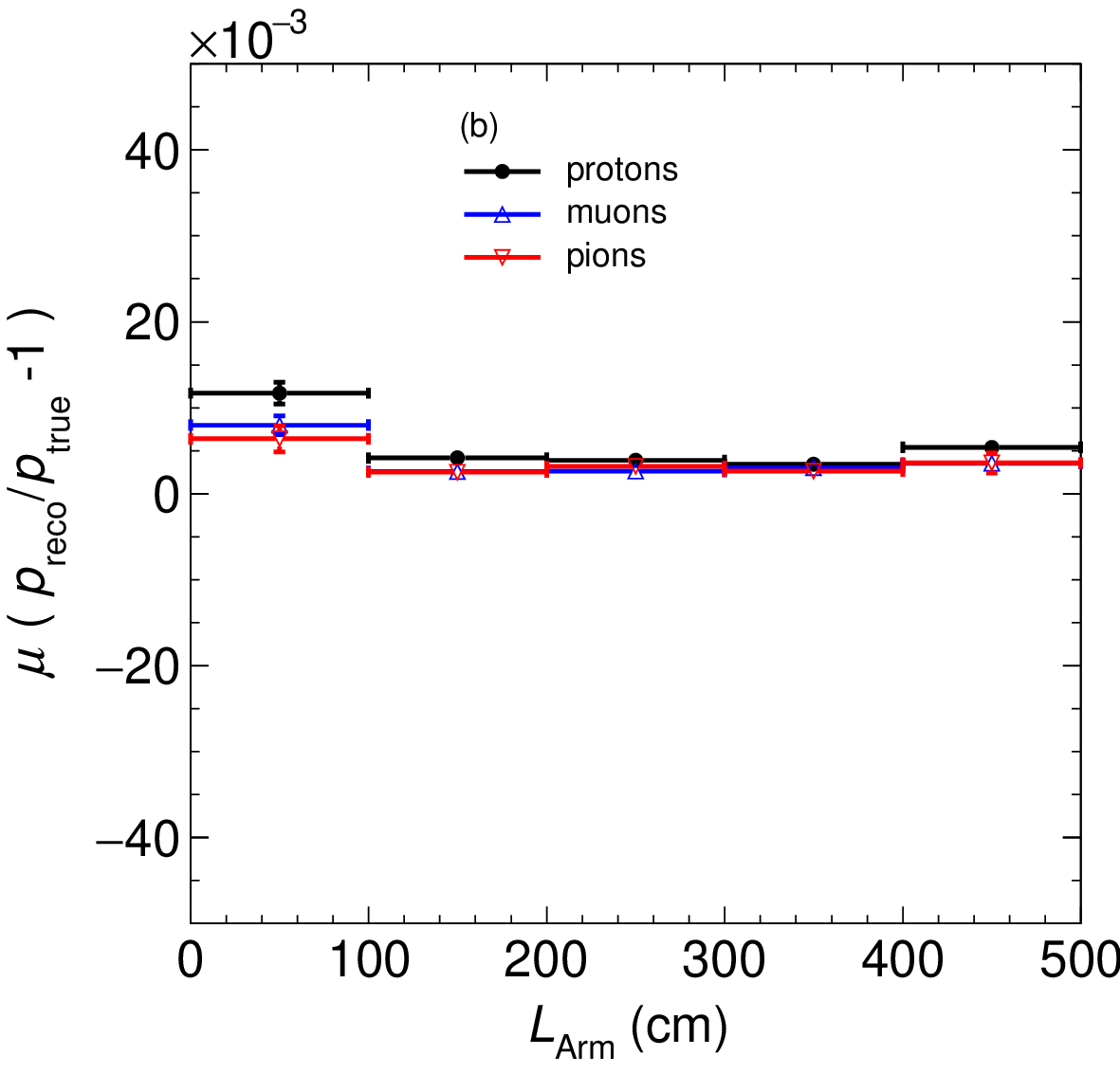}
         \caption{}
         \label{fig:BiasND-GArVSLArm}
     \end{subfigure}
        \caption{Relative momentum resolution (a) and bias (b) as function of the tracks' lever arm $L_\textrm{Arm}$ for the HP sample. The two properties are defined as $\mu$ and $\sigma$ of Gaussian fits of the momentum fractional residuals $p_{\text{reco}}/p_{\text{true}}-1$. The three particle types (protons, muons and pions) are drawn separately.}
        \label{fig:ND-GArVSLArm}
\end{figure}

\begin{figure}[!ht]
     \centering
     \begin{subfigure}[b]{0.46\textwidth}
         \centering
         \includegraphics[width=\textwidth]{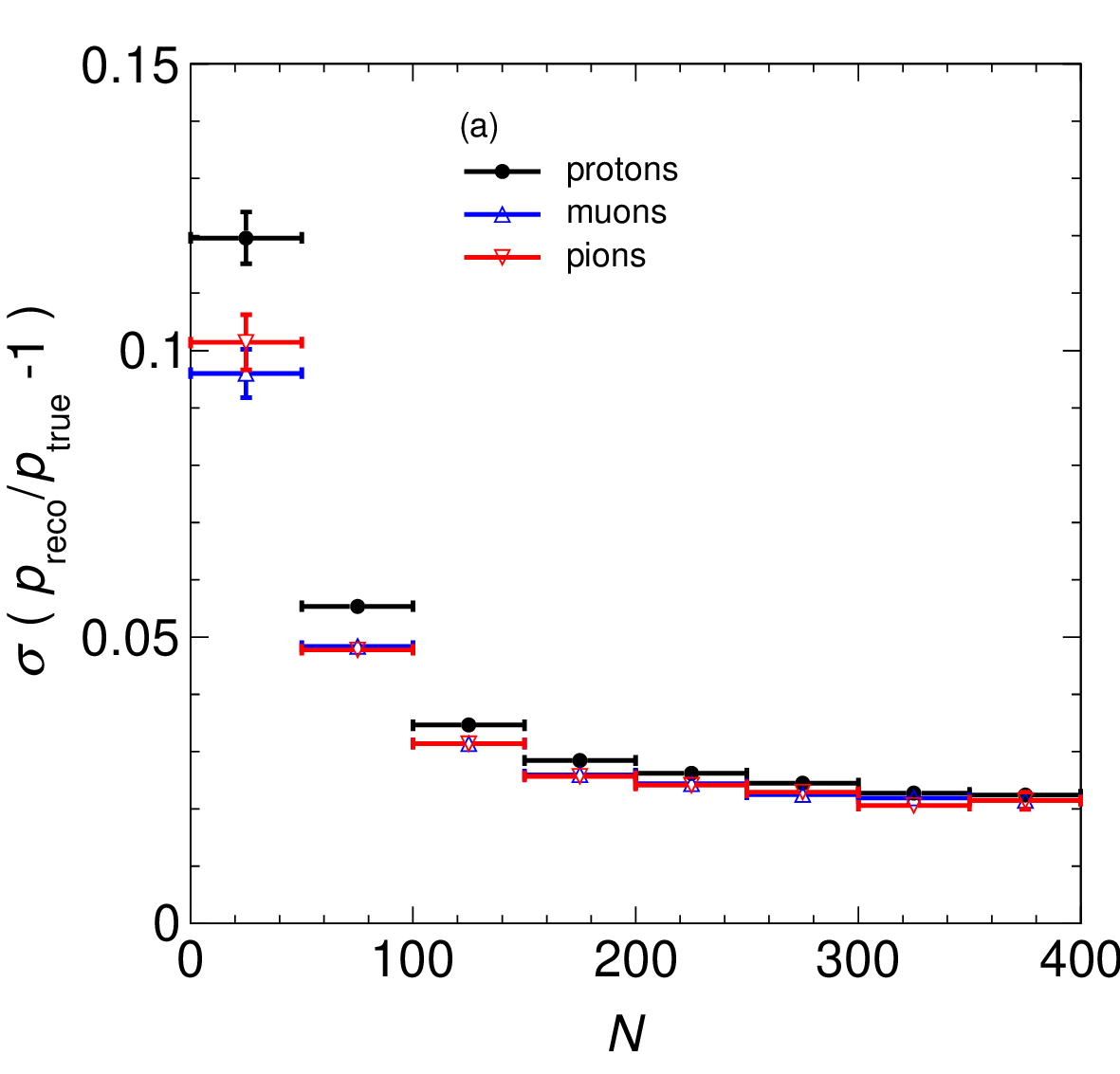}
         \caption{}
         \label{fig:ResND-GArVSNPoints}
     \end{subfigure}
     \begin{subfigure}[b]{0.46\textwidth}
         \centering
         \includegraphics[width=\textwidth]{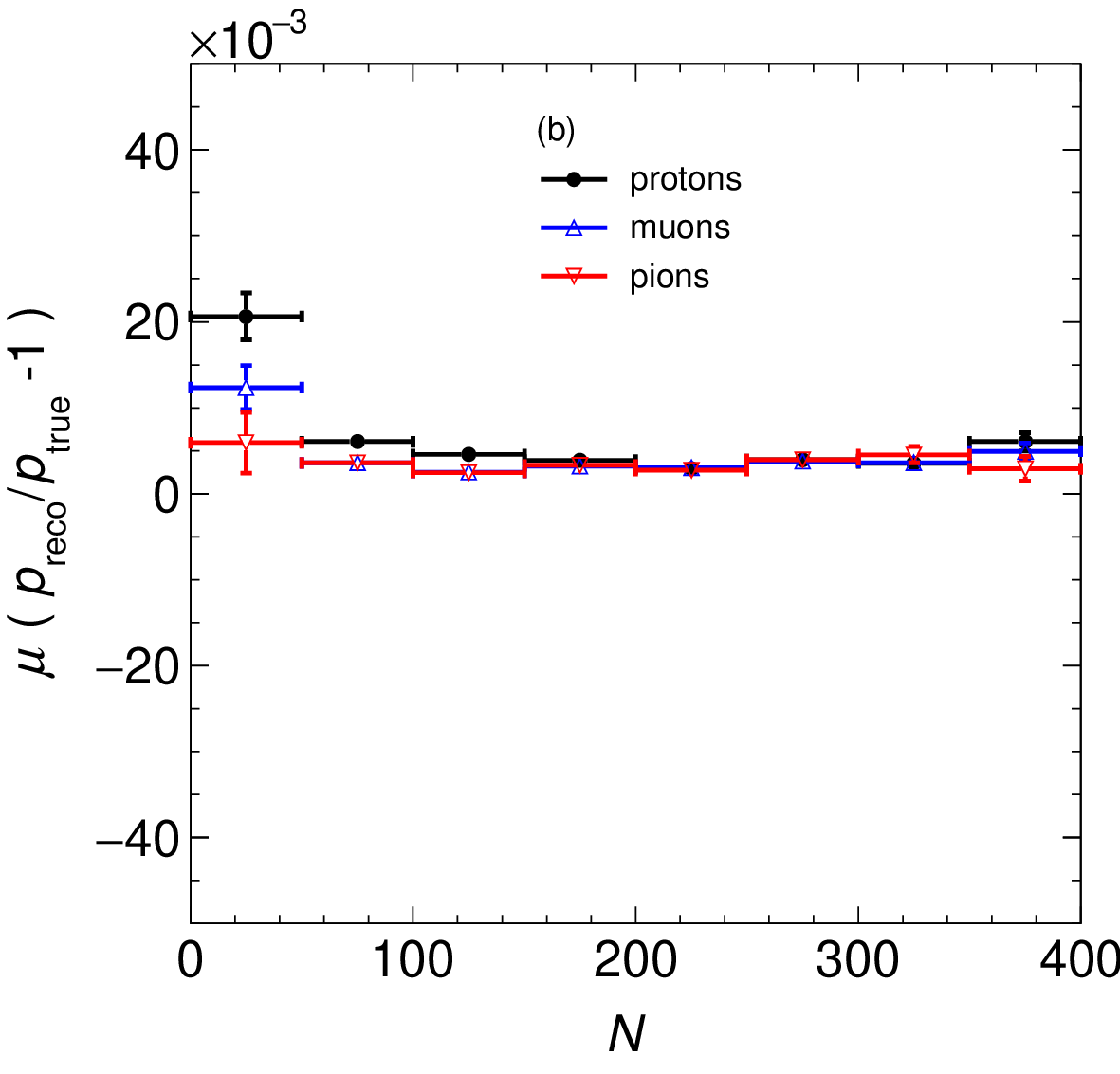}
         \caption{}
         \label{fig:BiasND-GArVSNPoints}
     \end{subfigure}
        \caption{Relative momentum resolution (a) and bias (b) as function of as a function of the number of points in the tracks $N$ for the HP sample. The two properties are defined as $\mu$ and $\sigma$ of Gaussian fits of the momentum fractional residuals $p_{\text{reco}}/p_{\text{true}}-1$. The three particle types (protons, muons and pions) are drawn separately.}
        \label{fig:ND-GArVSNPoints}
\end{figure}

\begin{figure}[!ht]
     \centering
     \begin{subfigure}[b]{0.46\textwidth}
         \centering
         \includegraphics[width=\textwidth]{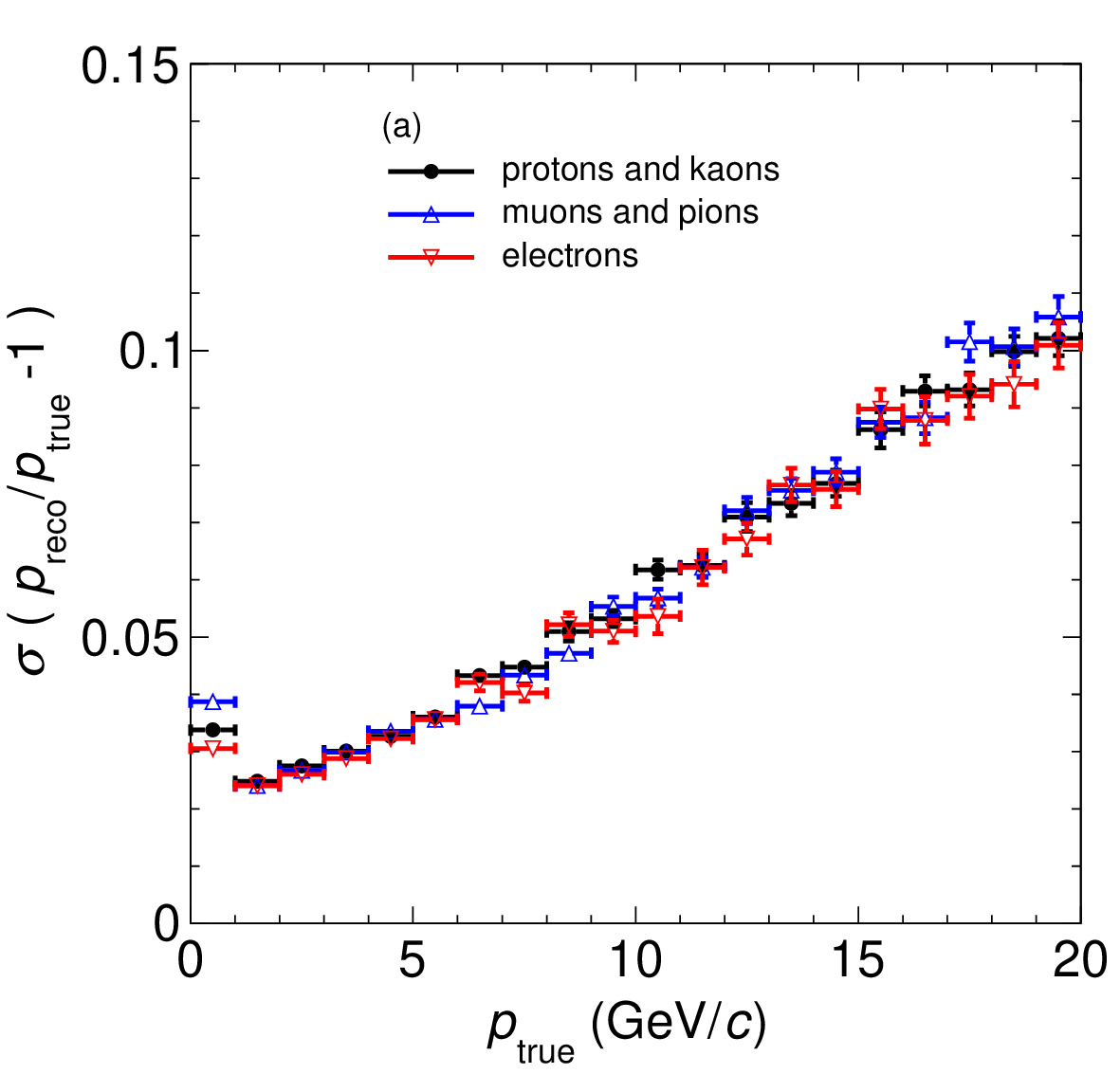}
         \caption{}
         \label{fig:ResPSVSp}
     \end{subfigure}
     \begin{subfigure}[b]{0.46\textwidth}
         \centering
         \includegraphics[width=\textwidth]{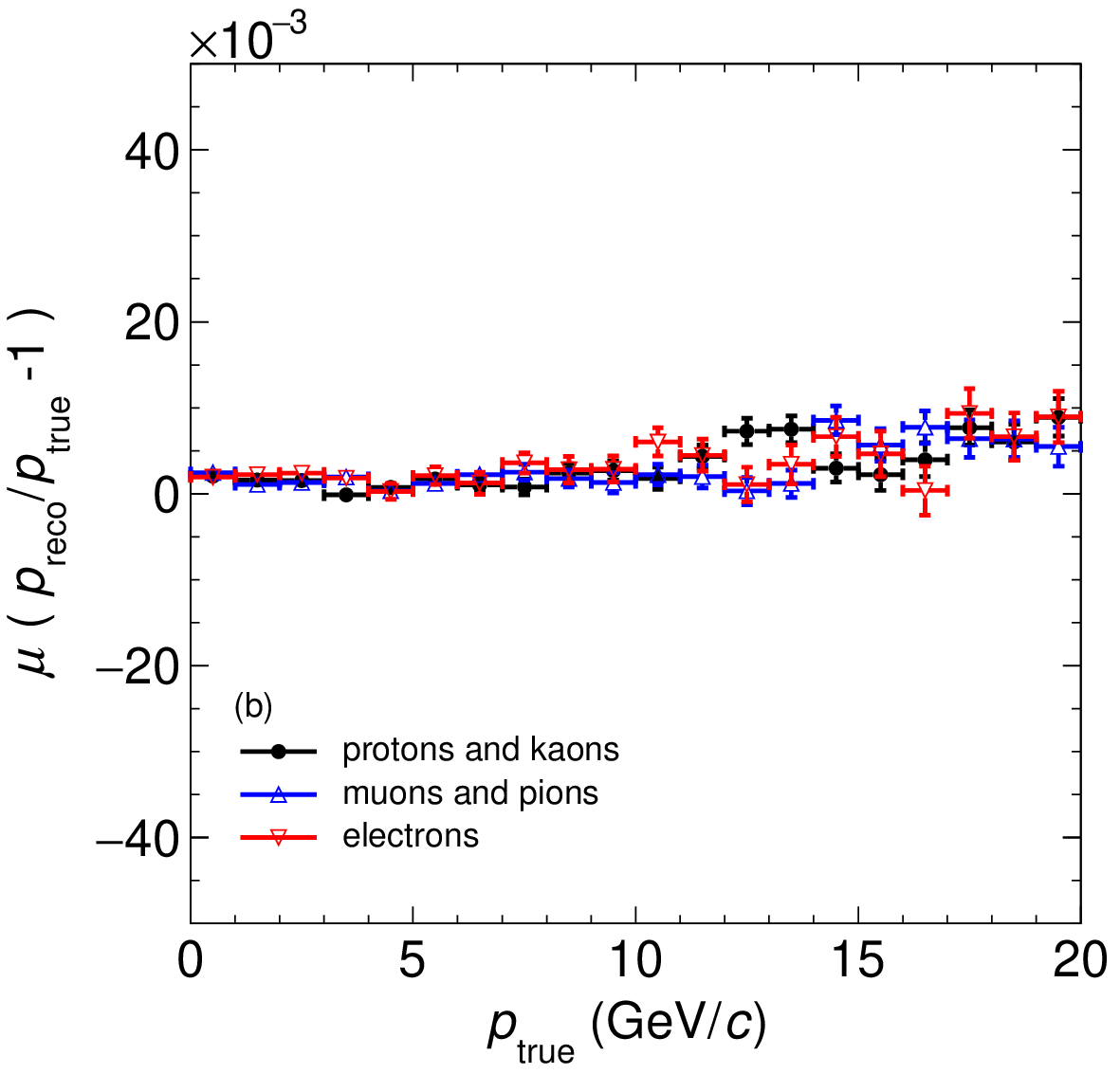}
         \caption{}
         \label{fig:BiasPSVSp}
     \end{subfigure}
        \caption{Relative momentum resolution (a) and bias (b) as function of the true momentum, $p_\textrm{true}$, for the PS sample. The two properties are defined as $\mu$ and $\sigma$ of Gaussian fits of the momentum fractional residuals $p_{\text{reco}}/p_{\text{true}}-1$. The particle types are divided based on their mass (protons and kaons, muons and pions, electrons) are drawn separately.}
        \label{fig:PSVSp}
\end{figure}

\begin{figure}[!ht]
     \centering
     \begin{subfigure}[b]{0.46\textwidth}
         \centering
         \includegraphics[width=\textwidth]{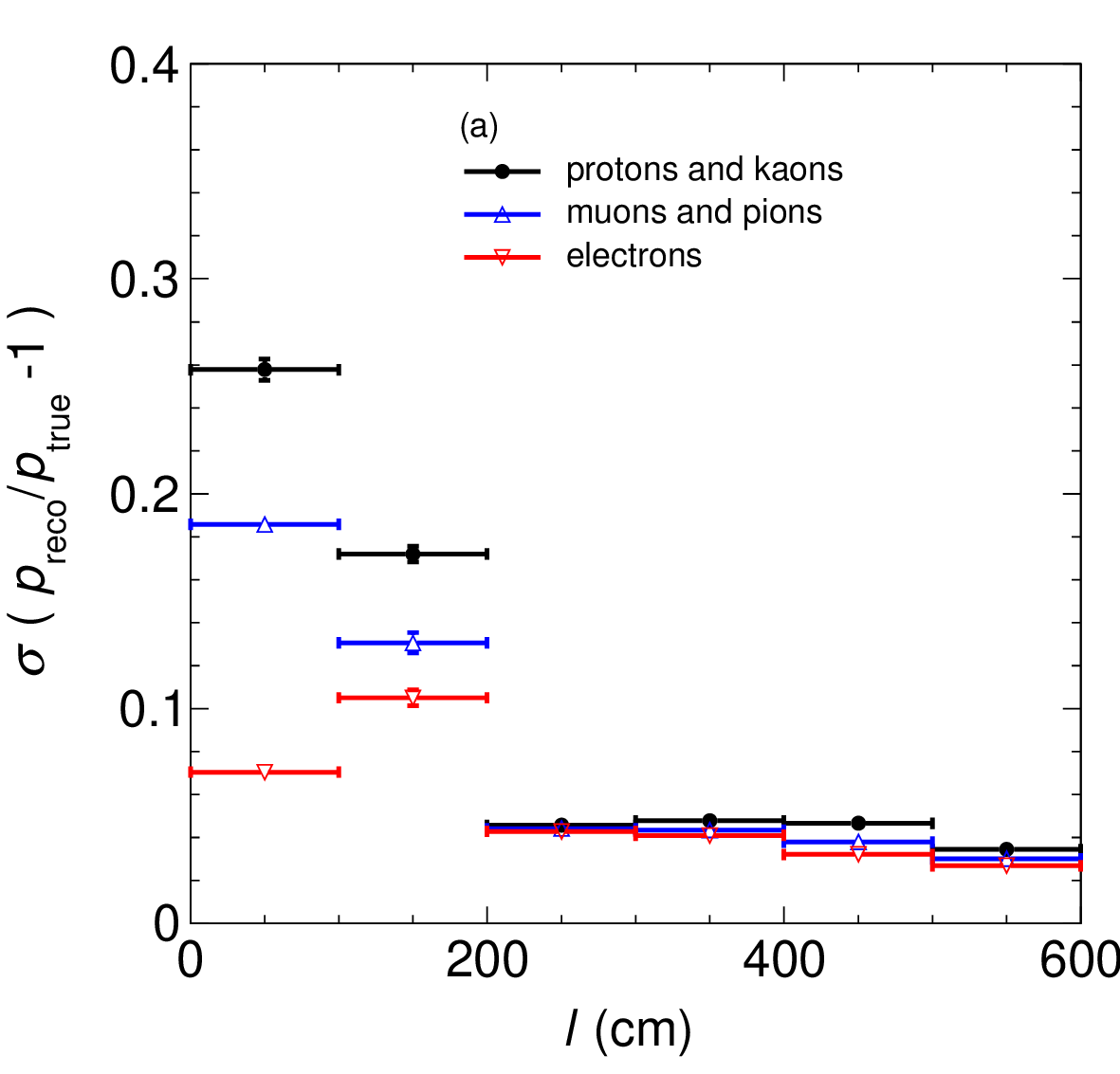}
         \caption{}
         \label{fig:ResPSVSl}
     \end{subfigure}
     \begin{subfigure}[b]{0.46\textwidth}
         \centering
         \includegraphics[width=\textwidth]{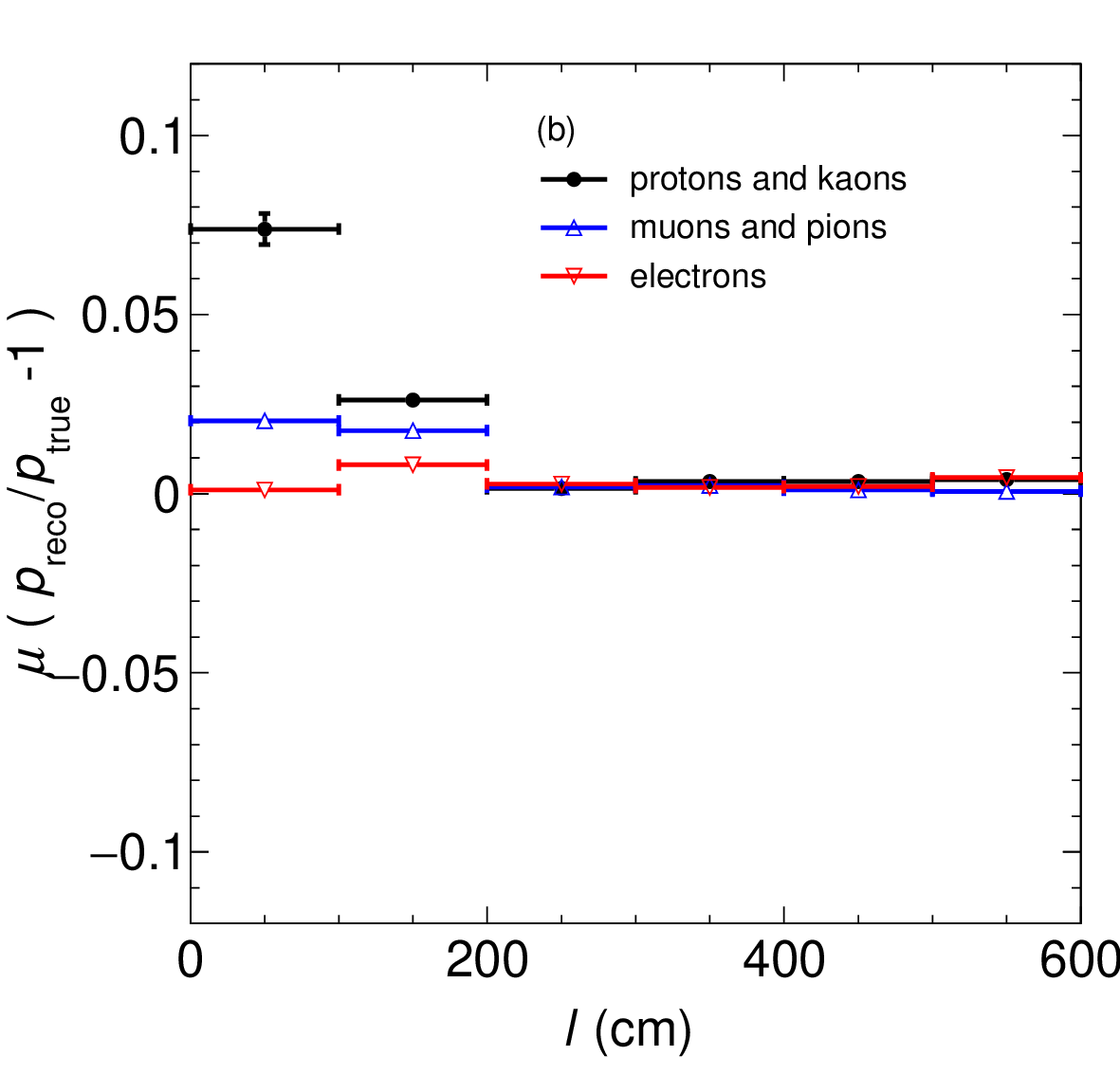}
         \caption{}
         \label{fig:BiasPSVSl}
     \end{subfigure}
        \caption{Relative momentum resolution (a) and bias (b) as function of the track length, $l$, for the PS sample. The two properties are defined as $\mu$ and $\sigma$ of Gaussian fits of the momentum fractional residuals $p_{\text{reco}}/p_{\text{true}}-1$. The particle types are divided based on their mass (protons and kaons, muons and pions, electrons) are drawn separately.}
        \label{fig:PSVSl}
\end{figure}

In Figs.~\ref{fig:ND-GArVSLArm} and~\ref{fig:ND-GArVSNPoints}, we show the relative momentum resolution and bias for the three particle types present in the HP sample, as a function of $L_\textrm{Arm}$ and $N$, respectively. These are analogous to Fig.~\ref{fig:ND-GArVSlength} in the main text. Similar plots are produced for the PS sample to allow for a direct comparison. In Figs.~\ref{fig:PSVSp} and~\ref{fig:PSVSl}, we show the relative momentum resolution and bias for the particles in the PS sample as a function of $p$ and $l$ respectively. The particle types are divided by mass into protons and kaons, pions and muons and finally electrons. Note that in this case the resolution and material properties are not uniform in the sample.

\clearpage

\end{document}